\documentclass[manuscript]{aastex}
\usepackage{natbib}
\usepackage{lineno}
\usepackage[figuresright]{rotating}
\usepackage{lscape}

\shorttitle{Evolving spectral  properties of  type-I X-ray bursts of 4U~1608--52 in 2022}
\shortauthors{Chen et al.}

\begin{document}


\title{Insight-HXMT observations on thermonuclear X-ray bursts from 4U~1608--52 in 2022:
the accretion rate dependent anisotropy of burst emission}

\author{Yu-Peng Chen\textsuperscript{*}}
\email{chenyp@ihep.ac.cn}
\affil{Key Laboratory for Particle Astrophysics, Institute of High Energy Physics, Chinese Academy of Sciences, 19B Yuquan Road, Beijing 100049, China}

\author{Shu Zhang\textsuperscript{*}}
\email{szhang@ihep.ac.cn}
\affil{Key Laboratory for Particle Astrophysics, Institute of High Energy Physics, Chinese Academy of Sciences, 19B Yuquan Road, Beijing 100049, China}

\author{Long Ji\textsuperscript{*}}
\email{jilong@mail.sysu.edu.cn}
\affil{School of Physics and Astronomy, Sun Yat-Sen University, Zhuhai, 519082, China}

\author{Shuang-Nan Zhang}
\affil{Key Laboratory for Particle Astrophysics, Institute of High Energy Physics, Chinese Academy of Sciences, 19B Yuquan Road, Beijing 100049, China}
\affil{University of Chinese Academy of Sciences, Chinese Academy of Sciences, Beijing 100049, China}

\author{Peng-Ju Wang}
\affil{Key Laboratory for Particle Astrophysics, Institute of High Energy Physics, Chinese Academy of Sciences, 19B Yuquan Road, Beijing 100049, China}
\affil{University of Chinese Academy of Sciences, Chinese Academy of Sciences, Beijing 100049, China}

\author{Ling-Da Kong}
\affil{Key Laboratory for Particle Astrophysics, Institute of High Energy Physics, Chinese Academy of Sciences, 19B Yuquan Road, Beijing 100049, China}
\affil{ Institut f\"{u}r Astronomie und Astrophysik, Kepler Center for Astro and Particle Physics, Eberhard Karls Universit\"{a}t, Sand 1, D-72076 T\"{u}bingen, Germany}

\author{Zhi Chang}
\affil{Key Laboratory for Particle Astrophysics, Institute of High Energy Physics, Chinese Academy of Sciences, 19B Yuquan Road, Beijing 100049, China}

\author{Jing-Qiang Peng}
\affil{Key Laboratory for Particle Astrophysics, Institute of High Energy Physics, Chinese Academy of Sciences, 19B Yuquan Road, Beijing 100049, China}
\affil{University of Chinese Academy of Sciences, Chinese Academy of Sciences, Beijing 100049, China}

\author{Qing-Cang Shui}
\affil{Key Laboratory for Particle Astrophysics, Institute of High Energy Physics, Chinese Academy of Sciences, 19B Yuquan Road, Beijing 100049, China}
\affil{University of Chinese Academy of Sciences, Chinese Academy of Sciences, Beijing 100049, China}

\author{Jian Li}
\affil{CAS Key Laboratory for Research in Galaxies and Cosmology, Department of Astronomy, University of Science and Technology of China, Hefei 230026, China}
\affil{School of Astronomy and Space Science, University of Science and Technology of China, Hefei 230026, China}

\author{Lian Tao}
\affil{Key Laboratory for Particle Astrophysics, Institute of High Energy Physics, Chinese Academy of Sciences, 19B Yuquan Road, Beijing 100049, China}

\author{Ming-Yu Ge}
\affil{Key Laboratory for Particle Astrophysics, Institute of High Energy Physics, Chinese Academy of Sciences, 19B Yuquan Road, Beijing 100049, China}

\author{Jin-Lu Qu}
\affil{Key Laboratory for Particle Astrophysics, Institute of High Energy Physics, Chinese Academy of Sciences, 19B Yuquan Road, Beijing 100049, China}
\affil{University of Chinese Academy of Sciences, Chinese Academy of Sciences, Beijing 100049, China}

\begin{abstract}

Thermonuclear X-ray bursts occur on the surface of an accreting neutron star (NS), and their characteristics and interplay with the surrounding circumstance could be a clue to understand the nature of the NS and accretion process.
For this purpose, Insight-HXMT has performed high cadence observations on the bright thermonuclear X-ray burster--4U~1608--52 during its outburst in July and August 2022; nine bursts were detected, including  
seven bursts with the photospheric radius expansion (PRE).
Time-resolved spectroscopy of the bright PRE bursts reveals that an enhancement of accretion rate or the Comptonization of the burst emission by the corona could reduce the residuals when fitting their spectra with the conventional model--blackbody.
The inferred energy increment rate of the burst photon gained from the corona is up to $\sim$40\%, even though the bursts have different peak fluxes and locate at different accretion rates.
Moreover, the flux shortage of the rising PRE is observed in the bursts
at a high mass accretion rate, but not for the burst with a faint persistent emission, which has been predicted theoretically but first observed in this work.  If the flux shortage is due to the disk obscuration, i.e., the burst emission is anisotropic, the phenomenon above could indicate that the anisotropy of the burst emission is accretion rate dependent, which could also be  evidence of the truncated disk in the low/hard state.

\end{abstract}
\keywords{stars: coronae ---
stars: neutron --- X-rays: individual (4U~1608--52) --- X-rays: binaries --- X-rays: bursts}

\section{Introduction}

In the lightcurves of a low-mass X-ray binary (LMXB) hosting a neutron star (NS), some spikes emerge with peak fluxes up to the Eddington luminosity and their timescale is seconds to minutes.
These events are named type-I X-ray bursts, or thermonuclear X-ray bursts because of the spectral softening during the decay phase and the $\alpha$ parameter (the fluence ratio of the burst and the outburst) which is consistent with the range of the relative efficiency of accretion and thermonuclear burning (for reviews, see \citealp{Lewin,Cumming,Strohmayer,Galloway}).
Since their discovery, the bursts have been one of the main observational targets for X-ray telescopes, e.g. RXTE, NICER, and Insight-HXMT, since their characteristics and the interplay with the surrounding circumstance could be a clue to understand the nature of the neutron star (NS) and the accretion process, e.g., taking the burst oscillation as the NS's spin, the flux of the touchdown time as the Eddington luminosity to constrain the NS's mass and radius, the enhancement of the persistent emission at soft X-ray band \citep{Worpel2013,Ji2014,Worpel2015,Bult2021}, the deficit of the persistent emission in the hard X-ray band \citep{maccarone2003, Chen2012, Ji2013, Chen2022} and a bump peaking at 20--40 keV and/or discrete emission \citep{int2013, Ball2004, Keek2014,Degenaar2018} to derive the influence of the burst emission on the accretion process.

 Several models are adopted to explain the enhancement of at soft X-ray band above, including  accretion rate increase \citep{1989ApJ...346..844W,1992ApJ...385..642W} by Poynting-Robertson effect \citep{1937MNRAS..97..423R, 1974ApJ...188..121B} ($f_{\rm a}$ model), the reflection of the burst emission by the accretion disk \citep{Ball2004,2022MNRAS.509.1736S} and double-blackbody emission \citep{Kashyap2022}. However, it is inconsistent with the deficit at hard X-ray band, since the soft X-ray excess and the hard X-ray deficit were simultaneously detected in some bursters \citep{Sanchez2020,2022MNRAS.512.6180K}.  The traditional models above are under the assumption that the persistent emission does not vary during an X-ray bursts. Several other models breaks the traditions which include the Comptonization of burst emission by the corona, both the corona temperature and the normalization changing \citep{Sanchez2020}.
However, given the quality of the data, no model above could be statistically favoured over the others.

The accretion circumstance could also have impacts on the burst emission, e.g., the obscuration of the burst emission by the disk \citep{Shaposhnikov2003,Shaposhnikov2004}, the Comptonization of the burst emission by the corona \citep{Chen2022a,Chen2022b}.
  The impacts above can be noticed more easily during the most luminous bursts--bursts with photospheric radius expansion (PRE):
the radiation pressure of the thermonuclear burning exceeds the NS gravitational force in the photosphere, resulting in an increase of the photosphere radius and a decrease of
 the photosphere temperature in an adiabatic expansion \citep{Grindlay1980}.

However, comparing the burst influence on accretion circumstance,
the consequences of the burst emission by the accretion circumstance are not well studied, e.g., the anisotropy of the burst emission is expected to be accretion state dependent, since the accretion disk in the low/hard state is believed to be truncated at a larger distance; however, there is no observational evidence for the accretion state dependent anisotropy of burst emission \citep{Galloway2021}.

 The emergent photons of bursts  suffer Compton scattering by the electrons of the NS photoshere,
(e.g., \citealp{London1986}),
resulting in the harden of the burst spectra by a factor of the color-correction factor $f_{\rm c}$.
 $f_{\rm c}$ is defined as the ratio of the color (observed) temperature to the effective temperature of the photosphere
 \begin{eqnarray}
 f_{\rm c}\equiv~T_{\rm bb}/T_{\rm eff},
 \label{color_factor}
\end{eqnarray}
which depends on the chemical compositions and is a strong function of the luminosity \citep{Titarchuk1994}.
Meanwhile, the apparent NS radius is caused to be smaller,
 \begin{eqnarray}
R_{\rm bb}=\frac{R_{\rm eff}}{f_{\rm c}^{2}},
 \label{R_bb}
\end{eqnarray}
%
 here $R_{bb}$ is the apparent (observed) radius of the photosphere,
$R_{\rm eff}$ is the effective radius of the photosphere ($R_{\rm eff}=R_{*}(1+z)$, $R_{*}$ is the photosphere radius observed at the NS surface);
The color temperature $T_{\rm bb}$ is typically higher than the effective temperature $T_{\rm eff}$ by the color-correction factor $f_{\rm c}$ ($T_{\rm eff}=T_{*}/(1+z)$ for the NS photosphere emission, $T_{*}$ is the NS photosphere temperature observed at the NS surface) \citep{Rybicki2004}.
The subscripts $\infty$ and ${\rm bb}$ denote that these values are detected by a distant observer, $*$ denotes that these values are detected by a local (at the NS photosphere) observer.


In theory, for the NS surface emission, according to the theory of general relativity,
the gravitational force is stronger by a factor of $(1+z)$,
and the observed emergent spectra of the photosphere are redshifted by a factor of (1+$z$)$^2$, i.e., the observed luminosity by a distant observer is diluted by a factor of $1/(1+z)$,
 \begin{eqnarray}
1+z = [1-2GM/(R_{*}c^{2})]^{-1/2}\simeq[1-0.416/r_{6}]^{-1/2},
  \label{redshift}
\end{eqnarray}
%
where $M$ is the gravitational mass of the neutron star, and $R_{*}$ is its radius as measured by a local observer on the NS surface; for $M=1.4M_{\odot}$, the equation is reducible to the left side of the equation above with $r_{6}=R_{*}$/(10 km).
Some bursts have luminosities reaching the Eddington limit, at which point radiation pressure exceeds the gravitational force and lifts the surface layer--photosphere from the star in the photospheric radius expansion (PRE) stage.
For isotropic radiation of the pure helium burning at the NS surface, e.g., at just the lift-up time and the touch-down time of the photosphere, the observed bolometric flux should be the same.
In the PRE phase, the flux observed should be higher than the above two values due to the redshift correction,
 e.g., the observed bolometric flux with $R_{*}$=30 km is $\sim$20\% higher than that with $R_{*}$=10 km.
In the observations of the time-resolved spectra of the PRE bursts \citep{Galloway,Galloway2008a}, the vast majority of the peak flux reach a (local) maximum close to the time of peak radius, which does follow the equation above.
However, in the bright PRE burst of 4U~1608--52 in 2020 detected by Insight-HXMT,
the bolometric flux of the just lift-up time and the touch-down time is different and the peak flux is not at the time when the photosphere reached its peak radius.
The flux shortage during the rising PRE phase is believed to correspond to the obscured lower part of the NS surface by the disk   \citep{Chen2022a},  and the similar phenomenon was first reported in 4U~1728--34 \citep{Shaposhnikov2003} and then 4U~1820--30 \citep{Shaposhnikov2004}. For the outburst of 4U~1608--52 in July and August 2022, Insight-HXMT performed high cadence observations on the source aiming to find more bursts to investigate the interaction between the burst and persistent emission, as well as the anisotropic degrees of the burst emission in different accretion rates.

Among thousands of observed bursts from the 118 bursters\footnote{https://personal.sron.nl/$\sim$jeanz/bursterlist.html},
apart from 4U~1636--536, 4U~1728--34 (Slow Burster) and Aql~X--1, 4U~1608--52 performs to be active every several months, and have $\sim$ 150 type-I X-ray bursts superposing on the outbursts in recent 20 years \citep{Galloway2020}; the inferred burst rate is $\sim$ 2 bursts per day.
Among these bursts, 1/3 of them are PRE bursts with a mean peak flux $\sim$1.7 $\times10^{-7}~{\rm erg}~{\rm cm}^{2}~{\rm s}^{-1}$ (e.g. \citealp{Galloway,Poutanen}), and the derived distance is $D\sim$2.9--4.5 kpc.
Its spin is around $\nu$=619 Hz \citep{Muno,Galloway}, based on the burst oscillation detection.

In this present investigation, using a broad energy band capabilities of Insight-HXMT in 1--100 keV, 
we provide a broad-band spectral
view of 4U~1608--52 during its 2022 outburst observed by 
Insight-HXMT, both for its outburst and burst emission.
We study nine bursts from 4U~1608--52, including seven PRE bursts. 
We first describe the data reduction procedure of 
 Insight-HXMT in Section 2. We then present an
in-depth spectral analysis and model parameters of its outburst emission in Section 3.1, burst lightcurves and spectral evolution in Section 3.2 and Section 3.3.
 Finally, we summarize our results and discuss their implications in Section 4 and give a summary in Section 5.

\section{Observations and Data Reduction}

Insight-HXMT was launched on the 15th of June 2017, which excels in its broad energy band (1--250 keV), large effective area in the hard X-rays energy band and little pile-up for bright sources (up to several Crab) \citep{Zhang2020}.
There are three main payloads, and all of them are collimated telescopes: the High Energy X-ray Telescope (HE; phoswitch NaI/CsI, 20–250 keV, $\sim$ 5000 cm$^2$), the Medium Energy X-ray Telescope (ME; Si pin detector, 5–40 keV, 952 cm$^2$) and the Low Energy X-ray telescope (LE; SCD detector, 1–12 keV, 384 cm$^2$).
For the three main payloads of Insight-HXMT, each has two main field of views (FoVs), i.e., LE: 1.6$^{\circ}\times6^{\circ}$ and 6$^{\circ}\times6^{\circ}$, ME: 1$^{\circ}\times4^{\circ}$ and 4$^{\circ}\times4^{\circ}$, HE: 5.7$^{\circ}\times1.1^{\circ}$ and 5.7$^{\circ}\times5.7^{\circ}$.
Moreover, they also have the blind FoV (fully blocked) detectors, which are used for the background estimation and energy calibration.
Under the quick read-out system of Insight-HXMT detectors, there is little pile-up effect at the burst peak, e.g.,
the fraction of pileup events of Insight-HXMT/LE is $<$ 1\% for the brightest sources such as Sco~X--1 with a maximum count rate $\sim$ 18000 cts/s ($>$ 10 Crab at 1--10 keV).
The dead times for ME and HE are $<$ 260 us and $<$ 10 us, respectively.
Insight-HXMT Data Analysis software (HXMTDAS) v2.06\footnote{http://hxmtweb.ihep.ac.cn/} is used to analyze the data.

\begin{figure}[t]
\centering
\includegraphics[angle=0, scale=0.8]{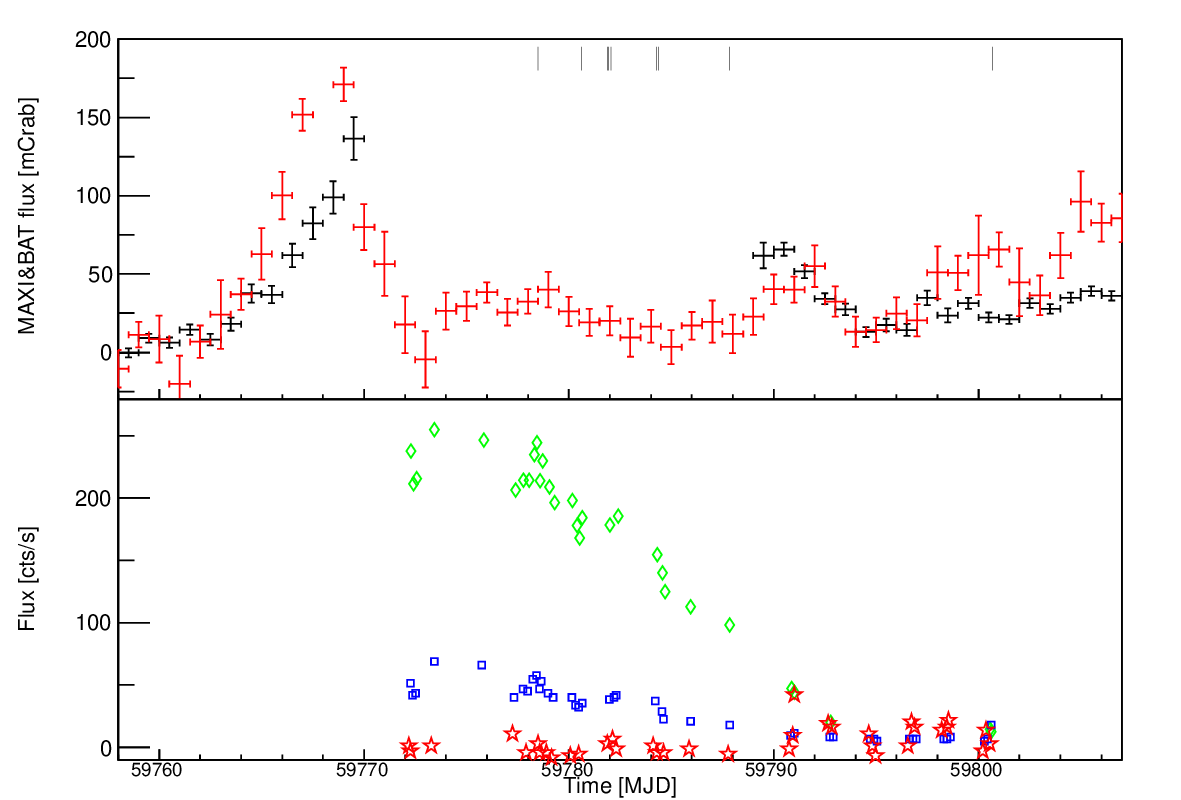}
  \caption{Top panel: daily lightcurves of 4U~1608--52 by MAXI (black, 2--20 keV) and Swift/BAT (red, 15--50 keV) during the  outburst  in  2022. The  bursts are indicated by vertical lines. Bottom panel: lightcurves of 4U~1608--52 by LE (green), ME (blue), and HE (red) which are rebinned by one obsid ($\sim$ 10000 s). Please note the error bars of the lightcurves are smaller than the size of the symbols.
  }
\label{fig_outburst_lc}
\end{figure}

As shown in Figure \ref{fig_outburst_lc}, for the outbursts in 2022,
Insight-HXMT observed 4U~1608--52 with 44 observations ranging from P040420700101-20220712-01-01 to P040420701503-20220809-01-01 with a total observation time of 432 ks.
These observations covered the peak/decay phase of the outburst in 2022, but missed the rising phase which is likely to be a low/hard state because of its high flux in hard X-ray band.

To avoid missing any bursts, lightcurves are extracted without filtering GTIs, as our previous procedures adopted in MAXI~J1816--195 \citep{Chen2022} and 4U~1730--22 \citep{Chen2022b}, since the GTI selection criteria of LE and ME are very conservative because of the influence of light leaks or the constraint of the background model.
The other results, e.g, the persistent spectra, background and net lightcurves are obtained following the recommended procedure of the Insight-HXMT Data Reduction, which are screened with the standard criterion included in Insight-HXMT pipelines: lepipeline, mepipeline and hepipeline.

 One may suspect that the bursts could be from a near-by source, since the fields of view of the LE, ME and HE are fairly large.
LE and ME each contains three independent detector boxes, each with an elongated rectangle FOV and different orientations. In the pointing observations of 4U~1608--52, the source was put at the center of
the FOVs. If bursts were from a nearby source, which is off-center, the fluxes detected by the three boxes should be different.
In this work, for each burst, the roughly same count rates in three boxes of LE or ME, and the roughly same values of the normalization constants derived from the spectral fitting denote that the emission is its X-ray's origin and  should originate from 4U~1608--52.

\begin{figure}[t]
\centering
\includegraphics[angle=0, scale=0.8]{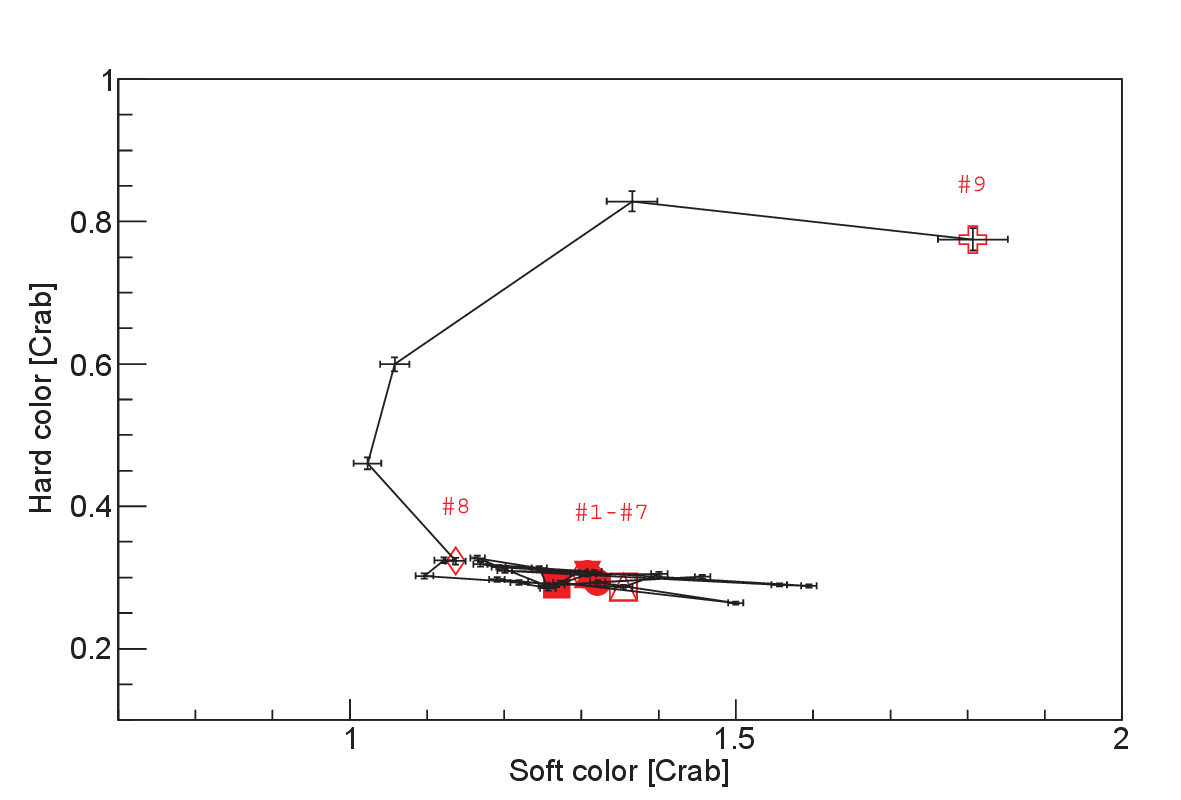}
  \caption{The color--color diagram (CCD) of this outburst. The soft color is defined as the ratio of count rate in 6--10 keV \& 1--6 keV of LE and the hard color is defined as the ratio of count rate in 10--15 keV \& 15--30 keV of ME. Both of them are normalized by the count rates of Crab at the same energy bands and in the same period.  The outburst moves in a clockwise direction: begins at the bottom right  and stops at the top right.}
\label{fig_outburst_ccd}
\end{figure}

\begin{figure}[t]
\centering
\includegraphics[angle=0, scale=0.3]{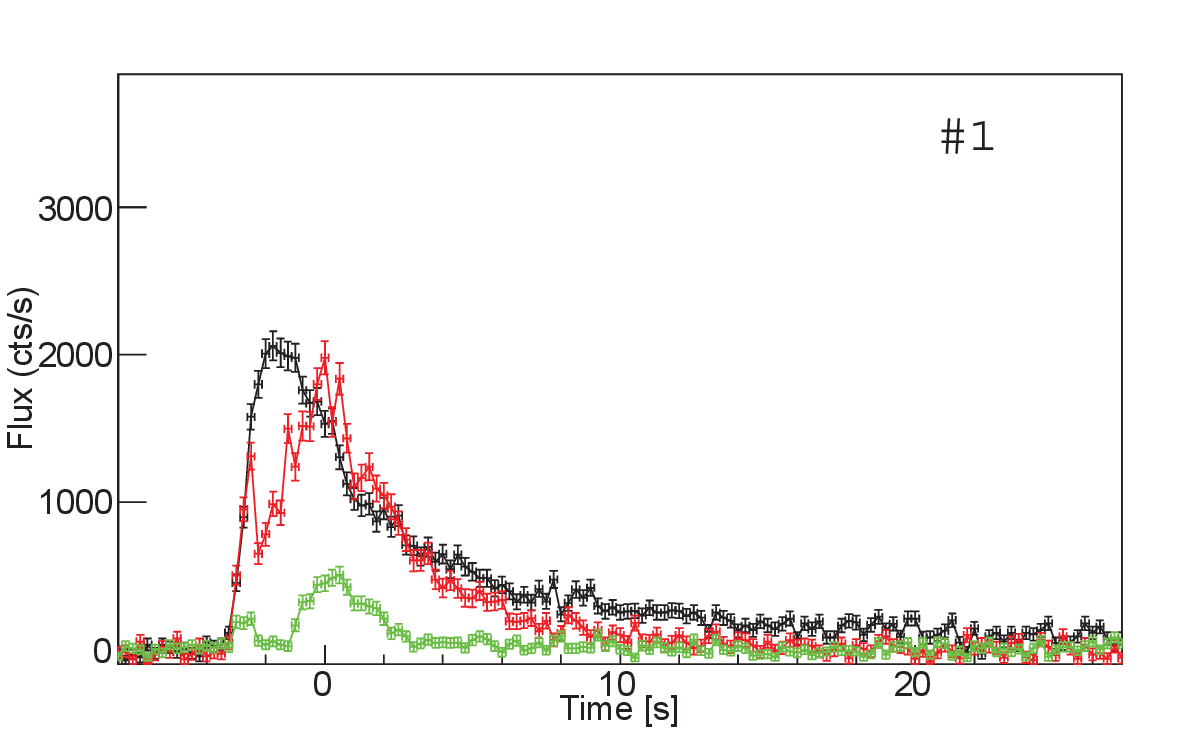}
\includegraphics[angle=0, scale=0.3]{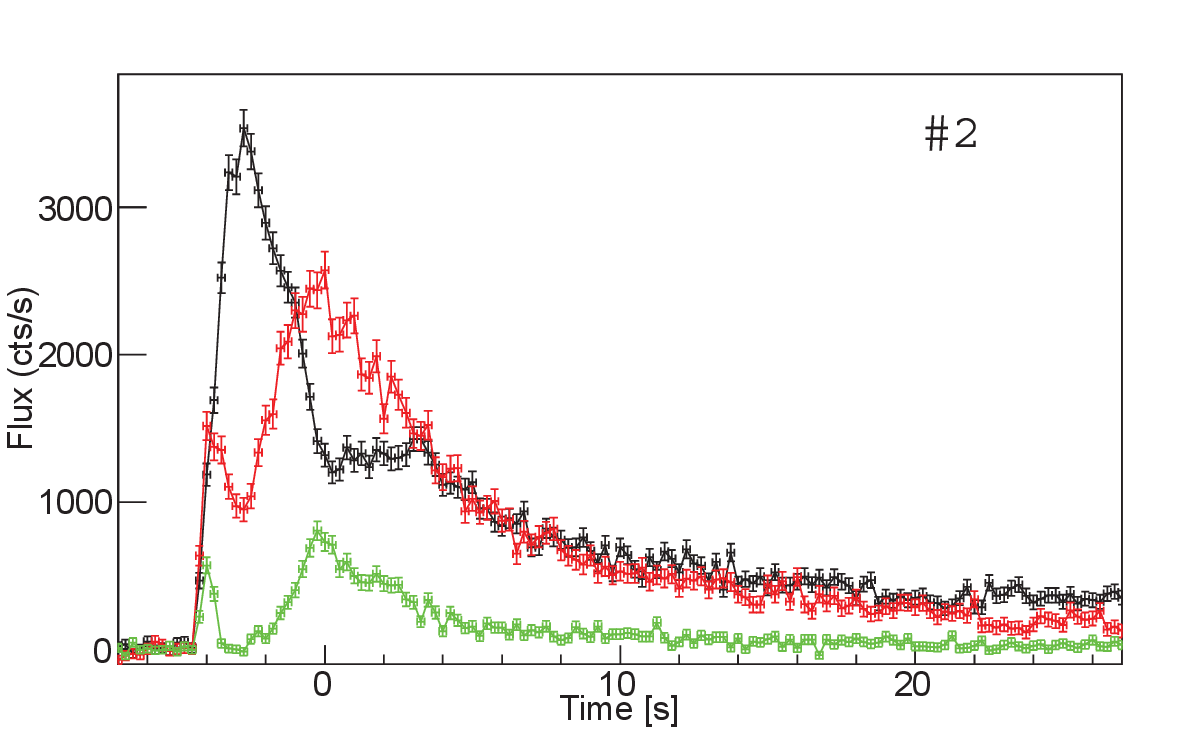}
\includegraphics[angle=0, scale=0.3]{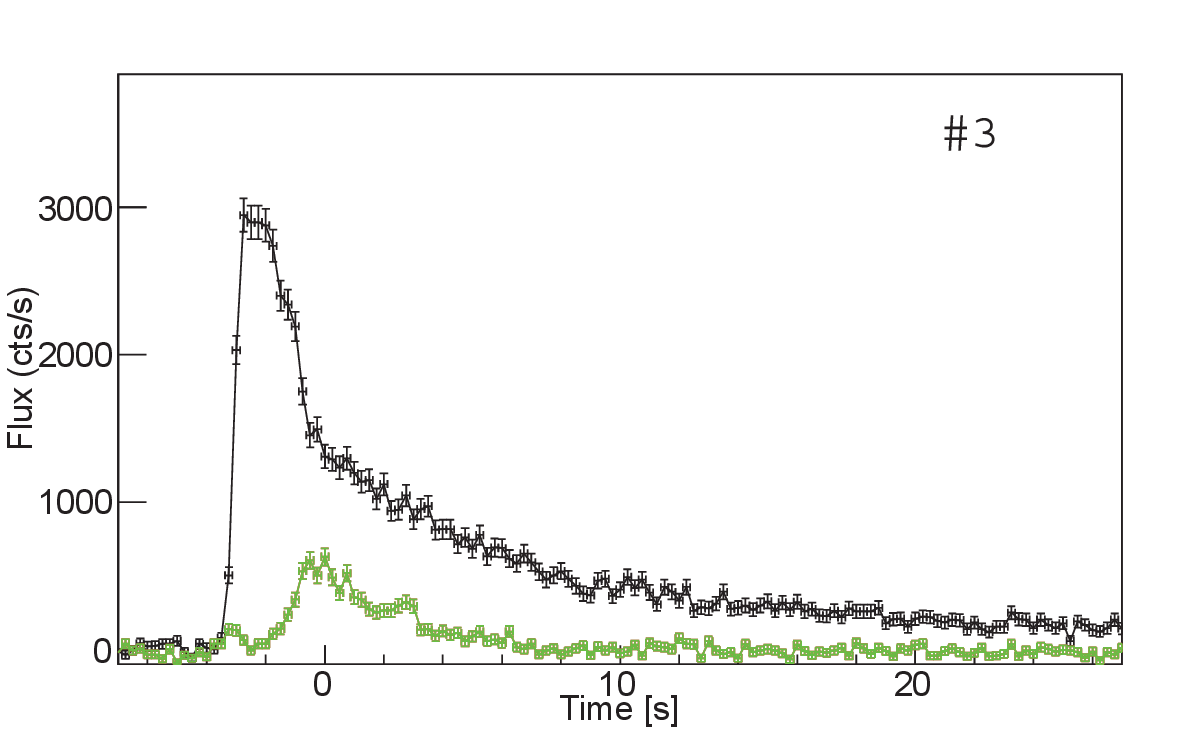}
\includegraphics[angle=0, scale=0.3]{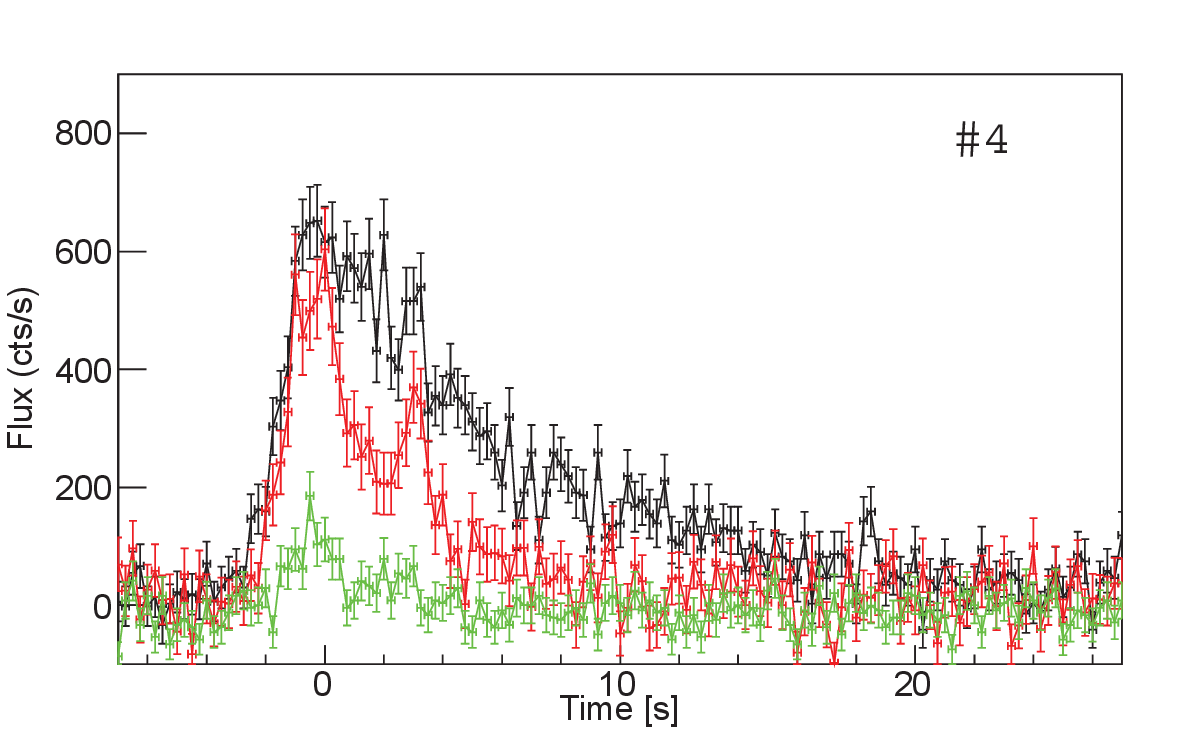}
\includegraphics[angle=0, scale=0.3]{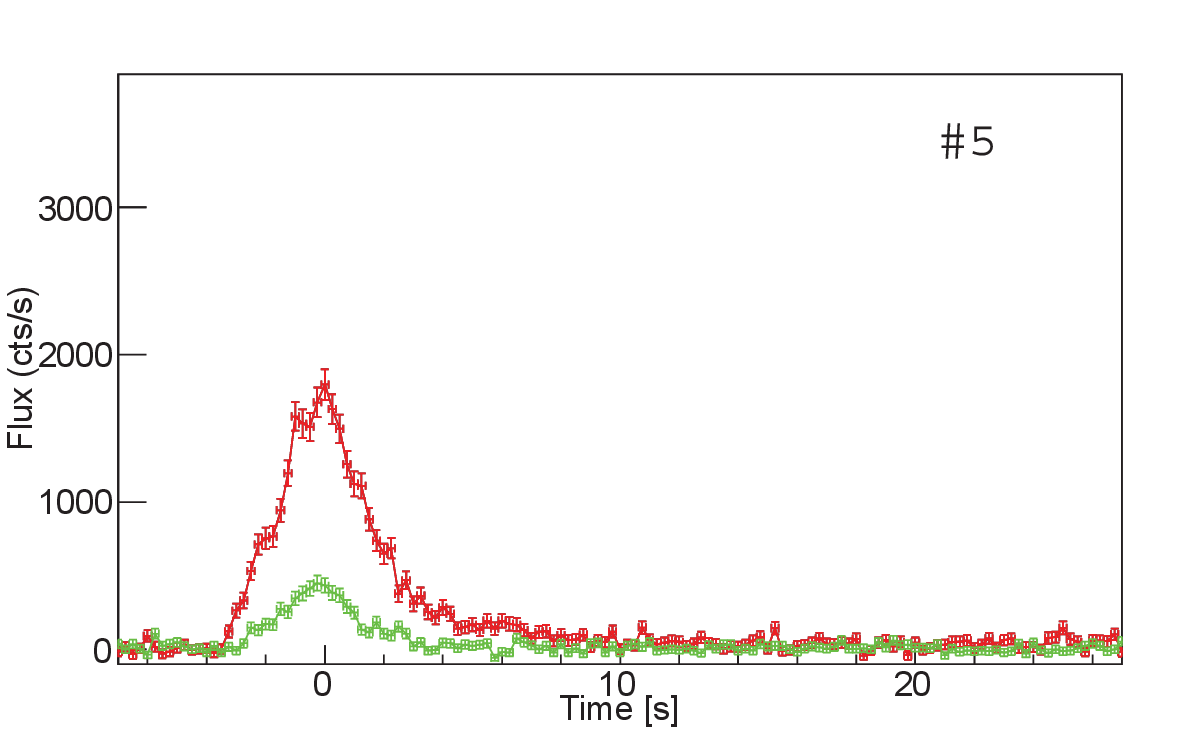}
\includegraphics[angle=0, scale=0.3]{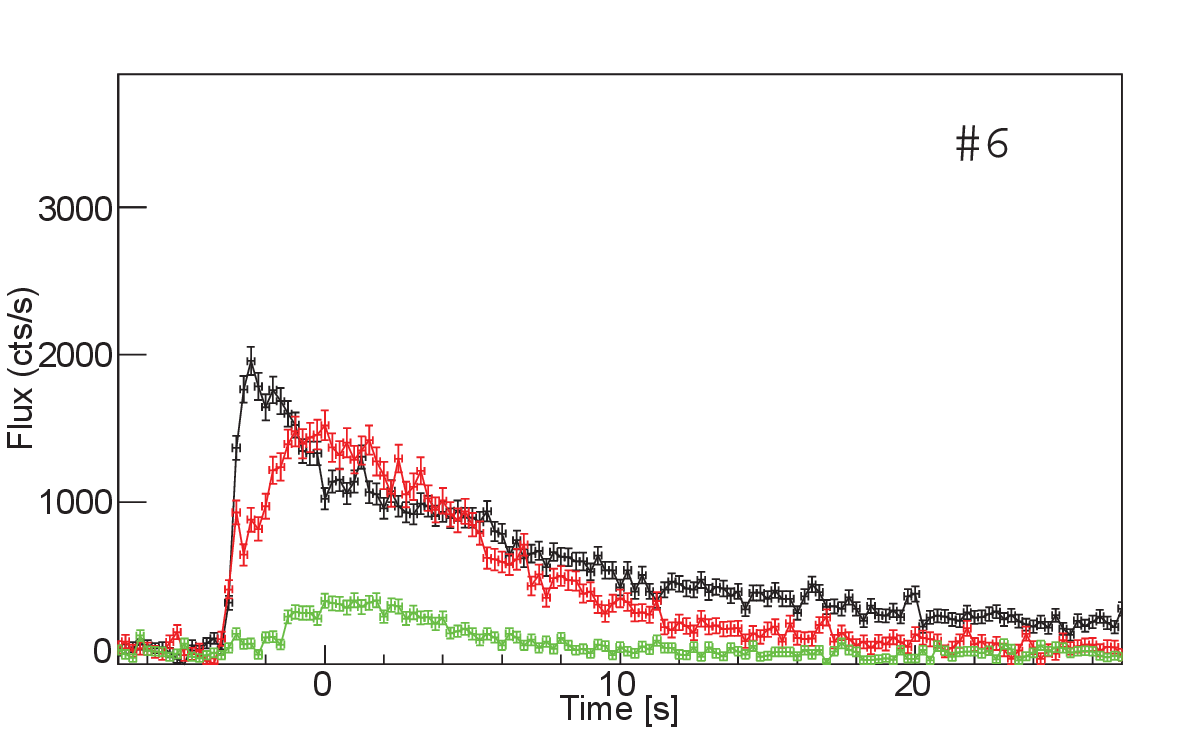}
\includegraphics[angle=0, scale=0.3]{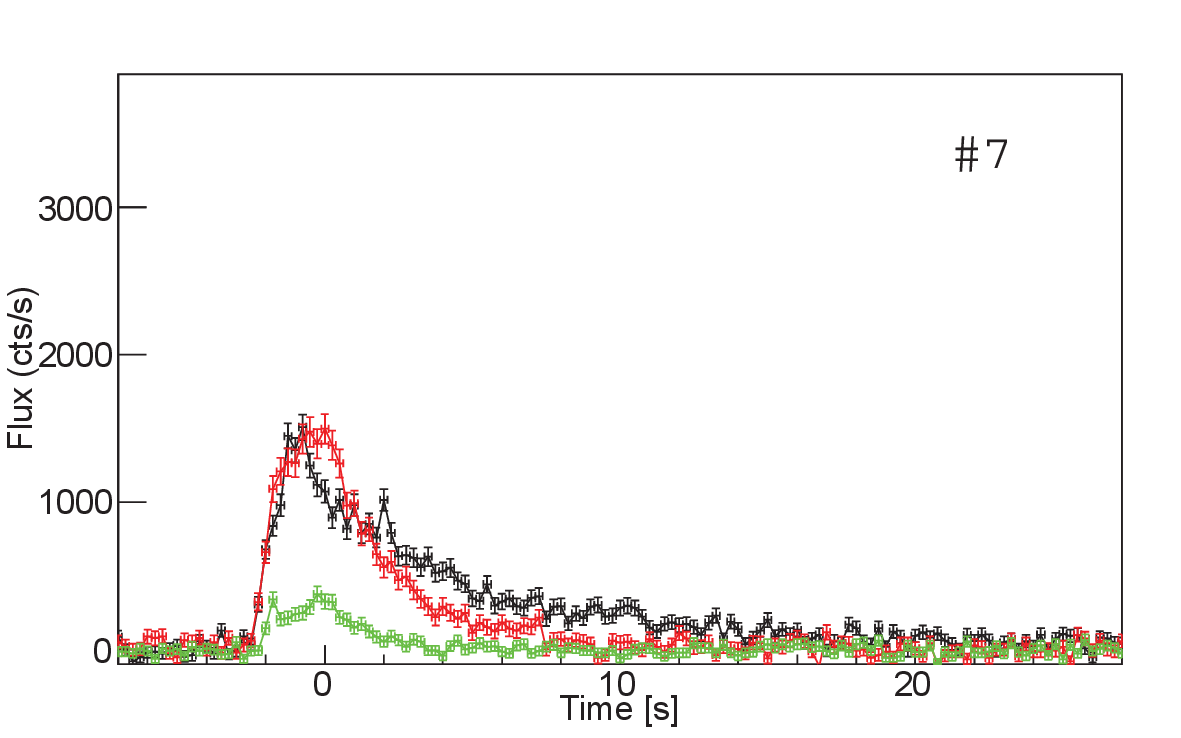}
\includegraphics[angle=0, scale=0.3]{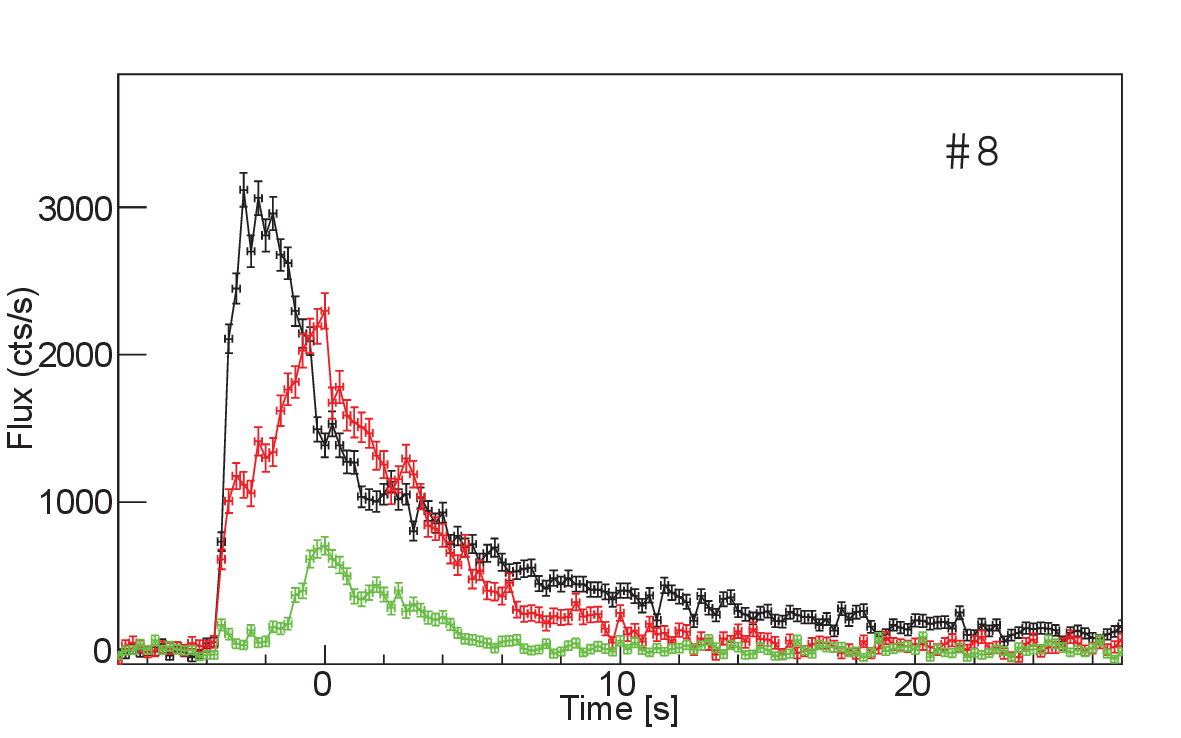}
\includegraphics[angle=0, scale=0.3]{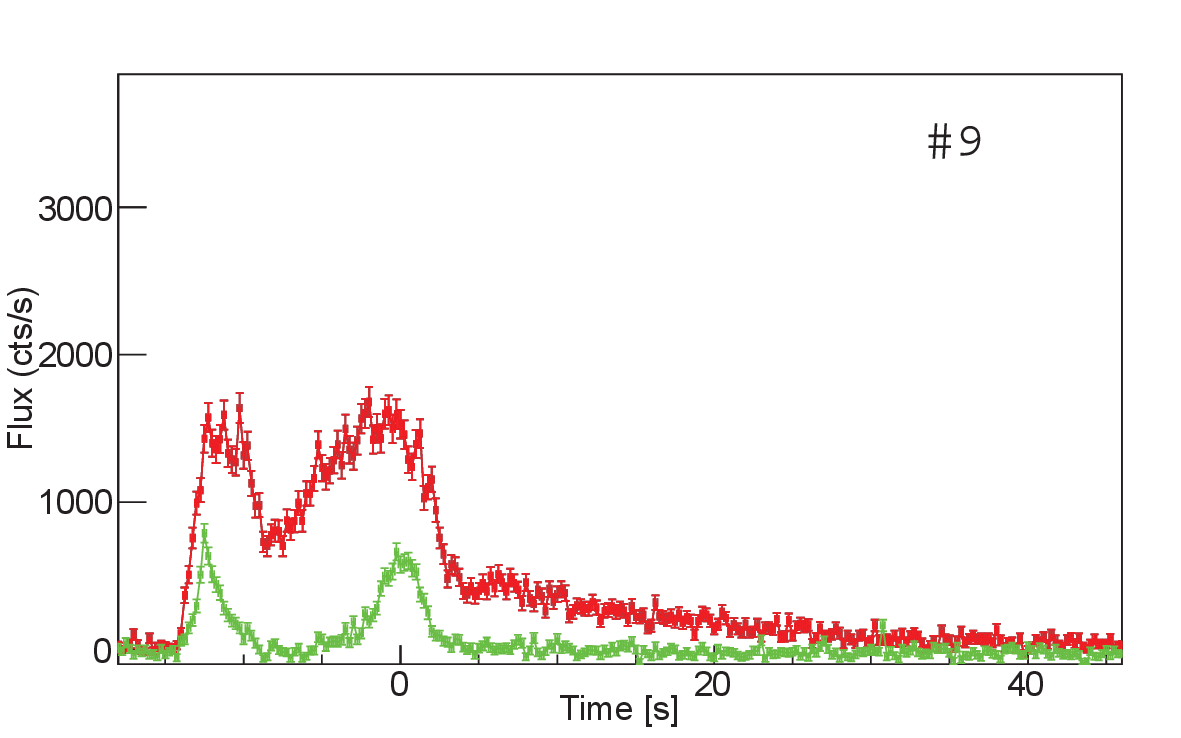}
 \caption{
 Lightcurves  with pre-burst emission subtracted of the 9 type-I X-ray bursts detected in the Insight-HXMT observation of 4U~1608--52 with time bin 0.25 s by LE (black), ME (red) and HE (green). The lightcurves of LE and ME are in  their full energy bands; the energy band of HE lightcurves is  20--50 keV.
  }
\label{fig_burst_lc}
\end{figure}

To ensure whether the 9 bursts belong to different accretion states, we plot a color-color diagram (CCD) in Figure \ref{fig_outburst_ccd}. The soft color is defined as the ratio of count rate in 6--10 keV \& 1--6 keV of LE and the hard color is defined as the ratio of count rate in 10--15 keV \& 15--30 keV of ME. Both of them are normalized by the count rates of Crab at the same energy bands and at the same period.
As shown in Table \ref{tb_burst} and Figure \ref{fig_burst_lc}, 9 bursts are found in LE, ME, and HE data with a peak flux $\sim$600--3500 cts/s, $\sim$600--2600 cts/s and $\sim$200--1000 cts/s, respectively; among them, 3 bursts are not found in LE data and 1 burst is not found in ME data.

For each burst, we use the time of the ME flux peak as a reference (0 s in Figure \ref{fig_burst_lc}) to produce lightcurves and spectra.
We extract time-resolved spectra of LE, ME and HE with a bin size of 0.25 s starting from the onset of each burst. As a conventional procedure, the pre-burst emission (including the persistent emission and the instrumental background) is extracted, which is taken as the background when fitting spectra during bursts. In practice, for each burst, we define the time interval between 70 and 20 seconds before the burst peak as the time window of the pre-burst emission, i.e., [-70 s, -20 s].


The slices of burst spectra of LE, ME and HE are rebinned by ftool grppha with a minimum of 10 counts per grouped bin, based on the limited photons of the burst slice spectra due to the short exposure time.  
The spectra of the persistent emission are rebinned by ftool ftgrouppha optimal binning algorithm with a minimum of 25 counts per grouped bin \citep{Kaastra2016}.
We add a systematic uncertainty of 1\% to the Insight-HXMT spectra to account for residual systematic uncertainties in the detector calibrations \citep{Li2020}.
We use a distance of 4 kpc to calculate the luminosity and the diskbb/blackbody radius.




\section{Analysis and Results}

\subsection{The evolving persistent emission}


As shown in Figure \ref{fig_outburst_lc},
the outburst first spent about ten days in the low/hard state (the island state) with a peak flux $\sim$ 175 mCrab detected by Swift/BAT in 15--50 keV. Just after the drop in the hard X-ray band, the soft X-ray flux reached its peak flux to $\sim$350 mCrab based on LE data (700 cts/s of LE $\sim$1 Crab) in 1--10 keV, although lacking the MAXI data.
The high/soft state (the banana state) lasted for $\sim$ 20 days, where the first 8 bursts are located.
At the tail of the outburst, there was a bump at MJD 59791.
After that, hard X-rays increased slowly near the last burst.
The CCD diagram, Figure \ref{fig_outburst_ccd}, is based on Insight-HXMT data;
apparently, it missed the rising phase of this outburst--the first Insight-HXMT observation is the right-lower point.
It slowly moved to the left-lower point and then moved to the right-lower point quickly and stopped at the right-upper point.
The first 8 bursts are located at the middle-lower points and the last burst is at the right-upper point, resulting in two groups for the 9 bursts: the first 8 bursts and the last burst.

For the 44 obsids, only the spectra of the burst-related obsids are extracted which will be used in the bursts' spectral fitting.
If the obsids where the burst occurred has LE and ME gti, the spectra of the obsids are taken as the pre-burst emission.
If not, the nearby obsids are taken as the pre-burst emission, which is only less than half day apart.

The hydrogen column (tbabs in XSPEC)
accounts for both the line-of-sight column density, as well as any intrinsic absorption near the source, and it is fixed at 1.33$\times10^{22}~{\rm cm}^{-2}$ \citep{Chen2022a}.
 Normalization constants are included during fittings to take into account the inter-calibrations of the instruments.
For the pre-burst spectra in the high/soft state,
the persistent emission detected by ME in
20--30 keV and HE in its whole energy band is very weak compared with the background. Most of these spectral channels are close to or fainter than the systematic uncertainty of the background. Thus only the LE data in
2--7 keV and the ME data in 10--20 keV are used to fit the pre-burst spectra.
We keep the normalization factor of the LE data with respect to the ME data to unity.
For the pre-burst spectra in the low/hard state, which behaves brighter in hard X-ray band,  the energy bands used in LE, ME and HE are extended to 2--7 keV, 8--30 keV, and 25--100 keV, respectively.

\begin{figure}[t]
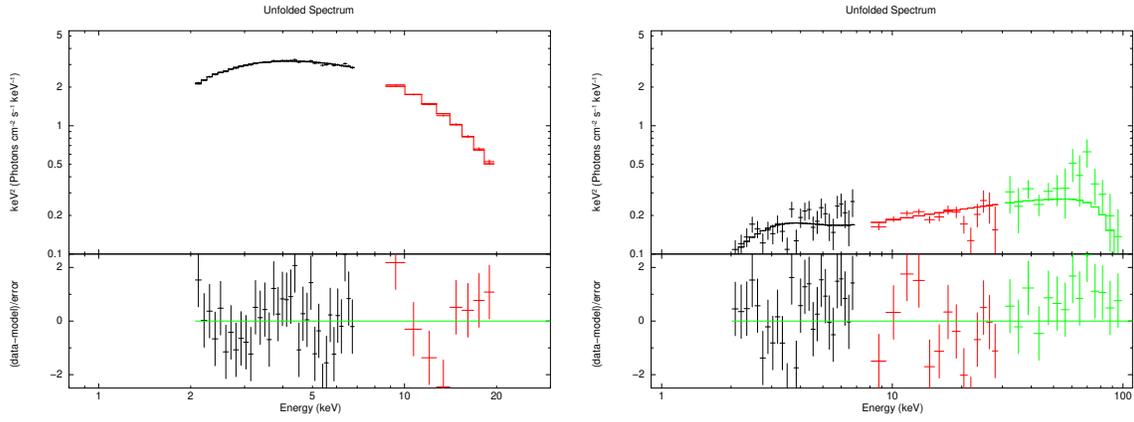

\centering
 \includegraphics[angle=270, scale=0.3]{xspec_le_me.eps}
            \includegraphics[angle=270, scale=0.3]{xspec_le_me_he.eps}
 \caption{Left: the spectral fit results of the  persistent emission of burst \#1   in the high/soft state by  LE (black) and ME (red)   with model cons*tbabs*thcomp*diskbb. Right: the spectral fit results of the  persistent emission of  burst \#9 in   the low/hard state  by  LE (black), ME (red) and HE (green)   with model cons*tbabs*thcomp*bbodyrad.
 }
\label{fig_outburst_spec}
\end{figure}

We fit the pre-burst/persistent spectra with an absorbed convolution thermal Comptonization model, available as thcomp (a more accurate version of nthcomp) \citep{Zdziarski2020} in XSPEC, which is described by the optical depth $\tau$, electron temperature $kT_{\rm e}$, scattered/covering fraction $f_{\rm sc}$.
For the high/soft state persistent spectrum, the seed photons of the thcomp are from the accretion disk and the corresponding model is diskbb, i.e., the disk corona scenario.
 Using the model above, we find an acceptable fit, as shown in Table \ref{tb_fit_thcomp_diskbb} and Figure \ref{fig_outburst_spec}.
The inferred bolometric flux in 0.01--1000 keV is $17.9\pm0.4$--$8.9\pm0.2\times10^{-9}~{\rm erg}~{\rm cm}^{-2}~{\rm s}^{-1}$, corresponding to 19.0\%--9.5\% $L_{\rm Edd}$ at distance of 4 kpc and $L_{\rm Edd}=1.8\times10^{38}$ erg/s (the Eddington limit of the hydrogen accretion material onto a NS with 1.4 solar mass).

For the low/hard state persistent spectrum, derived in the Insight-HXMT last observation, using the model above, we find an acceptable fit, but the derived inner disk radius is $<$ 3 km  (with the inclination angle $30^{\circ}$), which is not likely. Thus, substituting the diskbb component with a blackbody component in the aforementioned convolution model is attempted, corresponding to the spherical corona scenario.
The constants of ME and HE are fixed to unity compared with LE, respectively; if not, the constants swing during the spectral fitting which could be related to the heavy error bars in the hard X-ray band. The inferred bolometric flux in 0.01--1000 keV is $1.50\pm0.05\times10^{-9}~{\rm erg}~{\rm cm}^{-2}~{\rm s}^{-1}$
corresponding to 1.6\% $L_{\rm Edd}$ at a distance of 4 kpc and $L_{\rm Edd}=1.8\times10^{38}$ erg/s.

\subsection{Burst lightcurves}



\subsubsection{The lightcurve of the PRE bursts}

For bursts \#1, \#2, \#6--\#8, the lightcurves of LE present a one-peaked structure, but the hard X-rays (ME and HE) lag behind the soft X-rays (LE) by $\sim$ 2 s; the peak times of the ME and HE lightcurves are consistent with each other.
A closer look reveals that there is a double-peaked profile both for ME and HE lightcurves, which is a typical characteristic of PRE bursts.

Assuming the burst emission is isotropic, for the PRE phase, the photosphere emission at just lift-up time and the touch-down time have the same radii (NS radius) and the same temperatures; 
for lightcurves detected by the hard X-ray telescopes, e.g., HE, there should be two peak values at the start and end of the PRE phase, and a hard X-ray dip in the middle of the PRE phase because of the vast majority of the radiation falls temporarily outside the hard X-ray band. However, except for burst \#9, other bright PRE bursts (\#1-\#3, and \#8), the second peak is much brighter than the first one, deviating from the theoretical prediction.

Of course, the fast-rising timescale could also cause the hard X-ray shortage for the first peak: since the first peak time is too short for the time-bin of the lightcurves and averaged flux with nearby data points could drag down the count rate.
We also extract the hard X-ray lightcurves with 0.1 s, the flux ratio of the two peaks is very similar to the results derived with 0.25 s.
The last burst, burst \#9, both for ME and HE but lacking LE data,, shows a double-peaked profile in ME and HE lightcurves, and the peak values are roughly the same, which is consistent with the theoretical prediction.
Moreover, there is a much longer interval between the two spikes than the other PRE bursts, $\sim$ 10 s.
For the decay profile, the burst has the longest decay time, $>$ 10 s; and it seems that it evolves in two steps: the first step with a fast decay and the second step with a slow decay.


\subsection{Broad-band spectra of burst emission by Insight-HXMT}

Different from the persistent spectral fitting, the constants of LE, ME and HE are all fixed at unity for the combined-spectra fitting, because of the high flux of the burst emission and the negligible effect of the background's uncertainties \citep{Chen2022a}.
{\bf The energy band of LE is extended to
1–10 keV in the burst analysis because of no influence of the electronic noise when we
take the preburst spectrum as the background of burst spectra \citep{Chen2022}. }

\subsubsection{Fit the spectra of bursts by the blackbody model }

\begin{figure}[t]
\centering
\includegraphics[angle=0, scale=0.2]{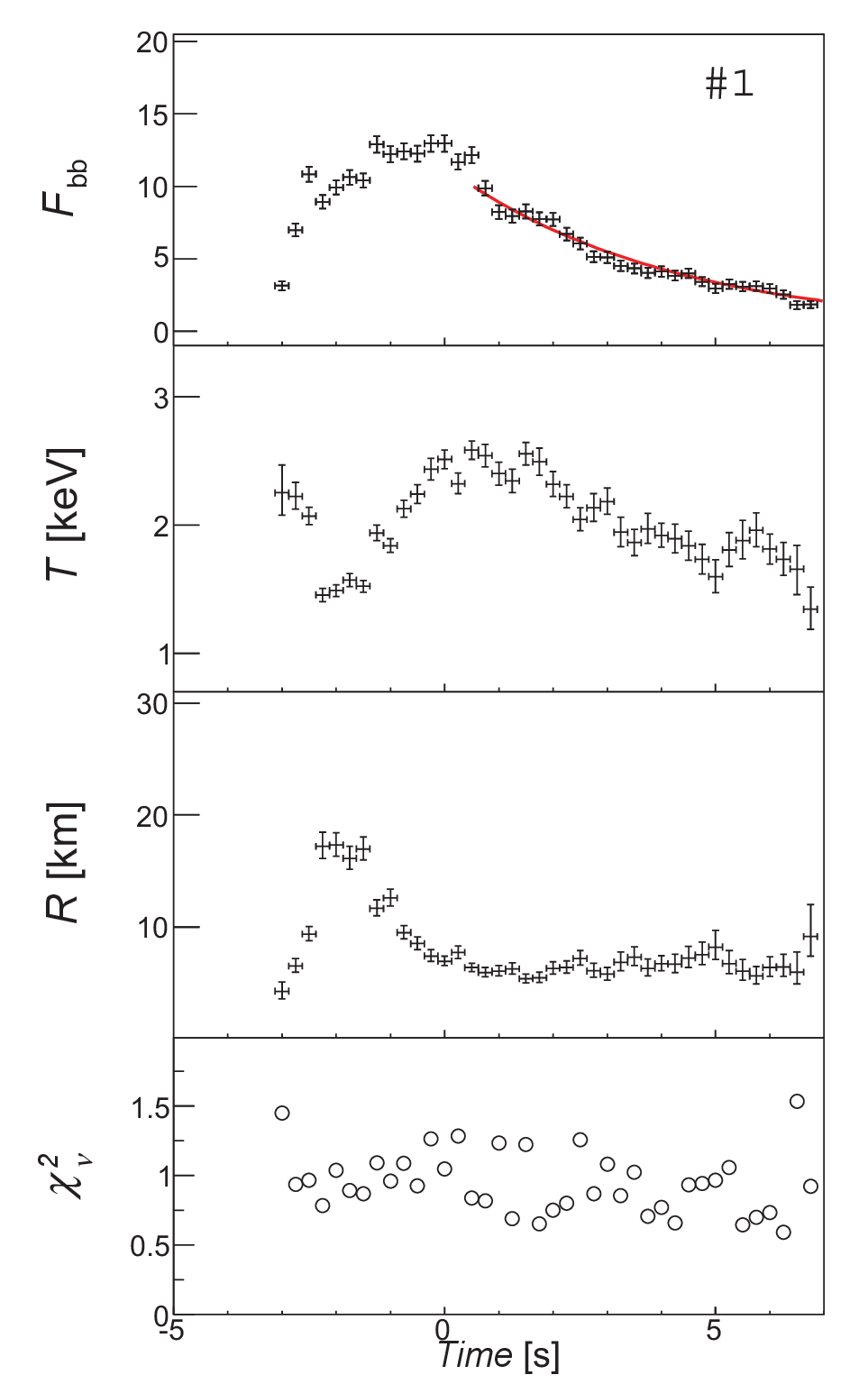}
\includegraphics[angle=0, scale=0.2]{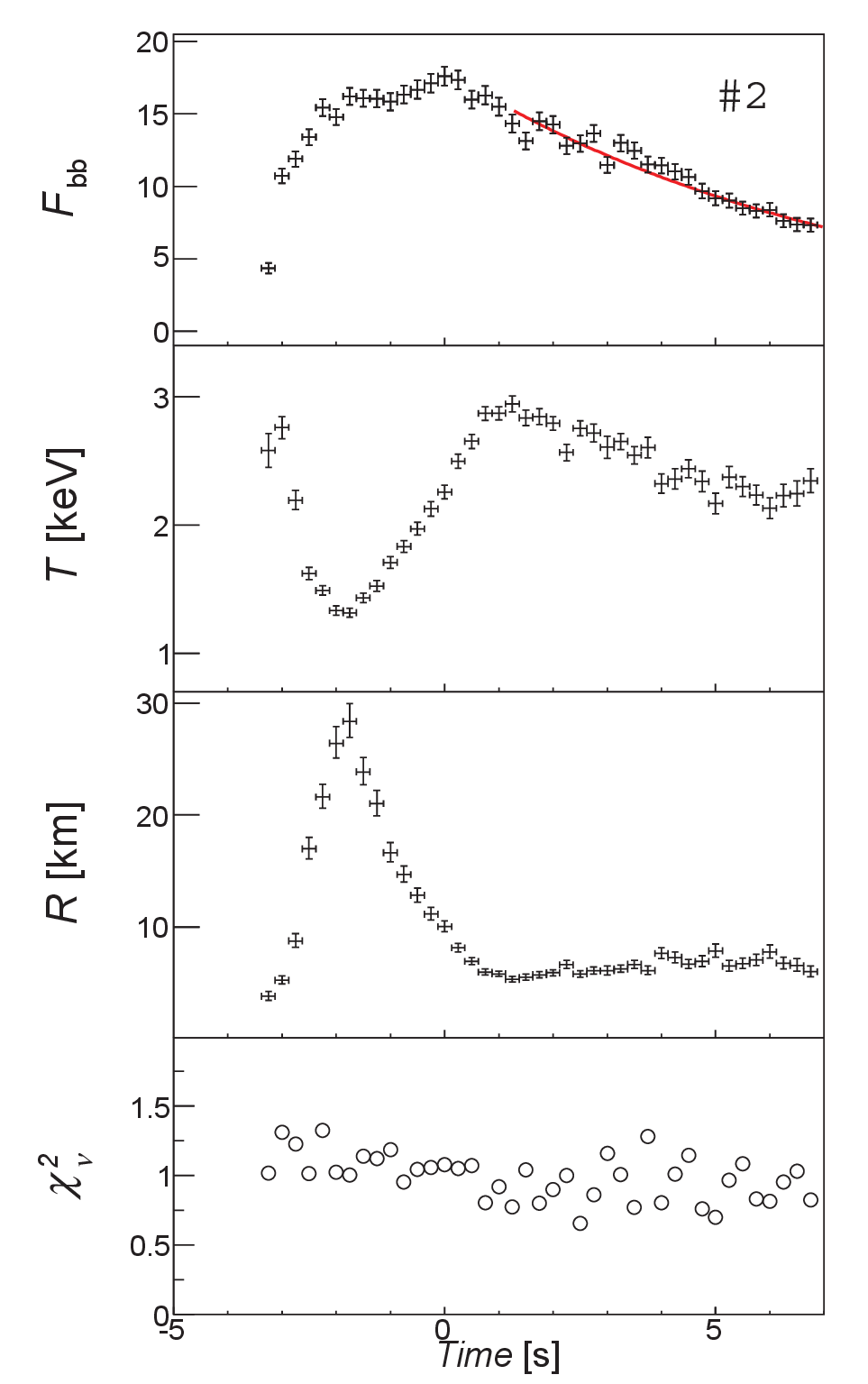}
\includegraphics[angle=0, scale=0.2]{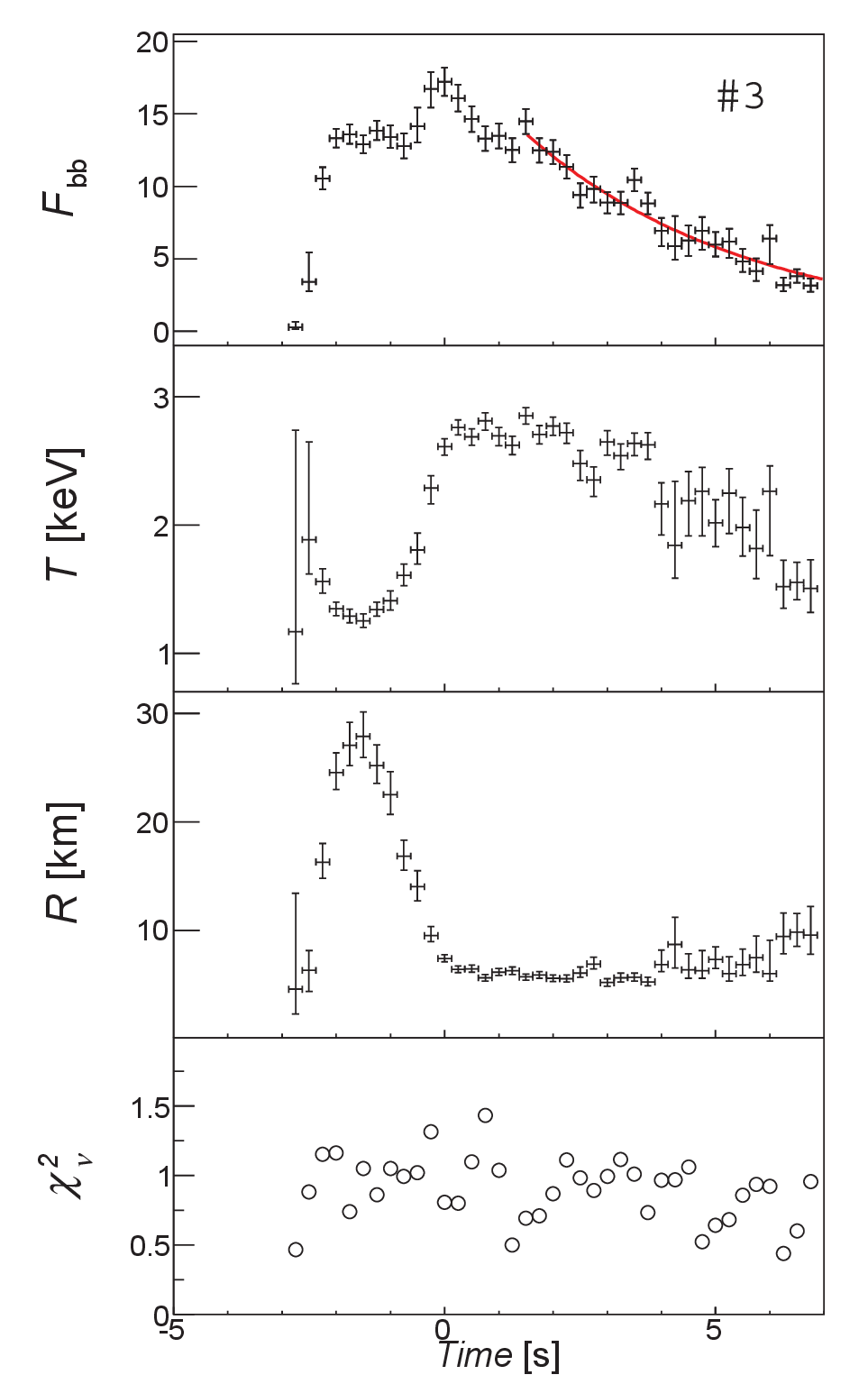}
\includegraphics[angle=0, scale=0.2]{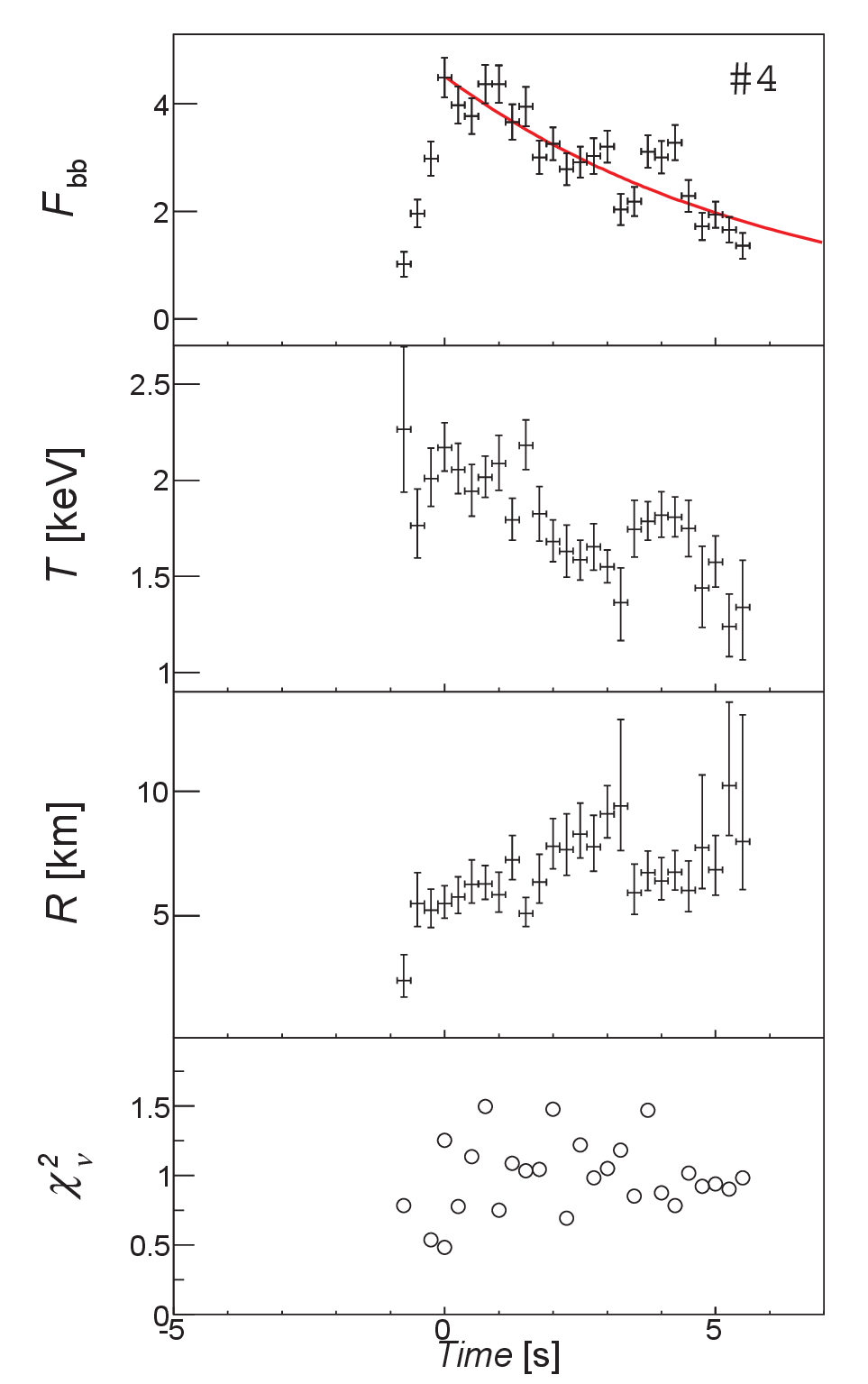}
\includegraphics[angle=0, scale=0.2]{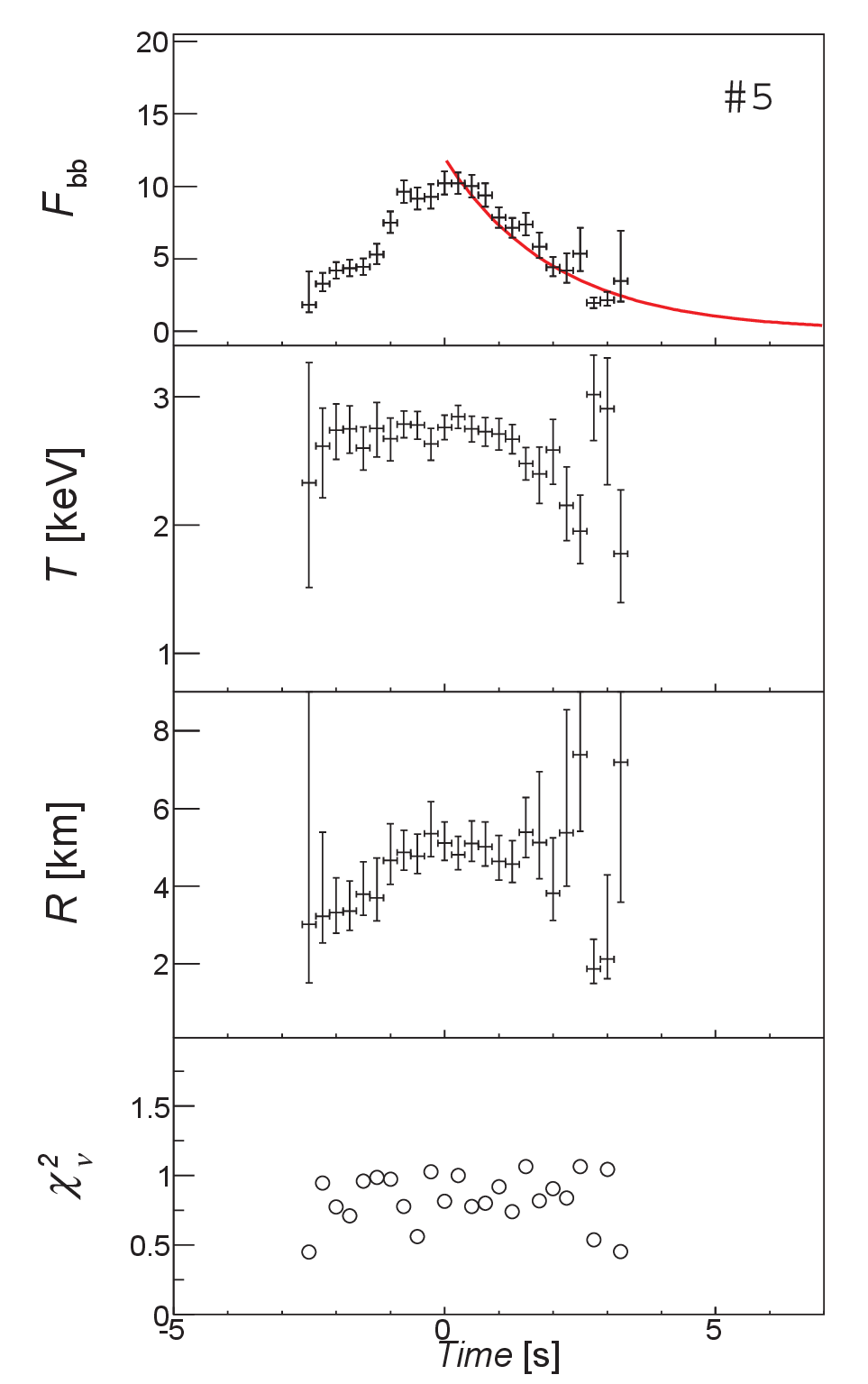}
\includegraphics[angle=0, scale=0.2]{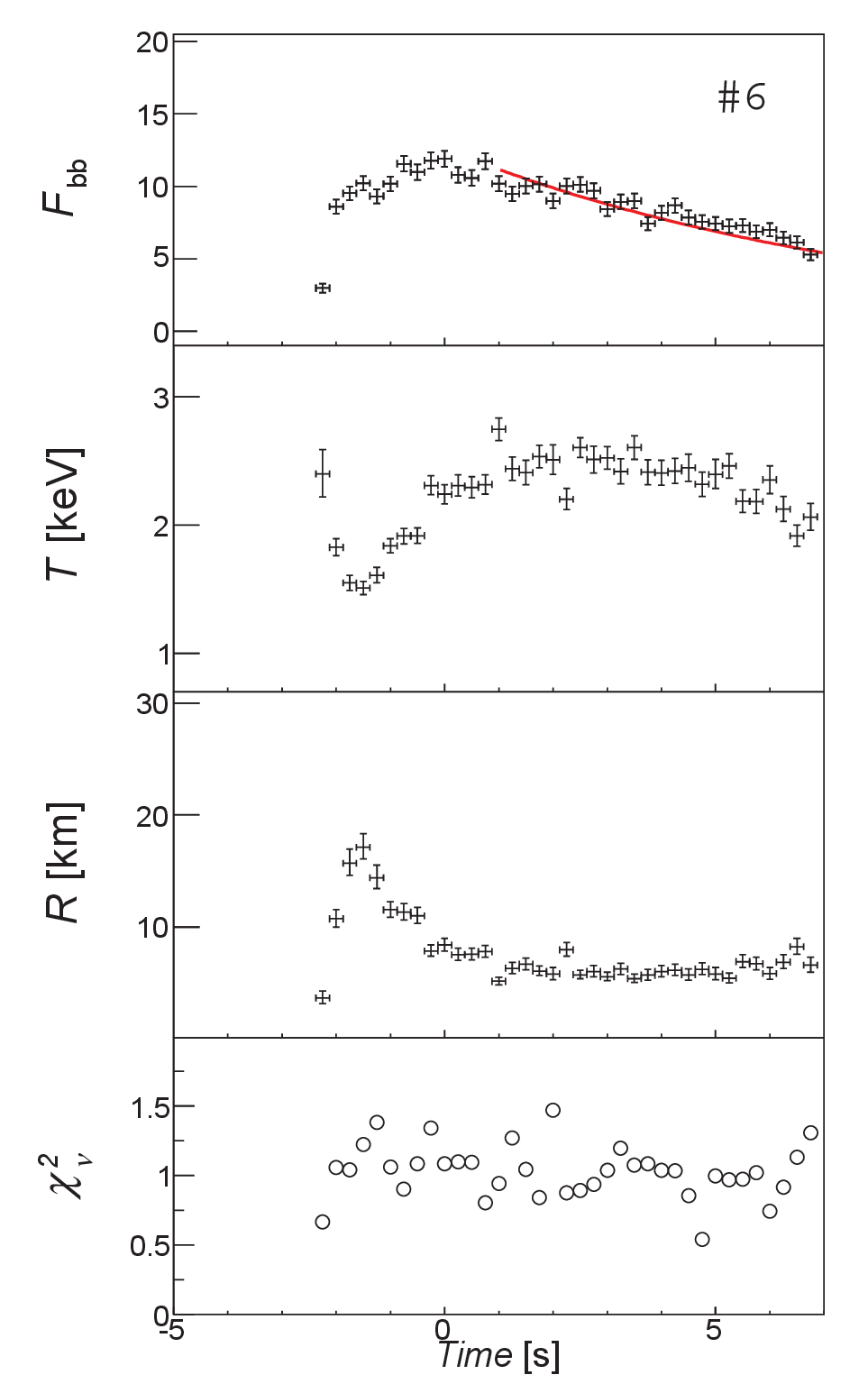}
\includegraphics[angle=0, scale=0.2]{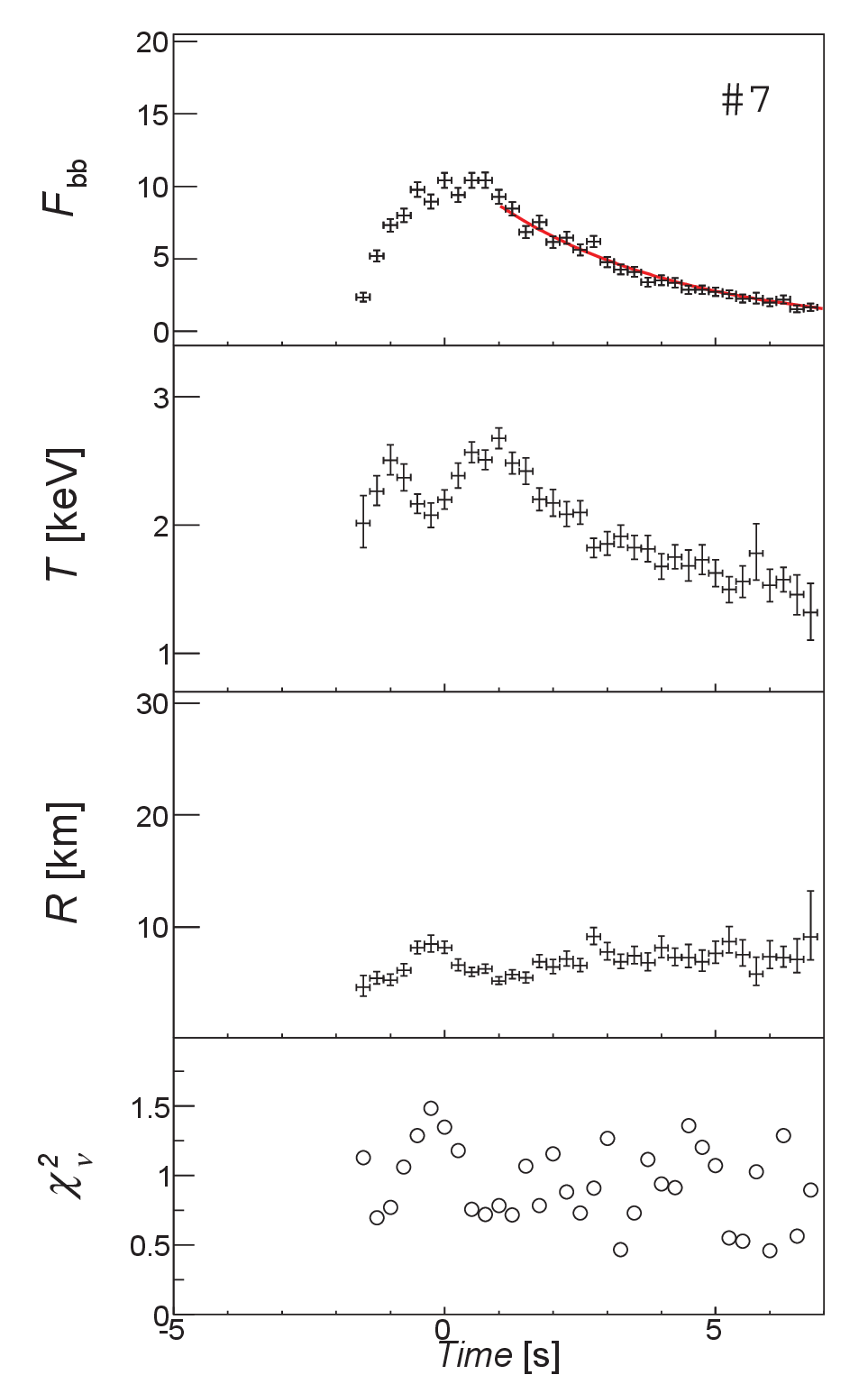}
\includegraphics[angle=0, scale=0.2]{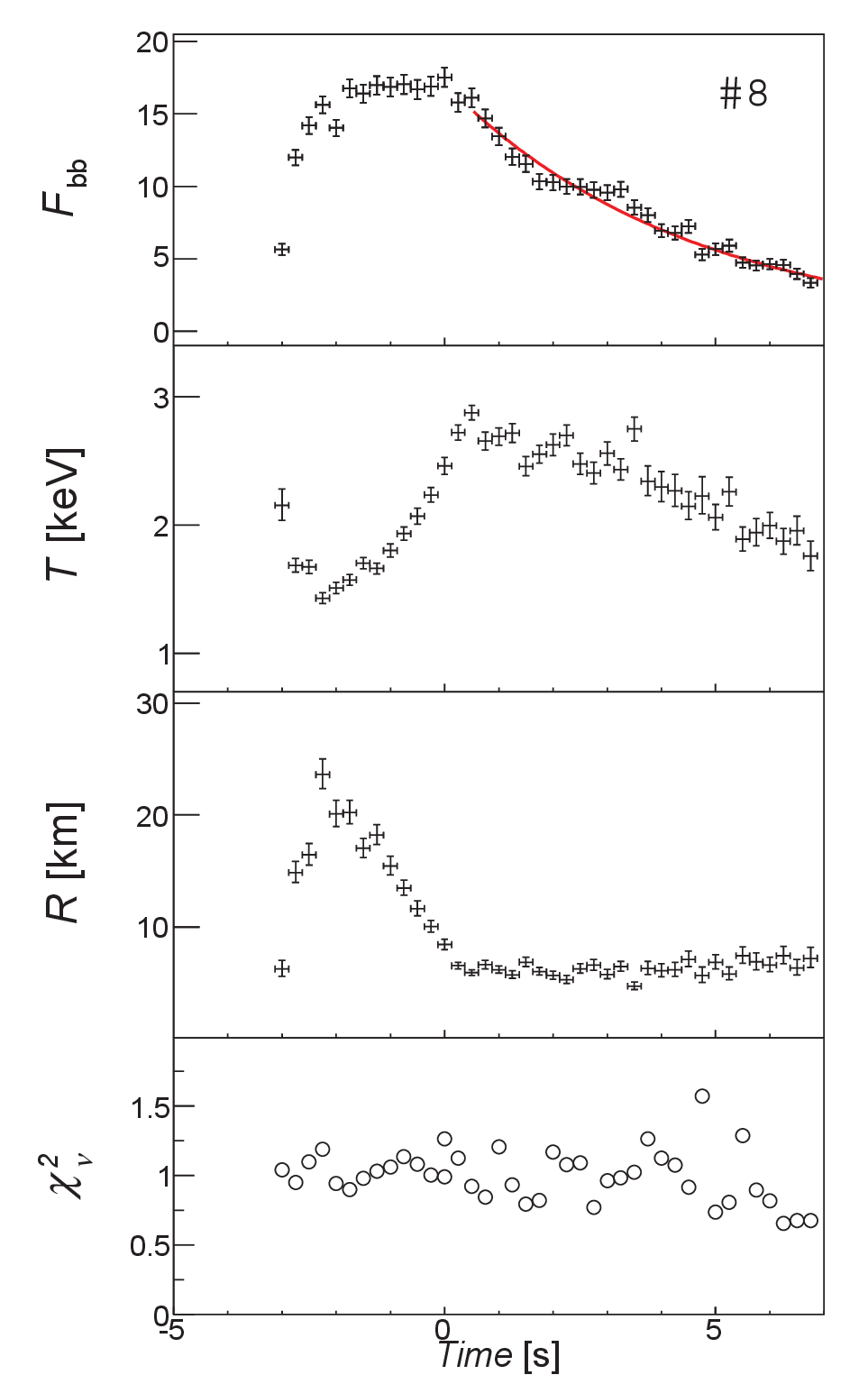}
\includegraphics[angle=0, scale=0.2]{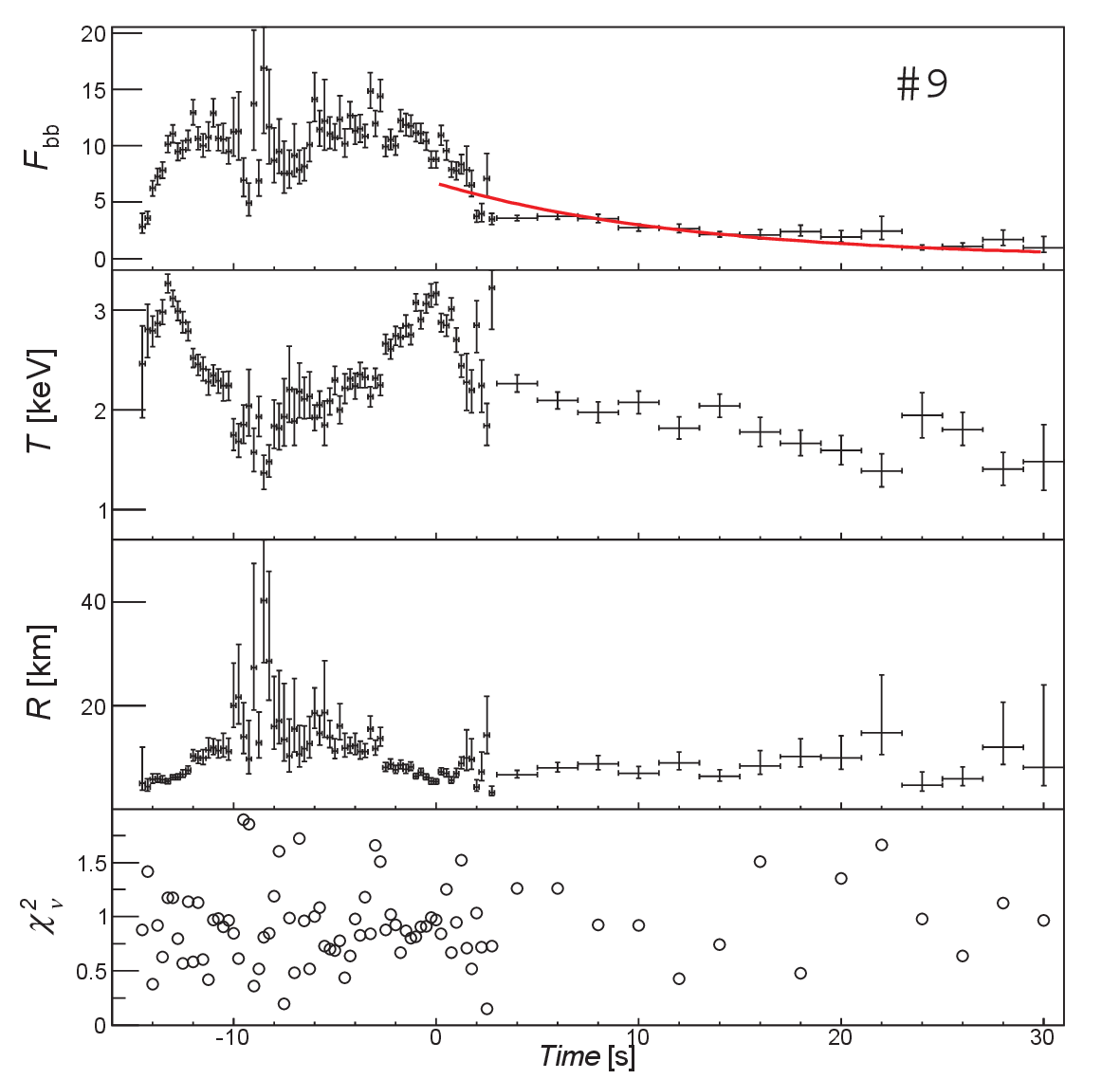}
 \caption{
 Results of the spectral fits of time-resolved spectra of the 9 bursts detected from 4U~1608--52 during its  2022 outburst, including the
  blackbody bolometric flux $F_{\rm bb}$, the temperature $kT_{\rm bb}$, the observed radius $R$ of the NS photosphere at 4 kpc, and the goodness of fit $\chi_{v}^{2}$.
 The red lines indicate the fitting results of the temporal evolution of the bolometric flux   with an exponential decay function.
  }
\label{fig_burst_fit_bb}
\end{figure}

We follow the classical approach to X-ray burst spectroscopy by subtracting the persistent spectrum
and fitting the net spectrum with an absorbed blackbody, as shown in Figure \ref{fig_burst_fit_bb}.
In the decay phase, such a spectral model generally results in an acceptable goodness-of-fit, with a mean reduced $\chi^{2}_{\upsilon}~\sim$ 1.0 (d.o.f. 20--60).
However, we note that significant residuals are shown below 3 keV and above 10 keV in Figure \ref{fig_spec_residual}, especially the spectra in the PRE phase.

From the fitting results by the absorbed blackbody, among the nine bursts, 7 bursts are PRE bursts with peak radii 12--30 km, peak temperatures $\sim$ 3 keV and peak bolometric fluxes 10--17$\times10^{-8}~{\rm erg}~{\rm cm}^{-2}~{\rm s}^{-1}$.
The model parameters of the bursts without LE or ME detection, i.e., bursts \#3, \#5 and \#9,
show greater errors than that of the bursts with LE and ME detection, which prevents us from adopting other models to fit.
A similar situation also fits for the faint bursts, bursts \#4 and \#5.

As shown in Table \ref{tb_burst}, the shortest burst interval (recurrence time) is 1.5--3 hours, the inferred $\alpha$--the ratio of the integrated persistent flux between bursts to the burst fluence is $\sim$ 150.
Given the short rise/decay timescale, these bursts are preferred as the helium-rich bursts.



\subsubsection{Fit the spectra of bursts by the $f_{a}$ model }

To reduce the residuals from the spectral fitting by the blackbody model, we first consider the $f_{a}$ model to fit the bright bursts which were detected simultaneously by LE, ME, and HE: bursts \#1, \#2 and bursts \#5--\#7. Following \citet{Worpel2013} we then include an additional component for fitting the variable persistent emission.
We use the model parameters derived from fitting results of the pre-burst emission of each burst.
We assume that during the burst the spectral shape of the persistent emission is unchanged, and only its normalization (known as a $f_{a}$ factor) is changeable.
As reported earlier with RXTE and NICER data,
the $f_{a}$ model provides a better fit than the conventional one (absorbed blackbody).
We compare the above two models using the F-test. 
In some cases, the $f_{a}$ model significantly improves the fits with a p-value $\sim10^{-5}$.

As shown in the left panels of Figure \ref{fig_fit_burst_P040420700407}  for burst \#1, 
and Figure \ref{fig_fit_burst_P040420700901} for burst \#8, the spectral fitting results from these two models have differences mainly around the PRE phase. By considering an additional factor $f_{a}$, the burst blackbody flux tends to slightly decrease, and the temperature becomes higher but the radius shrinks.

The $f_{a}$ factor reaches a maximum of $2.5\pm0.5$--$6\pm1$ when the radii reach their peaks, as shown in Table \ref{tb_burst} {\bf and Figure \ref{fig_fit_burst_fa}}.
We notice that there is a large scattering (by a factor of $\sim$2) for the maximum $f_{a}$ even though all of the them are PRE bursts, e.g., $f_{a,\rm p}$=$2.5\pm0.5$ and $f_{a,\rm p}$=$6\pm1$ for burst \#1 and burst \#8, respectively.
 Since the burst flux and accretion rate have a significant influence on $f_{a}$ (e.g., \citealp{Worpel2013,Worpel2015,2022MNRAS.510.1577G,2022ApJ...935..154G}),
the fractions of flux enhancement during bursts
defining as $\frac{f_{a}^{p}\times F_{\rm pre-burst}}{F_{d}^{\rm bb}}$ are given in Figure \ref{fig_fa}, in which $f_{a}^{p}$, $F_{\rm pre-burst}$ and $F_{\rm p}^{\rm bb}$ are the peak value of $f_{a}$, the bolometric flux of the pre-burst/persistent emission and the peak flux of bursts, respectively.
The brightest burst detected by NICER \citep{Jaisawal2019} in 2017, and two bursts detected by Insight-HXMT \citep{Chen2019,Chen2022a} in 2018 and 2020 are also plotted in Figure \ref{fig_fa}. The fraction of flux enhancement during bursts have a constant value $\sim$ 0.2--0.4, even though the bursts' persistent emission and bursts' emission has large spreads, e.g.,  $F_{\rm pre-burst}$=0.02--0.2 $L_{\rm Edd}$ and $F_{\rm p}^{\rm bb}$=10--16 $\times10^{-8}~{\rm erg/cm}^{2}/{\rm s}$.


During the PRE phase of the brightest burst (burst \#2), the radius is up to $\sim 30$ km, which is six times larger than the radius measured at touch-down time $\sim$ 5 km (assuming a distance of 4 kpc).
This is typical of a moderate photospheric expansion.
The other PRE bursts show a similar trend, but with a smaller PRE radius or shorter time interval of the PRE phase.

\subsubsection{Fit the spectra of bursts by the convolution thermal Comptonization model}

Another model, as we did in our previous work \citep{Chen2022a,Chen2022b}, the convolution thermal Comptonization model (with an input seed photon spectrum of blackbody),
with fixed thcomp parameters derived from the persistent emission, corresponding to the Comptonization of the burst photon by the surrounding corona/boundary-layer.
By taking the pre-burst emission as background emission, the burst spectra are fitted by the model tbabs*thcomp*bb, thus convolution thermal Comptonization model (with an input seed photon spectra blackbody) has the same d.o.f. as the canonical blackbody model, and a more d.o.f. than the $f_{a}$ model.

\begin{figure}[t]
\centering
      \includegraphics[angle=0, scale=0.5]{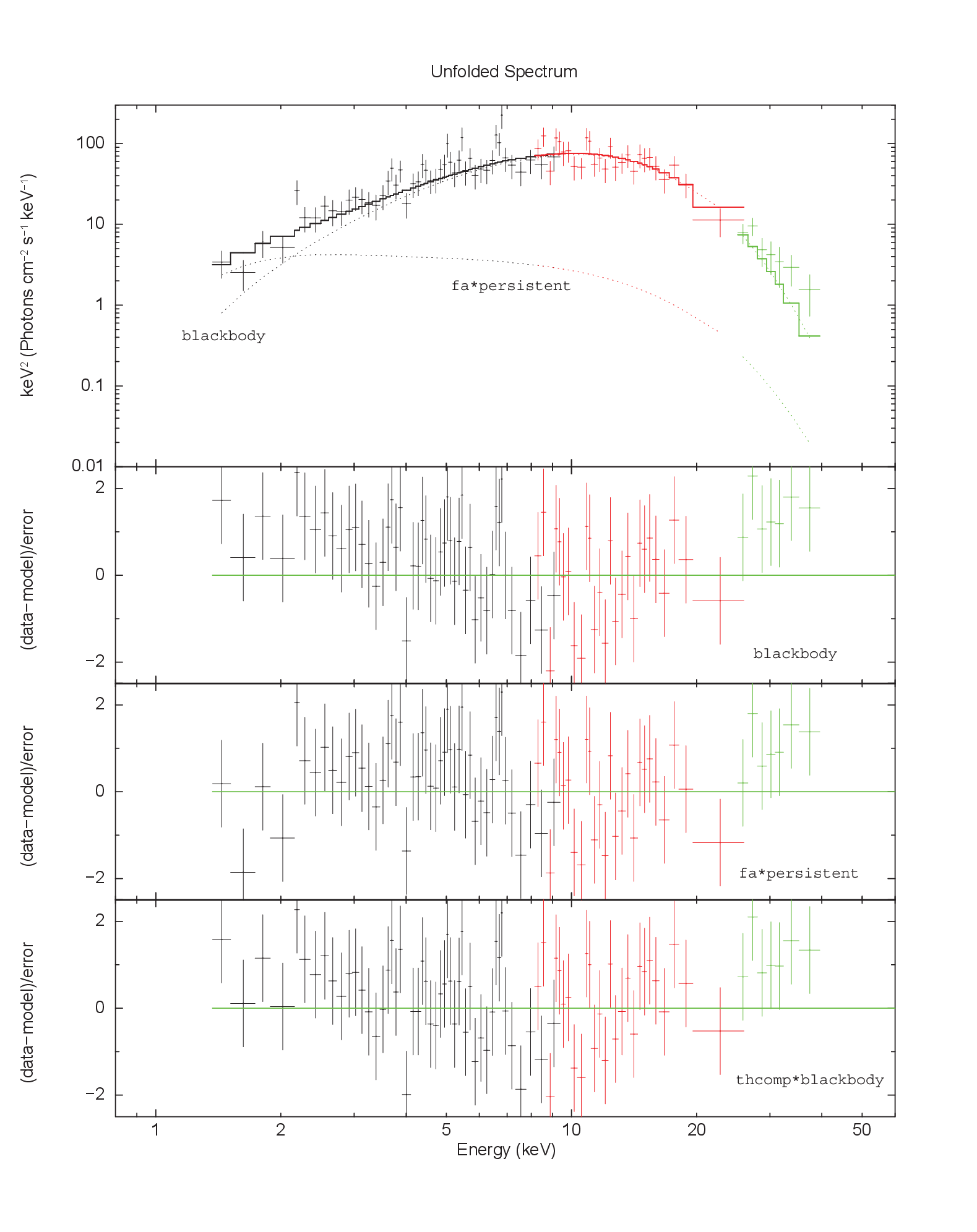}
 \caption{Top panel: the results of the spectral fits  by  LE (black), ME (red) and HE (green) of   burst \#8 at the peak flux by $f_{a}$ model, the blackbody model and enhancement of the persistent emission are labeled.
 The three panels below: residuals of spectral fits results by
 an absorbed black-body  model (the 2nd panel), $f_{a}$ model (the 3rd panel) and the  convolution thermal-Comptonization model (the bottom panel).  The reduced $\chi^2$ (d.o.f.) for the blackbody, $f_{\rm a}$ and thcomp model  are 1.27 (81),  1.09 (80), and 1.09 (81), respectively.
 }
\label{fig_spec_residual}
\end{figure}

As shown in the right panels of Figure \ref{fig_fit_burst_P040420700407}  for burst \#1,
and Figure \ref{fig_fit_burst_P040420700901} for burst \#8,
in the {\bf PRE} phase, this model provides
yields physically acceptable spectral parameters; the obtained best-fit parameters are given in the right panels. We find that, during the PRE phases, some data points by this convolved thermal-Comptonization model provides equally good results as the $f_{a}$ model but with a more d.o.f., as shown in Figure \ref{fig_spec_residual} {\bf and \ref{fig_fit_burst_chi}}. 
However, in the rising and decay parts, such model has a larger reduced $\chi^{2}_{\upsilon}$ than that of the $f_{a}$ model and even the canonical blackbody model, which may indicates that the burst emission suffers few Comptonization during this phase.

As mentioned above, the free/unfixed parameters include the blackbody temperature $kT_{\rm bb}$ and the normalization $N_{\rm bb}$. The trends of these parameters are similar to that of the $f_{a}$ model, but with a greater change. Compared to the $f_{a}$ model results, the maximum radius $R_{\rm bb}$ is up to $\sim$ 60 km, the minimum temperature $kT_{\rm bb}$ is down to $\sim$ 0.8 keV.

\begin{figure}[t]
\centering
      \includegraphics[angle=0, scale=0.50]{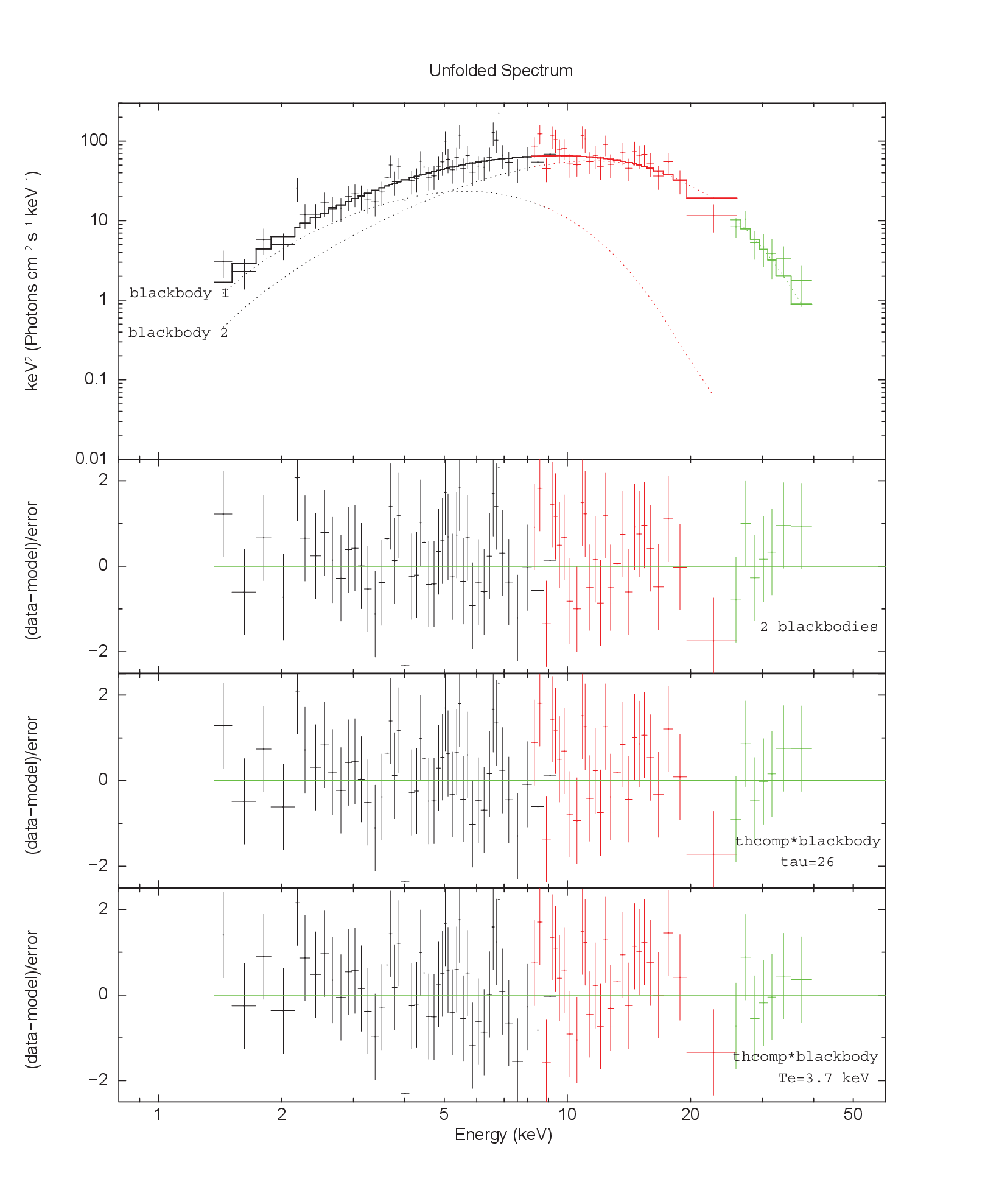}
 \caption{Top panel: the results of the spectral fits     by  LE (black), ME (red) and HE (green) of   burst \#8 at the peak flux by the two blackbodies model.
 The three panels below: residuals of spectral fits results by
 an absorbed two blackbodies model  (the 2nd panel),   the  convolution thermal-Comptonization model with   $\tau$ changed to $26\pm4$ (the 3rd panel) and $T_{e}$ changed to 3.7$\pm$0.1 keV  (the bottom panel).  
 }
\label{fig_spec_residual_1}
\end{figure}

\begin{figure}[t]
\centering
\includegraphics[angle=0,
scale=0.4]{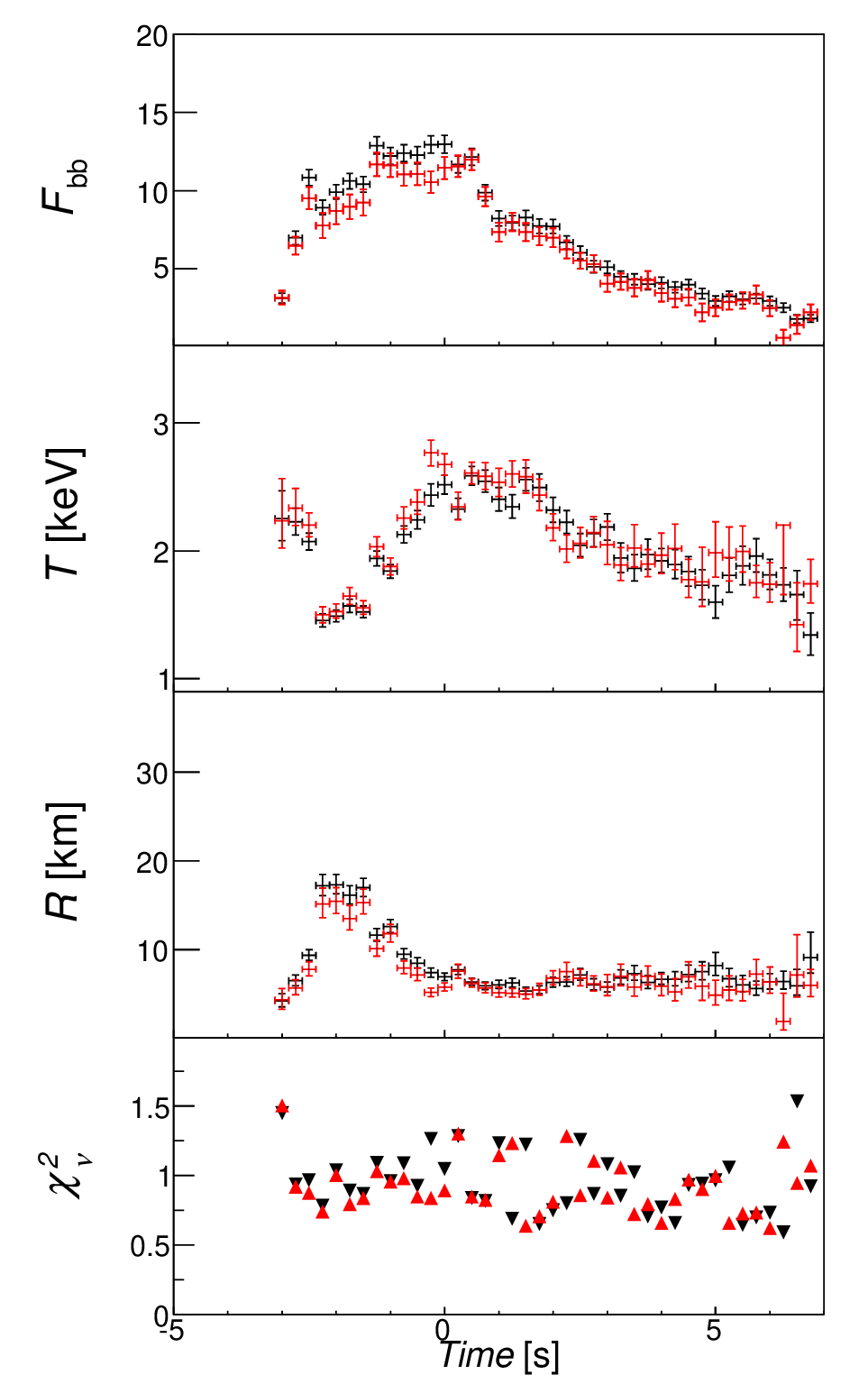}
\includegraphics[angle=0, scale=0.4]{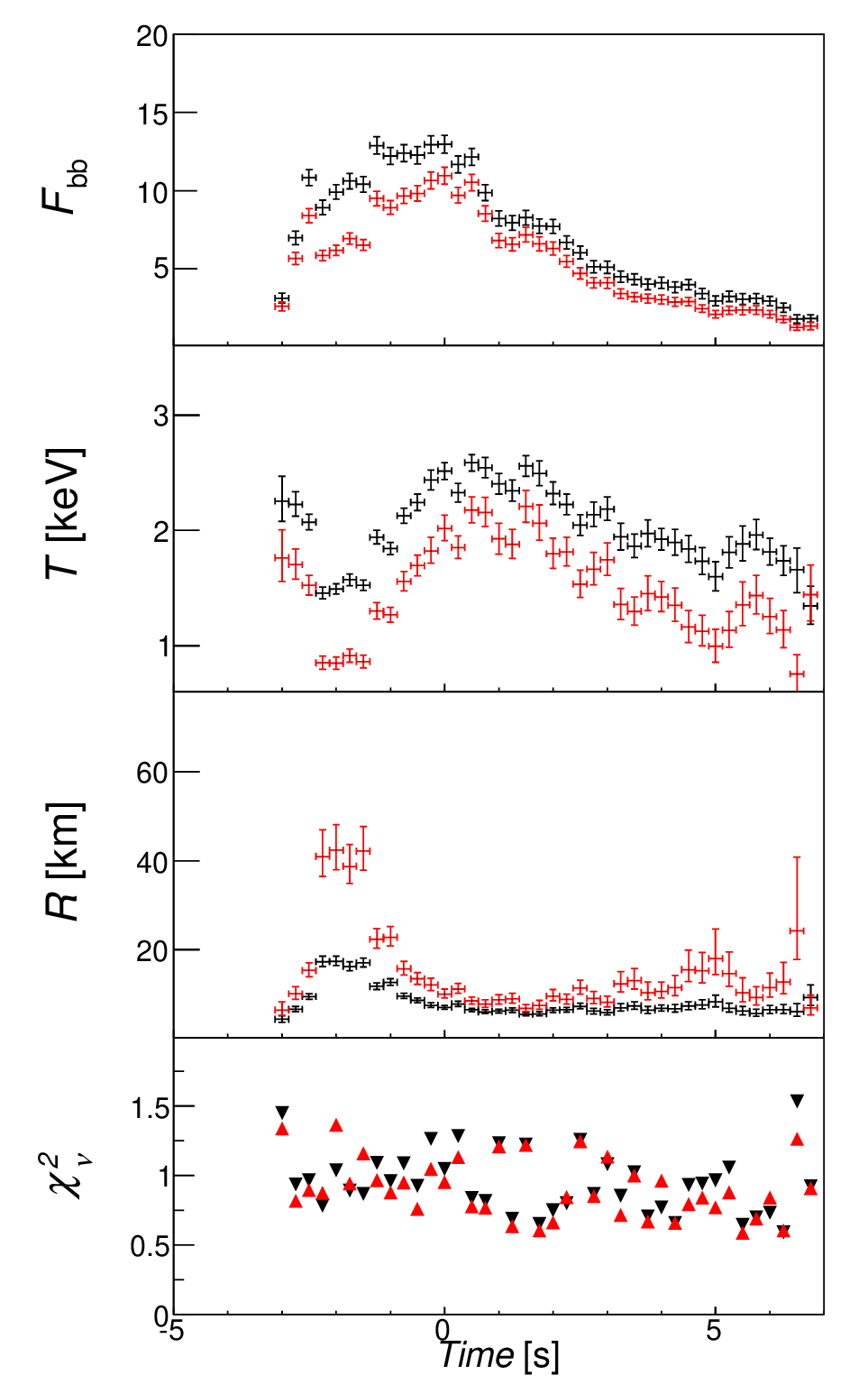}
  \caption{
Spectral fitting result of   burst \#1 with time bin 0.25 second  with  a pure blackbody (black), $f_{a}$ model (the left panel, red) and convolution thermal-Comptonization model (the right panle, red),  include the time evolution of the blackbody bolometric flux $F_{\rm bb}$, the temperature $kT_{\rm bb}$, the observed radius $R$ of the NS photosphere at 4 kpc, the goodness of fit $\chi_{v}^{2}$.
The bolometric flux of the blackbody model $F_{\rm bb}$ is in unit of $10^{-8}~{\rm erg/cm}^{2}/{\rm s}$.  }
\label{fig_fit_burst_P040420700407}
\end{figure}

\begin{figure}[t]
\centering
 \includegraphics[angle=0, scale=0.4]{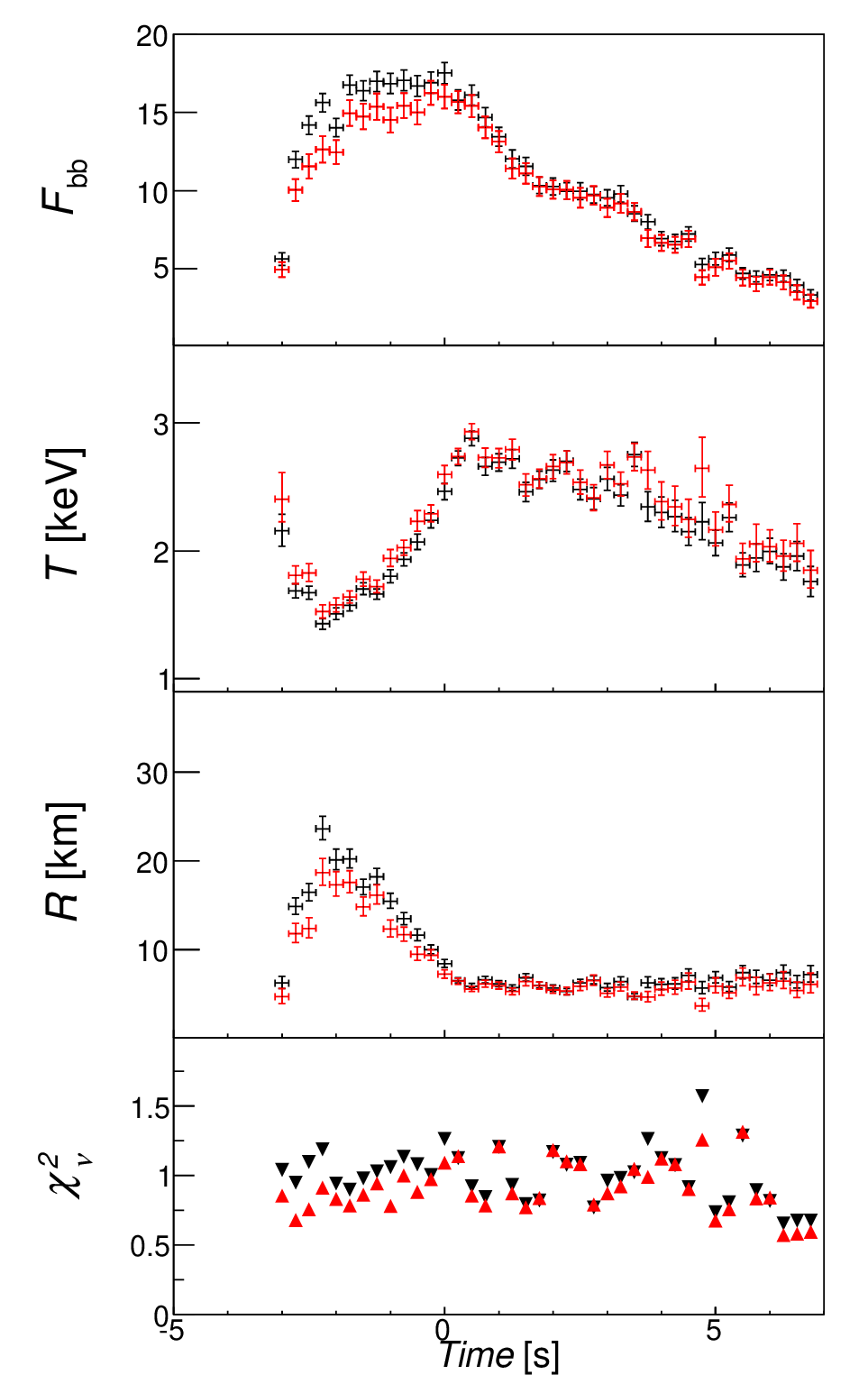}
\includegraphics[angle=0, scale=0.4]{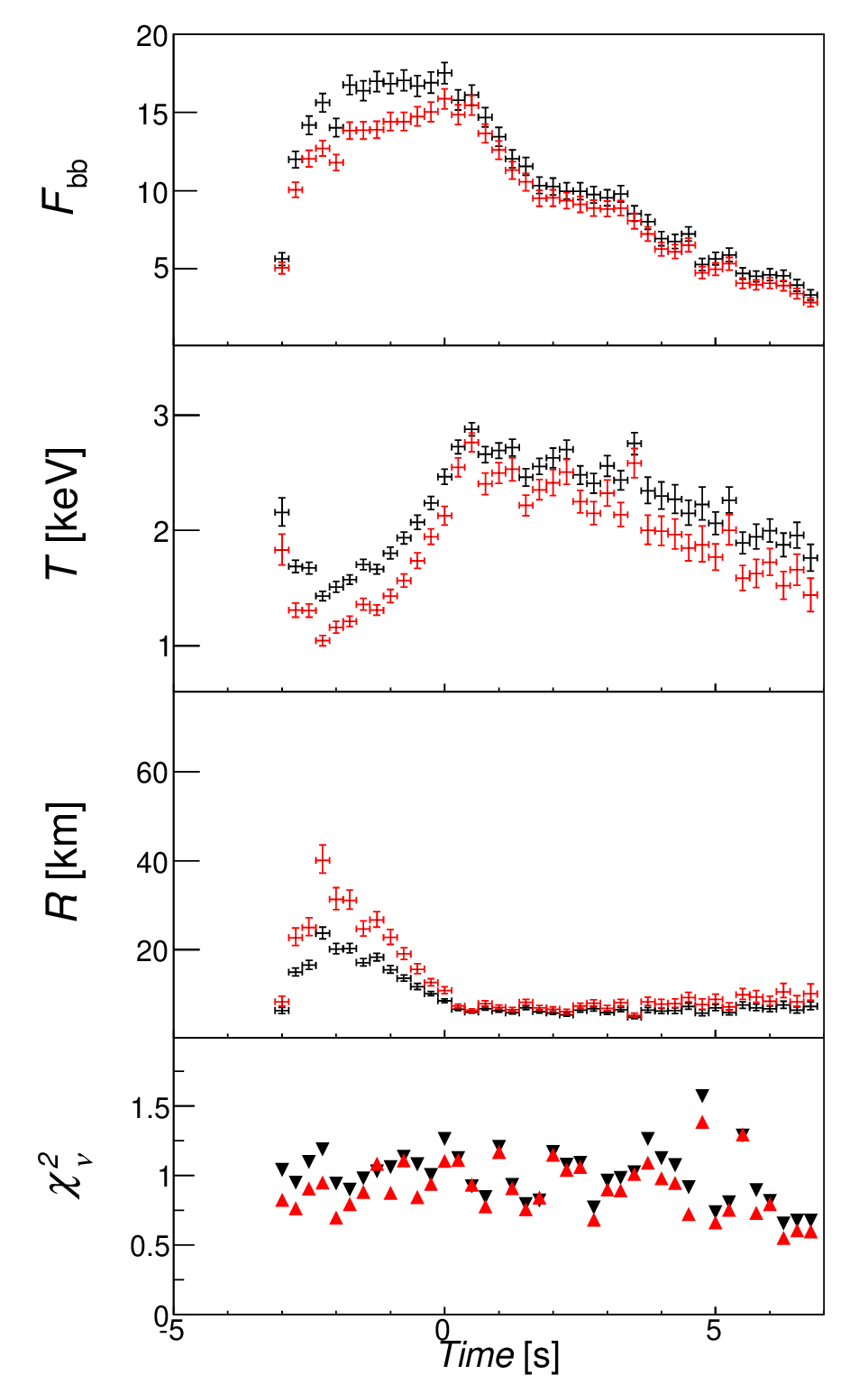}
 \caption{Spectral fitting result of   burst \#8.}
\label{fig_fit_burst_P040420700901}
\end{figure}

\begin{figure}[t]
\centering
 \includegraphics[angle=0, scale=0.4]{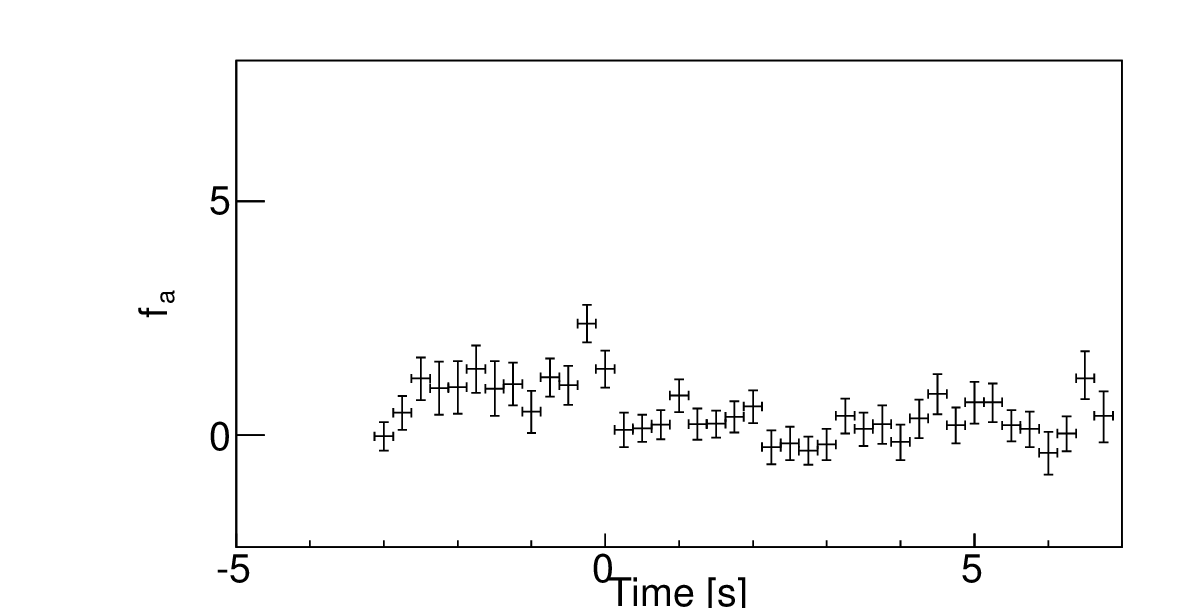}
\includegraphics[angle=0, scale=0.4]{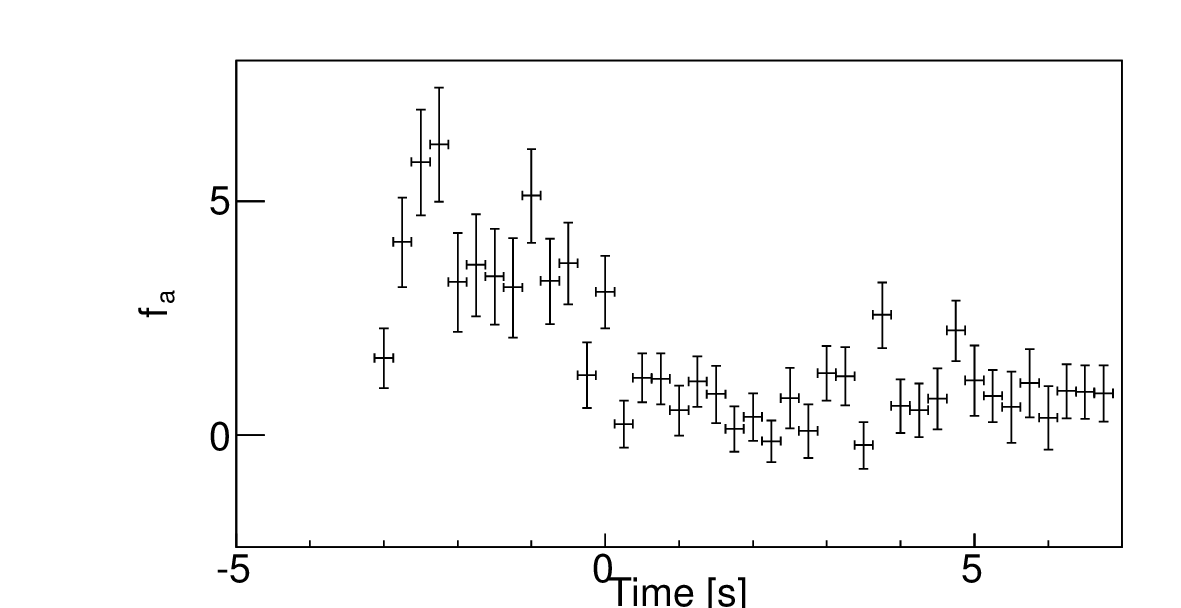}
 \caption{The evolution of $f_{a}$ for burst \#1 (left) and burst \#8 (right).}
\label{fig_fit_burst_fa}
\end{figure}

\begin{figure}[t]
\centering
 \includegraphics[angle=0, scale=0.8]{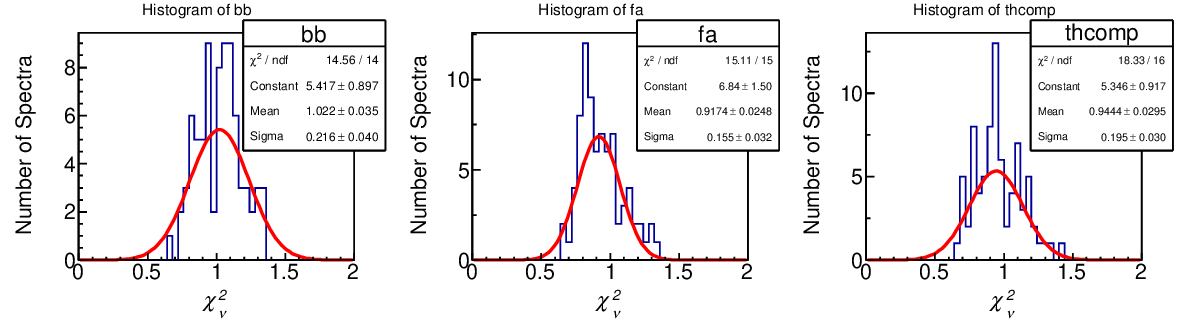}
 \caption{
 Histograms show the distributions $\chi^{2}_{\upsilon}$ values obtained from fitting X-ray burst spectra by
  a pure blackbody (left), $f_{a}$ model (middel) and convolution thermal-Comptonization model (right),
  respectively. The solid lines show the
  shows the Gaussian  model that best describes the data.}
\label{fig_fit_burst_chi}
\end{figure}

\begin{figure}[t]
\centering
\includegraphics[angle=0, scale=0.4]{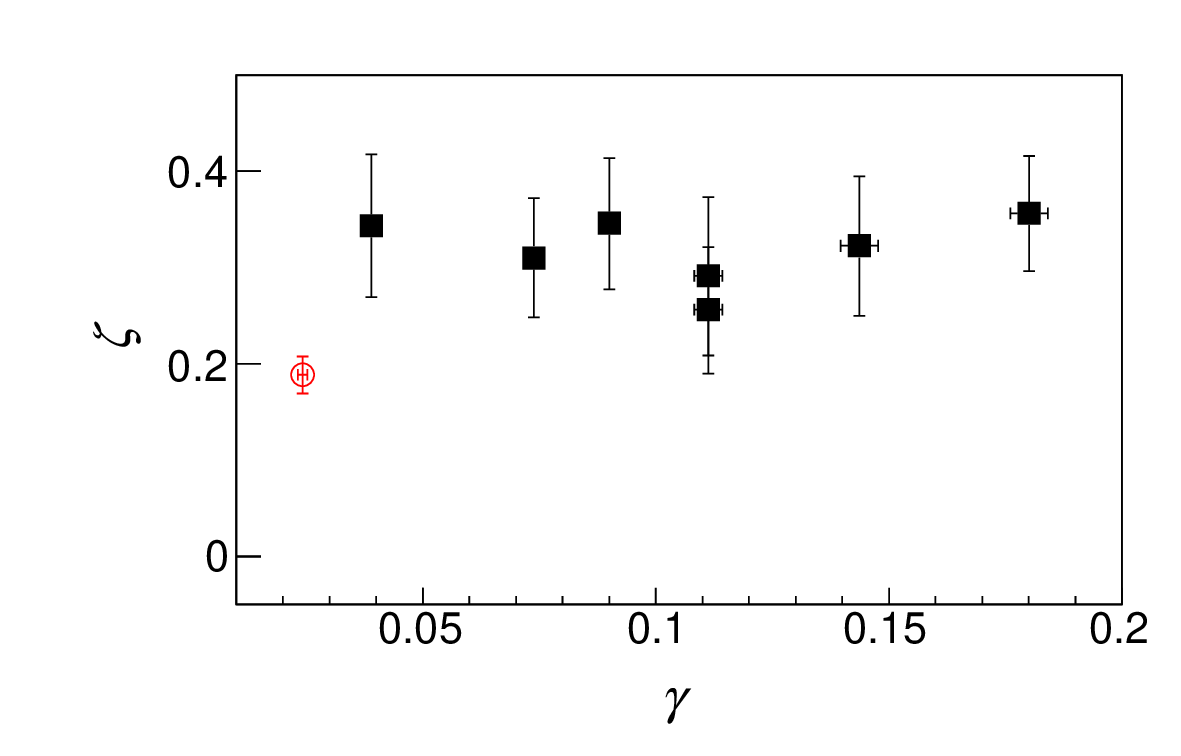}
\includegraphics[angle=0, scale=0.4]{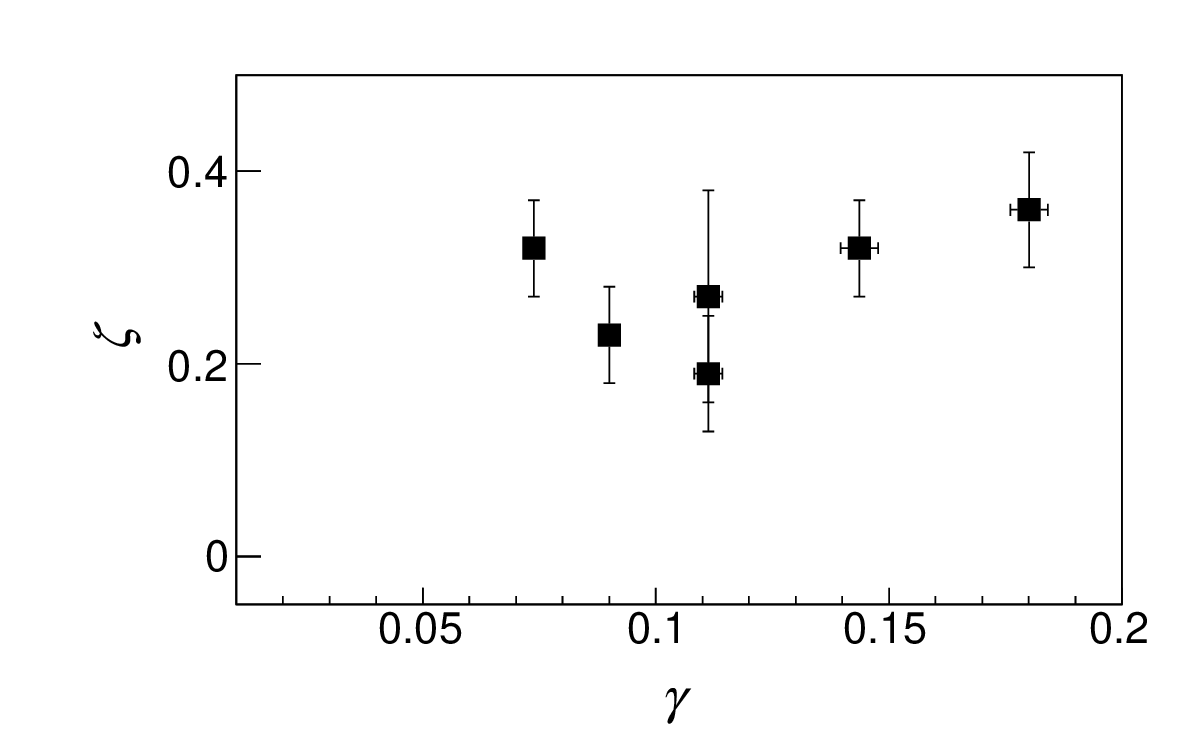}
\caption{  The fraction of flux enhancement $\zeta$  (Comptonization fraction) during bursts against $\gamma$ (accretion rate) for bursts detected by Insight-HXMT (black) and NICER (red). {\bf Left: }The fraction of the flux   increase $\eta$ during bursts is calculated by $\zeta$=$\frac{f_{\rm a}^{p}\times F_{\rm pre-burst}}{F_{p}^{\rm bb}}$, where $f_{\rm a}^{p}$, $F_{\rm pre-burst}$ and $F_{p}^{\rm bb}$ are the maximum of $f_{\rm a}$, the bolometric flux of the pre-burst/persistent emission and the peak bolometric flux of the bursts, respectively; $\gamma$ is the ratio of the pre-burst flux to the Eddington flux.
All bursts reported by \citet{Chen2019}, \citet{Chen2022a} and \citet{Jaisawal2019} are also included.
{\bf Right: The fraction of the flux   increase $\eta$ during bursts is calculated by $\zeta$=$\frac{F_{\rm thcomp}-F_{\rm bb}}{F_{p}^{\rm bb}}$, where $F_{\rm thcomp}$, $F_{\rm bb}$ and $F_{p}^{\rm bb}$ are the maximum of the flux of thcomp*bb derived by cflux, the bolometric flux of bb and the peak bolometric flux of the bursts, respectively; $\gamma$ is the ratio of the pre-burst flux to the Eddington flux.
The bursts reported by \citet{Chen2022a} is also included.}
}
\label{fig_fa}
\end{figure}

We notice that there are still some residuals in the bursts' spectral fitting by the thcomp model (Figure \ref{fig_spec_residual}), and then we take this spectrum as an example and thaw one of the thcomp parameters ($\tau$, $T_{e}$, $f_{sc}$), resulting in reducing these residuals and smaller reduced $\chi^{2}_{\upsilon}$ (Figure \ref{fig_spec_residual_1}); e.g., $\tau$=26, $T_{e}$=3.7 keV and $f_{sc}$=1 result in $\chi^{2}_{\upsilon}$=0.92, 0.93, and 1.01 respectively (all with d.o.f. 80).
The large optical depth is reminiscent of the two blackbodies model, which was also used for burst spectral fitting \citep{Kashyap2022}.
We also attempt to use this model to fit the burst spectra, and result in $\chi^{2}_{\upsilon}$=0.93 (with d.o.f. 79) and the inferred radii are $5\pm1$ km and $12\pm3$ km, as shown in Figure \ref{fig_spec_residual_1}.
The data prevent us from distinguishing which model could fit the burst spectra best, thus the joint observation of NICER and Insight-HXMT could solve this problem.
Moreover, recalling that the temperature change of the corona could explain the hard X-ray shortage of the persistent emission during bursts in the low/hard state, the two blackbodies model can not explain this phenomenon.
Other models, e.g., burst reflection by the disk and NS atmosphere model carbatm/hatm \citep{Suleimanov2011,Suleimanov2012,Suleimanov2018} in Xspec, are also tried to fit the burst spectra, as we did in \citet{Chen2019}. However, neither could alleviate the residuals at soft X-ray and hard X-ray bands simultaneously.

\subsubsection{Different profiles of the rising bolometric flux during the PRE phase for the two groups bursts}

\begin{figure}[t]
\centering
    \includegraphics[angle=0, scale=0.28]{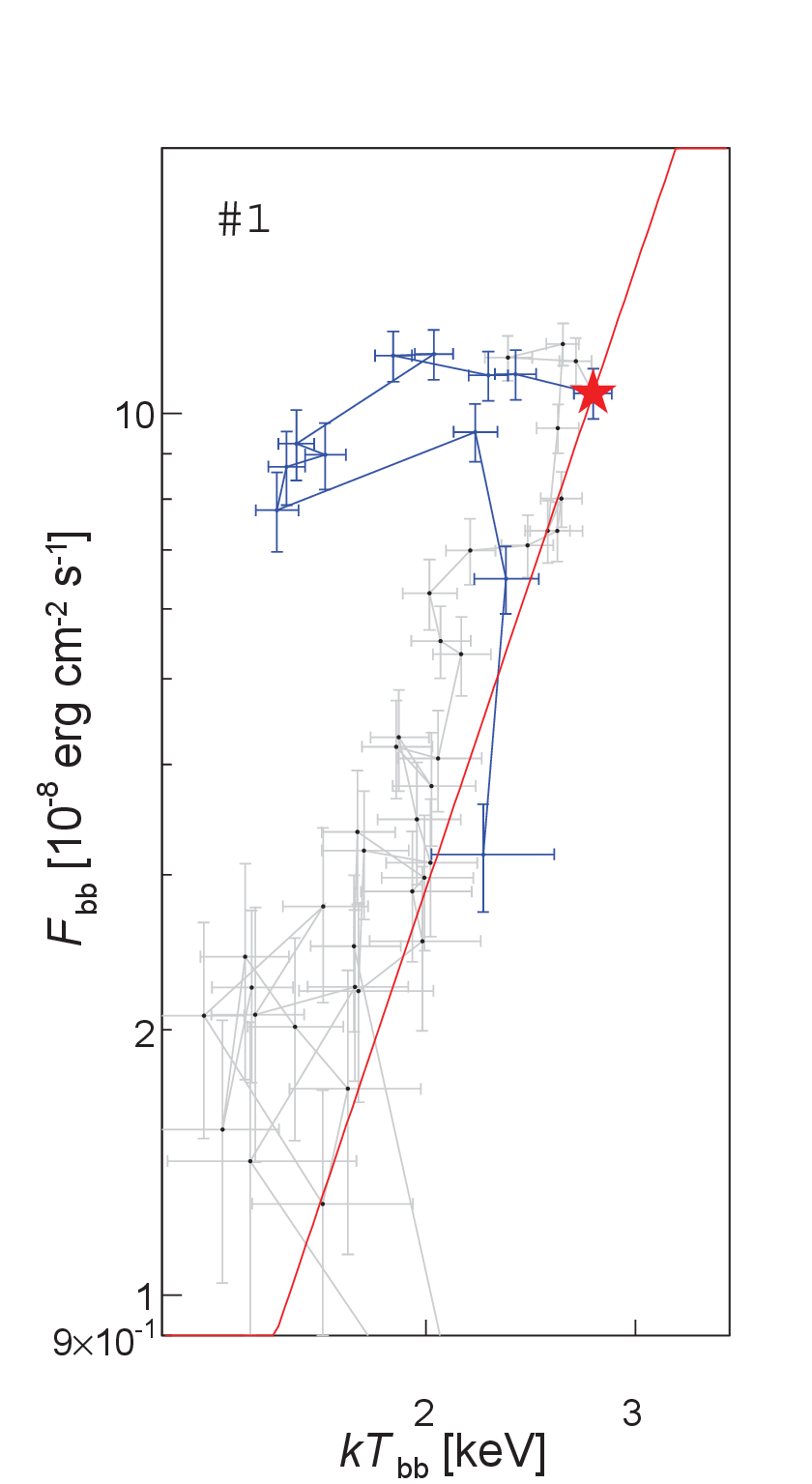}
    \includegraphics[angle=0, scale=0.28]{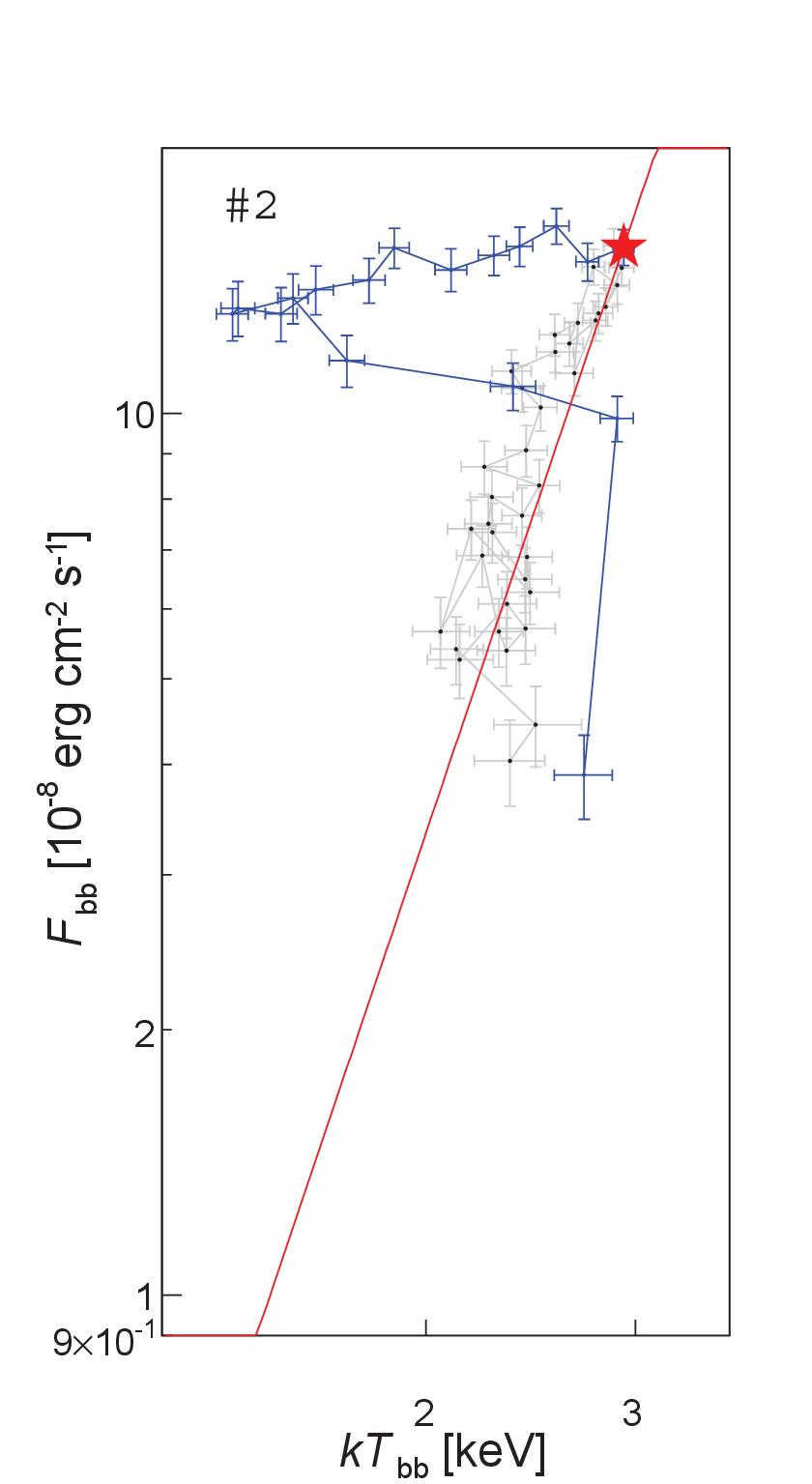}
    \includegraphics[angle=0, scale=0.28]{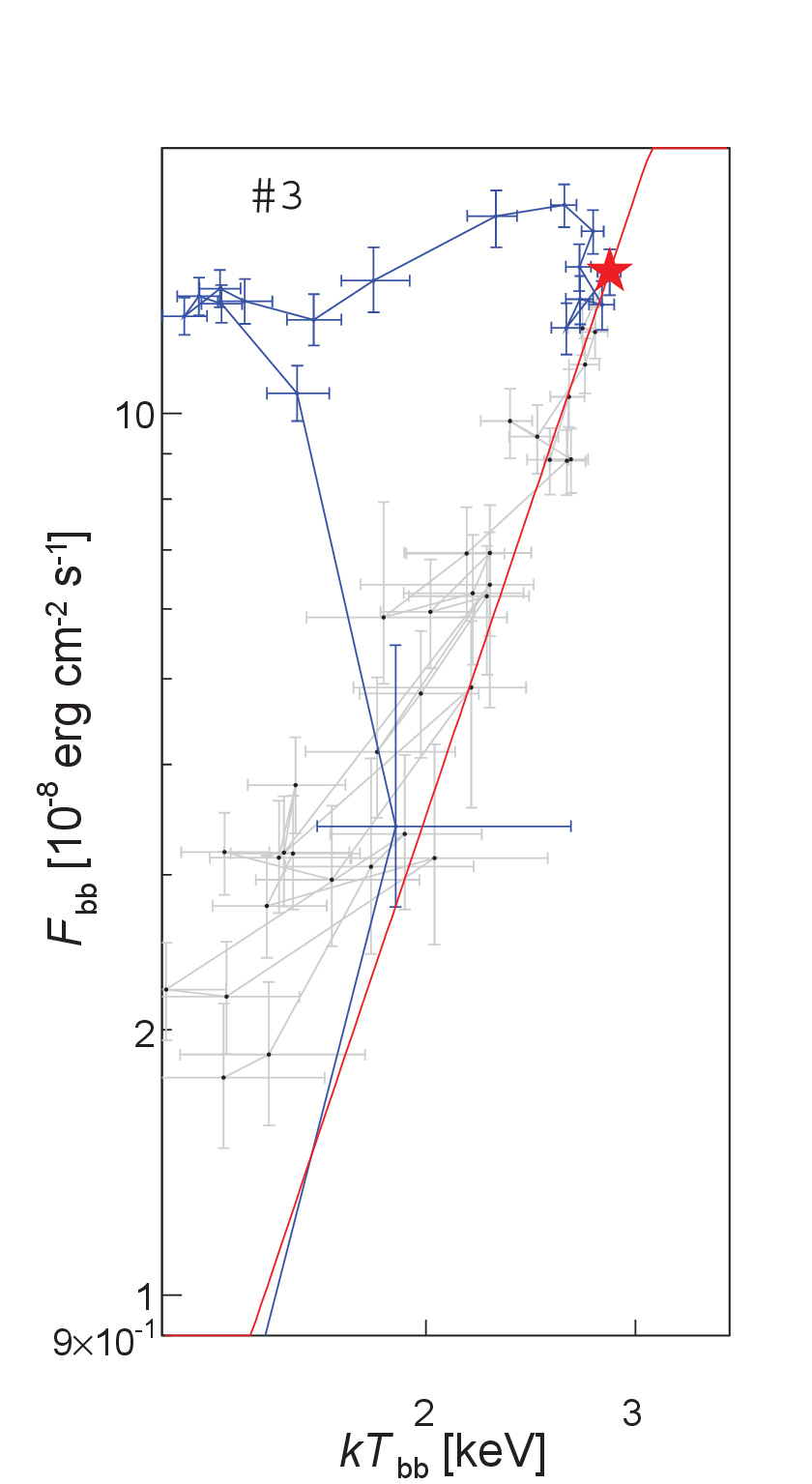}
    \includegraphics[angle=0, scale=0.28]{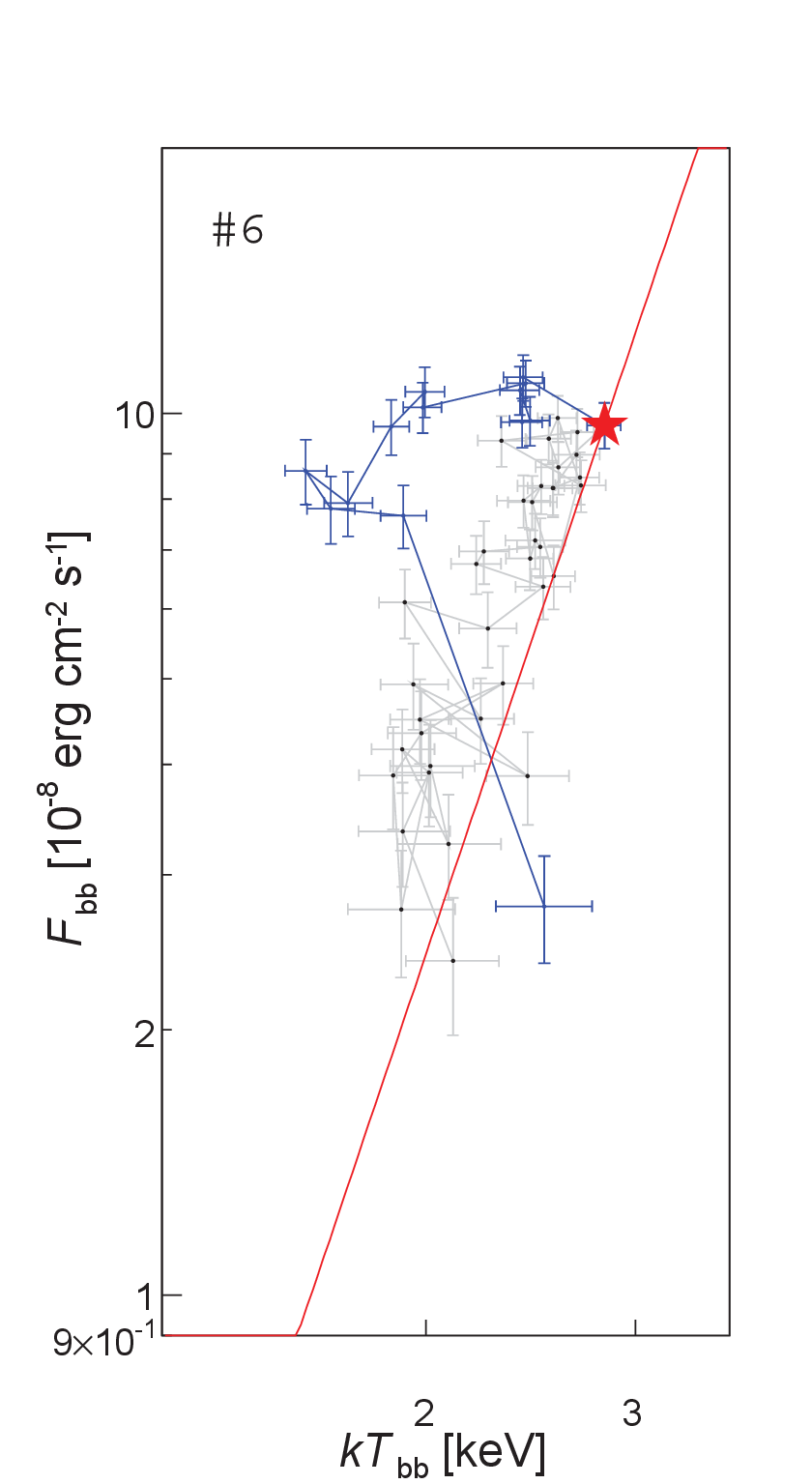}
    \includegraphics[angle=0, scale=0.28]{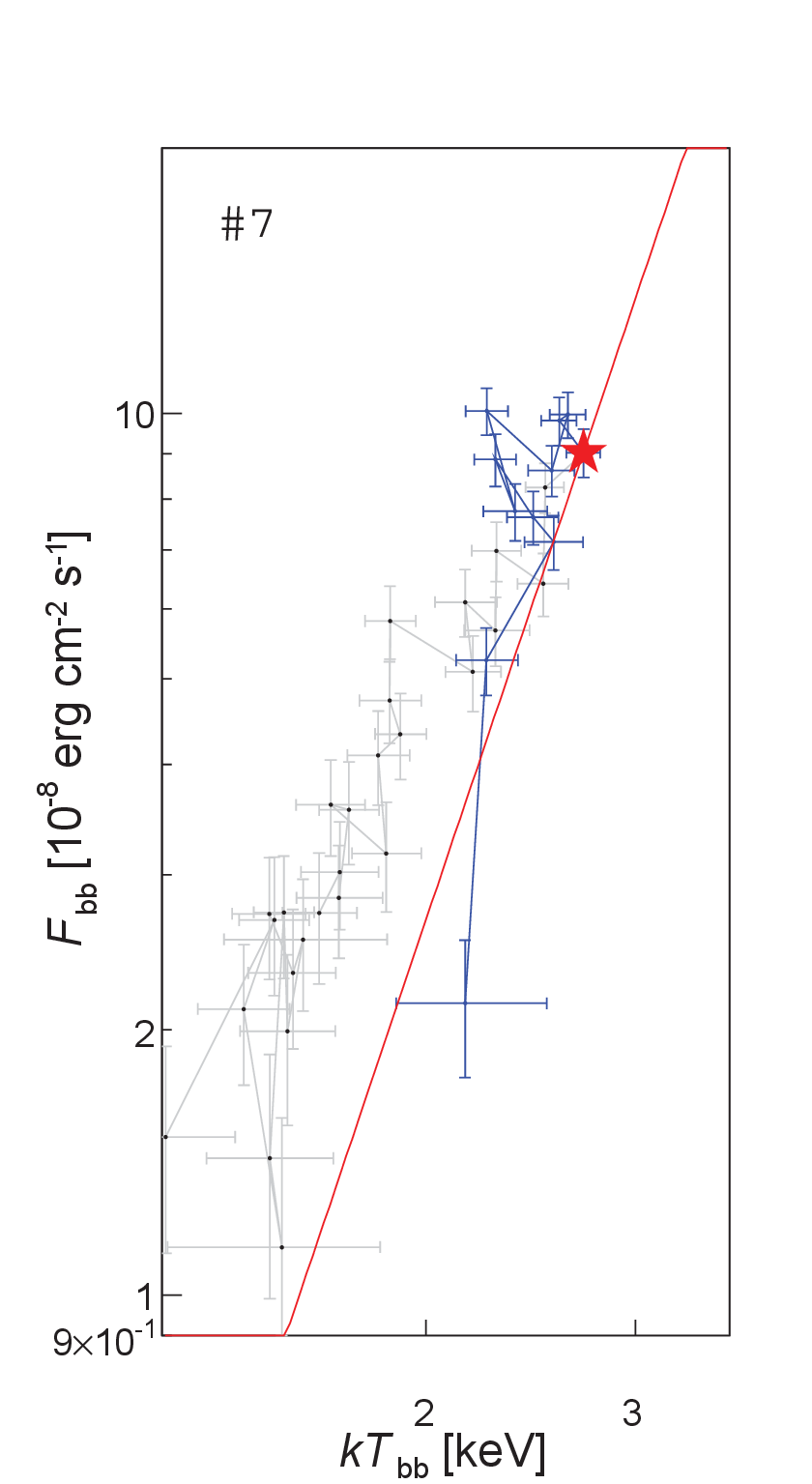}
    \includegraphics[angle=0, scale=0.28]{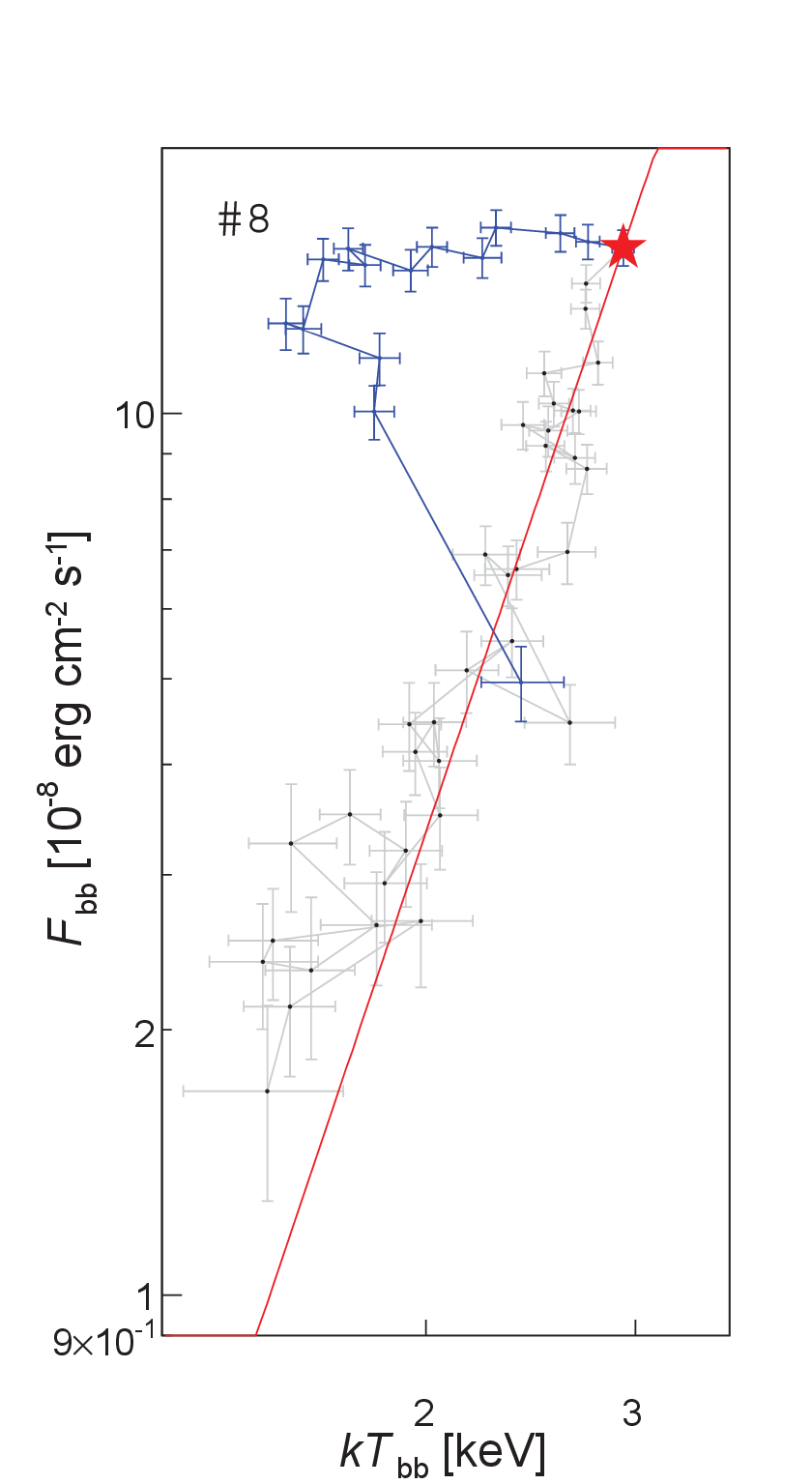}
    \includegraphics[angle=0, scale=0.28]{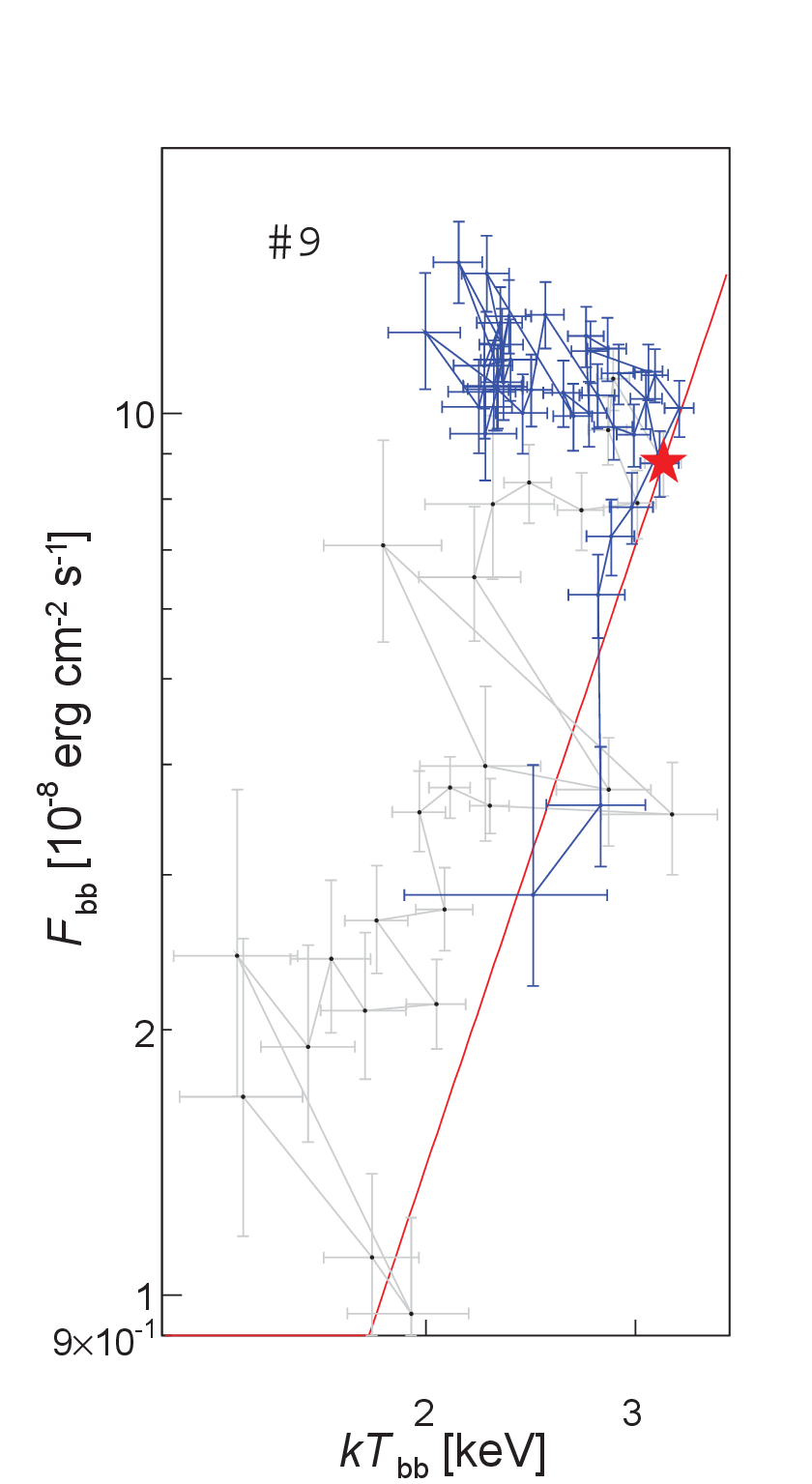}
        \caption{
        Burst flux vs. blackbody temperature of the PRE bursts \#1, \#2, \#3, \#6, \#7, and \#8, and \#9. The red diagonal line in the plot represents the line of constant radius   under a distance $d$=4 kpc to the source, which is derived from the touchdown time of each burst (marked by the red star); the green  and gray points indicate the data points before and after the touch-down times. Please note, for bursts \#3 and \#8, the parameters are derived from the blackbody  model, and the others are from the $f_{a}$ model; for burst \#8,  lacking of LE detection, we ignore the data points with larger error bars in the middle of the PRE phase with temperature $<$ 2 keV, which are beyond the ME \& HE's detection.}
\label{fig_t_f_6.918km}
\end{figure}

A useful way to display the burst spectral evolution is a diagram of $\log F_{\rm bb}$ versus $\log kT_{\rm bb}$, as shown in Figure \ref{fig_t_f_6.918km},
using the parameters derived from the $f_{a}$ model for the PRE bursts. If the whole NS surface emits as a single-temperature blackbody and a constant color correction factor, the burst flux $F$ should scale as $kT_{\rm bb}^{4}$ in the flux-temperature diagram, and the slope represents the emitting area in the double logarithmic coordinates \citep{Guver2012}.
The diagonal line in the plot represents the line of constant radius assuming a distance $d$=4 kpc to the source, which is derived from the touchdown time (marked by the red star in Figure \ref{fig_t_f_6.918km}) of the bursts.

From the diagram, in the decay phase (gray points in Figure \ref{fig_t_f_6.918km}), it is apparent that the bursts approximately follow the expected relation $F_{\rm bb}\propto~T_{\rm bb}^{4}$ but the radius shows a trend of increase as the bursts decay (after the touch-down time).
In the PRE phase (blue points in Figure \ref{fig_t_f_6.918km}), the bursts depart from the $F_{\rm bb}\propto~T_{\rm bb}^{4}$ relation and locate at the left of the red line, which indicates the larger radii.
There are two junctions between the blue points and the red line: the upper one corresponds to the touch-down time, and the lower one corresponds to the time when the atmosphere is just lifted from the NS surface.

We notice that for the bright PRE bursts located at the high/soft state--bursts \#1, \#2, \#3, \#6, \#8, and \#9, the fluxes of the two junctions are different, i.e., the upper one is $\sim$ 1.5--1.8 times bright as the lower one, e.g. the two fluxes for burst \#3 are $10.41_{-0.71}^{+0.72}$ and $16.08_{-0.81}^{+0.81}$, which should be the same value since both of them are the values of the Eddington limit.
This could be related to the occlusion by the disk, which closely resembles the behavior of 4U~1730--22 \citep{Chen2022b}.
However, the fast-rising timescale prevents us from getting the flux/radius when the burst just spreads over the whole NS surface, by which the inclination angle could be derived based on the two values.

The last burst, \#9, behaves that it follows the theoretically predicted trajectory---the tracks of the expansion and contraction phase overlap each other in the diagram, and the junctions of the blue points and the red line are located at the roughly the same point in the diagram, which may indicate that there is no occlusion by the disk.



\section{Discussion}
 \subsection{X-ray continuum evolution}

The CCD indicates that the outburst evolved from the high/soft state (the banana state) to the low/hard state (the island state) during Insight-HXMT's observation in July and August 2022.
Adopting the thermal-Comptonization model thcomp in XSPEC, the spectral fitting results indicate that the thermal components are different for the two states above, i.e.,
the emissions of the accretion disk and the NS surface respond to the high/soft state and the low/hard state, respectively.
Furthermore, the accretion disk is close to the NS surface with an inner disk radius
$<$ 20 km, which is consistent with the conclusion below based on the burst flux evolution of the PRE phase.
For the parameters of the thermal-Comptonization model, both states have a large scattering/covering factor which indicates most of the thermal emission is involved in the Comptonization process.
The large scattering/covering factor of the thermal emission by the corona makes us favor that the corona has a slab/sandwich geometry, rather than the lamp-post geometry.
Given the obtained temperature and optical depth which deviates from the corona's canonical value ($\tau < 1$, see the review by \citealp{Done2007}), we prefer the corona model--a so-called warm layer \citep{Zhang2000} with temperature $\sim$2--3 keV and optical depth $\sim$5--10, which is around the accretion disk rather than around the NS surface.
For the low/hard state, an optically thin plasma near the NS surface and a hot, quasi-spherical flow is favored, as shown in Figure 2 of \citet{Kajava2014}.

For the bursts in the high/soft state, in the burst decay phase, e.g., after the touch-down time of the PRE bursts, the apparent radius $R_{\rm bb}$ increases as the burst decays;
e.g., $R_{\rm bb}$ of burst \#2 increases from $\sim$ 5 km to $\sim$ 8 km along with the burst decay from the touch down time to the tail,
which corresponds to the color factor $f_{\rm c}$ decrease by a factor of 0.8.
The enhancement of the apparent radius during the burst decay phase in the high/soft state is inconsistent with previous observations and models, in which the spreading layer on the NS surface will cause a roughly constant $f_{\rm c}$ during the burst's decay \citep{Kajava2014}.

\subsection{Implications of the flux enhancement during bursts}

Similar to our previous work \citep{Chen2022a,Chen2022b},
 adopting the convolution model with the parameters in the persistent emission fitting to convolve the blackbody emission to fit the burst spectra,
the scenario is also applied to the five bursts from 4U~1730--22, resulting in an equally good fit compared with the $f_{a}$ model during the bright/PRE phase. Under this scenario, the radius of the photosphere is underestimated with the canonical blackbody model or the $f_{a}$ model.

We notice that there is a large scatter of the peak $f_{a}$,  e.g., 2.5--6 in this work for 4U~1608--52 by Insight-HXMT, 1--7 for 4U~1636--536 by NICER \citep{2022ApJ...935..154G} and 1--10 for Aql X-1 by NICER \citep{2022MNRAS.510.1577G}, but a mild scatter of the  fraction of flux enhancement during bursts (ratio of the maximum flux increment against the burst peak flux, as shown in Figure \ref{fig_fa}).
This means that the energy  gain of each burst photon 
 is up to $\sim$20\%--40\%, even though the bursts have different peak fluxes and locate at different accretion rates.
A reasonable explanation is that the  probability/Comptonization-fraction of the burst photons up-scattered by the corona is roughly the same for every PRE burst, which is consistent with the little changes of the corona parameters ($\tau$ and $T_{\rm e}$) for the pre-burst emission of the PRE bursts in the high/soft state.
The result above is on some level consistent with the PRE bursts detected by RXTE/PCA with $f_{a}\gamma<0.84$ for all these bursts ($\gamma$ is the preburst accretion
flux as a fraction of the Eddington flux) \citep{Worpel2013,Worpel2015}.
 Moreover, we notice that there is a larger scatter of $f_{a}\gamma$ for the PRE bursts given by \citet{Worpel2013,Worpel2015} based on RXTE/PCA data.
However, the above estimation does not take the burst peak flux into account; since the difference of the peak flux of the PRE bursts is up to 2, and the burst flux  should affect $f_{a}$ (e.g., \citealp{2022MNRAS.510.1577G,2022ApJ...935..154G}).

Furthermore, if the flux enhancement during bursts is related to the increased accretion rate  induced by the Poynting-Roberson effect, i.e., the increased accretion rate $\dot{M}$ \citep{1989ApJ...346..844W}during a burst with luminosity $L_{\rm burst}$ is
 \begin{eqnarray}
  \label{accretion-rate}
   \dot{M}\sim\frac{L_{\rm burst}}{c^{2}}\times{\rm Max}(R_{\rm NS}/R_{\rm disk}, {\rm d}H/{\rm d}R_{\rm disk})
  \end{eqnarray}
  for the NS radius $R_{\rm NS}$, inner disk radius $R_{\rm disk}$, and height of inner disk $H$.
   Among them, ${\rm Max}(R_{\rm NS}/R_{\rm disk}, {\rm d}H/{\rm d} R_{\rm disk}$) is the geometric factor of the disk for the NS surface.
  Since the inner part of the accretion disk is geometrically thin, optically thick, the ${\rm d}H/{\rm d}R_{\rm disk}$ is negligible compared with $R_{\rm NS}/R_{\rm disk}$.
  Assuming 
  the efficiency of transforming gravitational energy to radiation is constant for different bursts,
    the energy gain of each burst photon
    $\zeta$ is inversely proportional to  $R_{\rm disk}$.
  As a result, bursts in the low/hard state with a larger $R_{\rm disk}$ should have a smaller $\zeta$.
 Obviously, the observed constant  $\zeta$ is inconsistent with the deduction above of the Poynting-Roberson effect.

 Additionally, another  deduction of the Poynting-Roberson effect is that the NS with a bigger spin frequency $\nu$  has a smaller $\zeta$ \citep{1992ApJ...385..642W}.
 Study of the relation between  $\zeta$ and $\nu$ in different bursters could be another test of the Poynting-Roberson effect.  There are, of course, many simplistic assumptions above, e.g., $R_{\rm disk}$ is invariable during bursts. Further studies of detailed time-dependant burst emission which including both the Poynting-Roberson effect and the Comptonization by the corona could be needed.

 If the soft excess during burst is caused by the reflection by the accretion disc, the inner disk radius should significantly affect the reflection flux  \citep{2022MNRAS.509.1736S}.
 Similar to the case of the increased accretion rate  induced by the Poynting-Roberson effect, an inference is that the irradiation should be inversely proportional to $R_{\rm disk}$, i.e., $\zeta$ should be much small than the  bursts locating at the high/soft state with $R_{\rm disk}\sim$ 10 km, which is inconsistent with the observed constant $\zeta$.

 The above analysis prefers that the Comptonization of the burst emission by the corona causes the soft excess, it seems to be reasonable because that the emergent burst emission firstly encounter with the corona since the corona is thought to be closed to the NS and have a larger geometric factor for the burst emission than the disk.
 For bursts in low/hard states with corona temperature $kT_{\rm e}$ tens keV, the burst could have a significant effect on the temperature of the corona, e.g., the bursts in low/hard state of GS~1826--238 immediately cool $kT_{\rm e}$  from $\sim$20 keV to $\sim$3.4 keV \citep{2005ApJ...634.1261T},  which is reminiscent of the persistent spectral transition from low/hard state to high/soft state with $kT_{\rm e}$  $\sim$2--3 keV.
 Under this condition,
 for the bursts both in low/hard states and high/soft states,
  the corona property seems to be similar,  and the similarity of the Comptonization fraction could also be predictable.

 We also notice that the bursts above used for estimation of $\zeta$  are all PRE bursts,
 dozens of non-PRE and several PRE bursts in 4U~1636--536 detected by Insight-HXMT have also used to calculated $\zeta$, which also shows a roughly constant $\zeta\sim$ 0.3 (Yan Z, et al. in preparation).
 Thus may indicate that the constant of $\zeta$  is universally valid for bursters.

The hard X-ray shortage during bursts was detected in several NS X-ray binaries during these low/hard states (e.g., \citealp{Chen2012,Chen2022}), which is thought to be due to the cooling of the corona by the burst.
However, in this work, for the last burst which was located at the low/hard state, the hard X-ray shortage was not detected.
Now we will estimate the detection significance if the hard X-ray shortage during the burst exists.
Assuming the average shortage fraction is 50\% during the PRE phase,
considering persistent emission in 40--100 keV $F_{\rm pre}\sim$ 17 cts/s, the instrument background $\sim$182 cts/s, and the interval of the PRE phase is 20 s, the significance of the
shortage should be
 \begin{eqnarray}
  \label{excess}
  \sigma=\frac{0.5F_{\rm pre}}{F_{\rm bkg}^{1/2}}\times t^{1/2}=2.8.
  \end{eqnarray}
 The estimation above indicates that the shortage is very hard to detect, 
 even if the hard X-ray shortage exists.
 The non-detection of the hard X-ray shortage could be due to the low flux level of the persistent emission in hard X-ray band, e.g., $\sim$ 50 mCrab in the Swift/BAT detection.
Several bursts in observations of Insight/HXMT on the brighter ($>$ 100 mCrab in the Swift/BAT detection) low/hard state of 4U~1608--52 were detected; among them, the hard X-ray shortage is significantly detected with $\sim$ 6 $\sigma$ from one single burst, which will be given in detail in our forthcoming work (Chen, et al. in preparation).

\subsection{Evidence of obscured NS surface during outbursts}

The deviation from the model-predicted flux peaks at the time when the photospheric radius also reaches its peak value, in other words, the flux shortage at the PRE phase, is reminiscent of the disk obscuration on the burst emission, which was also observed in 4U~1728--34 \citep{Shaposhnikov2003}, 4U~1820--30 \citep{Shaposhnikov2004}, 4U~1730--22 \citep{Chen2022b} and a burst \citep{Chen2022a} in 2020 detected from this source.
  Under the standard disk theory, the evaporation of the inner disk by the burst emission could be evaporated less than 0.1 s \citep{Shaposhnikov2003}.
The anisotropy factor $\xi_{\rm b}$ is defined as
 \begin{eqnarray}
\xi_{\rm b}=\frac{L_{\rm bb, model}}{L_{\rm bb,\infty}},
 \label{anisotropy}
\end{eqnarray}
where $L_{bb, \infty}$ is the observed bolometric flux by a distant observer, $L_{\rm bb, \rm model}$ is the model-predicted (obscuration-free) bolometric flux of the photosphere; it is supposed to be unity for an obscuration-free burst emission and to be larger than unity for a geometric anisotropy, e.g., the NS's lower hemisphere is screened by the disk.
Since the lift-up time and the touch-down time of the PRE burst should have the same bolometric flux, we suggest that the flux shortage of the lift-up time is due to the obscuration by the disk.
However, for the PRE bursts in the high/soft state in this work, lacking data just at the lift-up time prevents us from deriving the anisotropy factor of the burst onset time, which was used to estimate the inclination angle of the system derived from a PRE burst with lift-up time detected in this source \citep{Shaposhnikov2003,Chen2022b}.
The fast rise time scale could also dilute the flux derived from the spectral fitting of the spectra at the burst onset time.
Considering this condition, the first data point of the burst spectral evolution is not used to infer $\xi_{\rm b}$.

For $\xi_{\rm b}$ in the rising PRE phase, it could be obtained from the ratio between the model-predicted area and the detected area of the photosphere.
Here, we present the observed (apparent) photosphere radius $R_{\infty}$ inferred from the observed (color) photosphere temperature $kT_{\infty}$ or $kT_{\rm bb}$ in the PRE phase.
Based on $kT_{\infty}$ as a function of several input parameters (see equation 1 in \citet{Shaposhnikov2003} or equation 3 of \citet{Shaposhnikov2004}), under the assumption that the burst is helium-rich with $Y_{\rm He}=1.0$\footnote{
The burst durations $\tau$ are less than 10 s, which indicates that the bursts are helium-rich. Under this condition, $Y_{\rm He}=1.0$ is taken. Please note that the different value of $Y_{\rm He}$, e.g., $Y_{\rm He}$=0, 0.73,  should not change the trend of the derived burst temperature and the anisotropic degree below.}, the mass of NS is $1.4M_{\odot}$,
 and the color factor $f_{\rm c}$ is a constant of 1.7 during the PRE phase,
$R_{\rm bb}$ could be derived. We copy this equation as below:
\begin{eqnarray}
 kT_{\infty}=2.1f_{\rm c}\bigg[\frac{lm}{(2-Y_{\rm HE})(1+z)^{3}r_{6}^{2}}\bigg]^{1/4} ~{\rm keV}.
  \label{kt_bb}
\end{eqnarray}
We only care about the parameters of the PRE phase, thus the dimensionless luminosity in unit of the Eddington luminosity $l$ is unity; $m$ is the NS mass in unit of $M_{\odot}$, $r_{6}$ is the NS photosphere radius in unit of 10 km measured by a local observer.
Assuming that $f_{\rm c}$ is a constant value--$f_{\rm c}$=1.7 during the PRE phase, the above estimation is reduced to solving a quadratic equation with one variable (also for the equation 10 of \citet{Titarchuk2002}).

\begin{figure}[t]
\centering
\includegraphics[angle=0,scale=0.40]{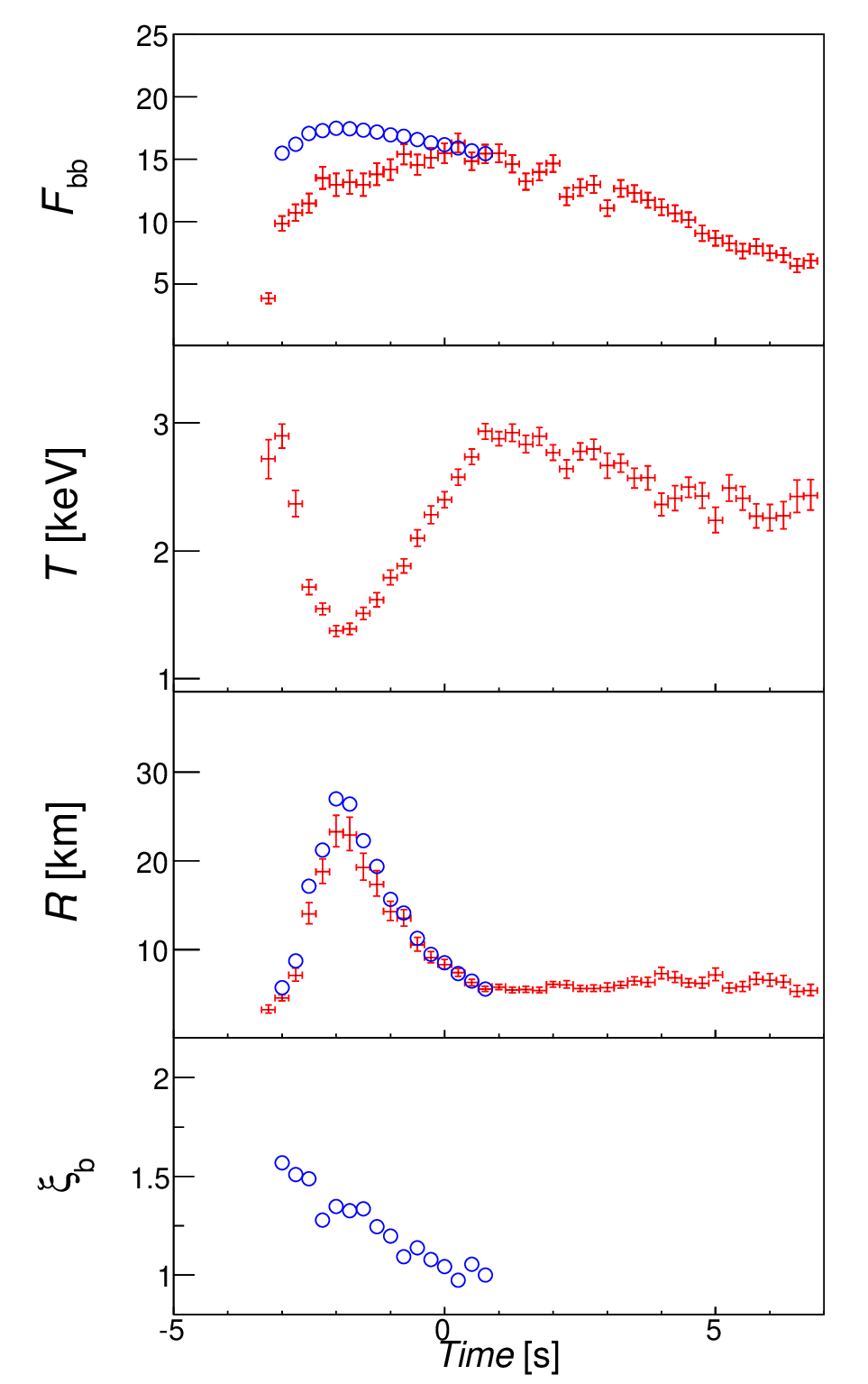}
\includegraphics[angle=0,scale=0.40]{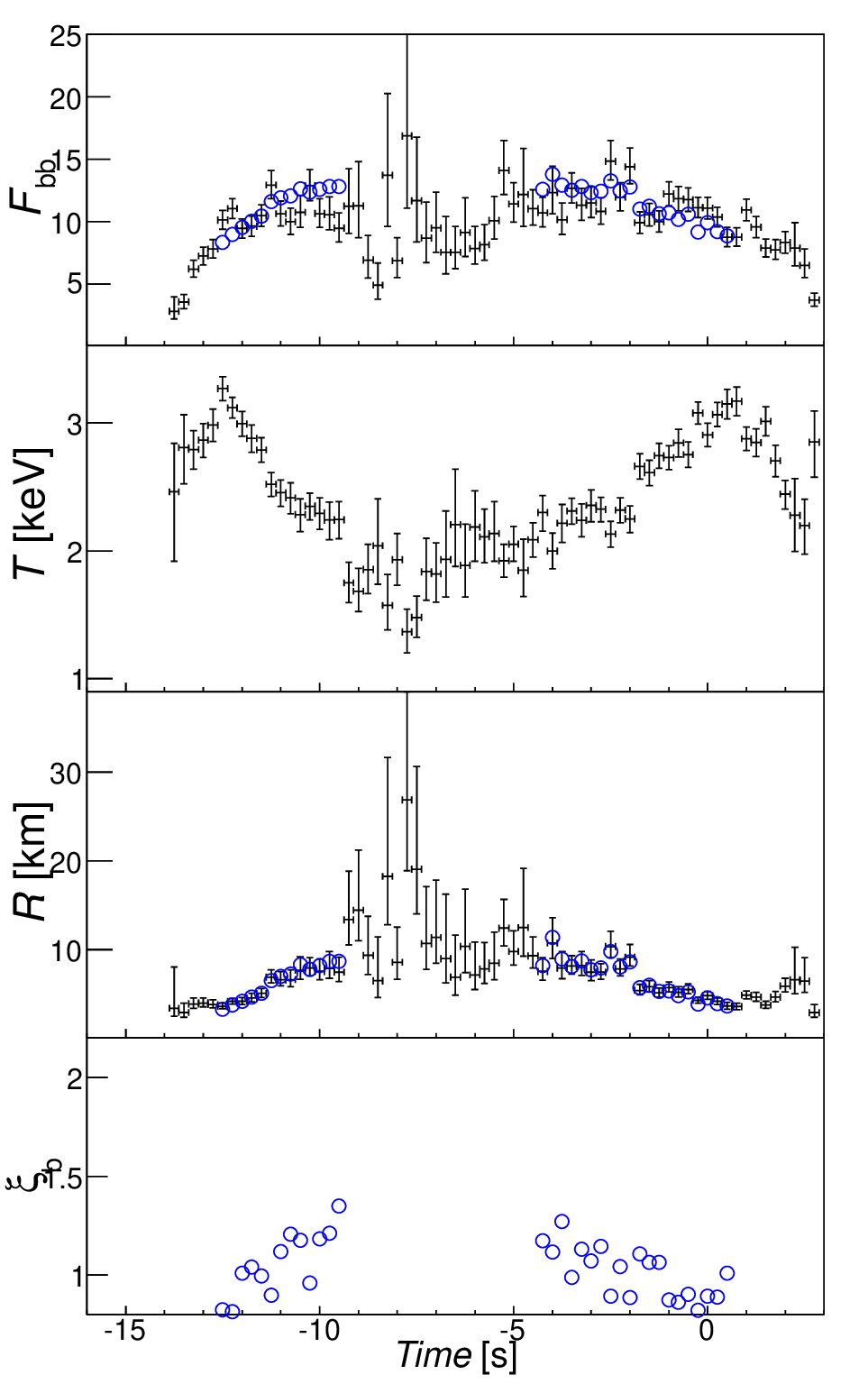}
         \caption{ For burst \#2 (left) and burst \#9 (right), the model predicted   time evolution of   the
       blackbody bolometric flux $F_{\rm bb}$,    the radius $R$ of NS photosphere  at 4 kpc,  anisotropic degree  $\xi_{\rm bb}$ are marked in blue; for comparison, the corresponding observed parameters, including the temperature $kT_{\rm bb}$, are also given.
       Please note for burst \#9,  lacking of LE detection, we ignore the data points with larger error bars in the middle of the PRE phase with temperature $<$ 2 keV, which are beyond the ME \& HE's detection.
         }
\label{fig_anisotropy}
\end{figure}

We solve the equation above and obtain $r_{6}$.
It is important to note that the color temperature $kT_{\infty}$ is a geometry independent quantity, i.e., the disk obscuration would not change the observed spectral shape (the color temperature $kT_{\infty}$), only changes the observed flux.
Substituting $r_{6}$ ($R_{*}$=$r_{6}\times10$ km) into equation \ref{R_bb}, we obtain the model predicted apparent radius $R_{\rm bb, \rm model}$, as shown in the third panel of Figure \ref{fig_anisotropy}.
For blackbody emission
 \begin{eqnarray}
 L_{}=4\pi R_{}^{2} \sigma T_{}^{4},
  \label{l_bb_apparent}
\end{eqnarray}
where $\sigma$ is the Stefan-Boltzmann constant; since the
color temperature $kT_{\infty}$ is a geometry independent quantity, i.e.,
the model predicted temperature $T_{\rm bb, \rm model}$ and the observed temperature $T_{\rm bb}$ is the same, then $\xi_{\rm b}$ is derived based on equation \ref{anisotropy}: 
 \begin{eqnarray}
\xi_{\rm b}=\frac{L_{\rm bb, model}}{L_{\rm bb,\infty}}=\bigg(\frac{R_{\rm bb, model}}{R_{\rm bb,\infty}}\bigg)^{2}.
 \label{anisotropy_1}
\end{eqnarray}
Thus, the model predicted flux $L_{\rm bb, \rm model}$ of the PRE phase and the anisotropy factor $\xi_{\rm b}$ are calculated, as shown in the top panel and the bottom panel of Figure \ref{fig_anisotropy}.

In theory, for each burst of a given NS with the same $Y_{\rm He}=1.0$, the inferred $R_{*}$, $T_{*}$ and $L_{*}$ in the touch-down time should be the same.
However, there is slight diversion between the detected values and model predicted values, e.g., $F_{\rm d}$ is different for each burst, as shown in Table \ref{tb_burst}.
To study the evolution of $\xi_{\rm b}$,
we assume $\xi_{\rm b}$=1 for the touch-down time, i.e., the model predicted values of $R_{*}$ are normalized by the radius in the touch-down time.
As shown in the left panel of Figure \ref{fig_anisotropy}, for burst \#3, at the beginning of the PRE phase, $\xi_{\rm b}$ reaches maximum $\sim$ 1.7 then decreases in the whole PRE phase, at last reaches minimum of unity, which is consistent with the qualitative analysis above.

\begin{figure}[t]
\centering
     \includegraphics[angle=0, scale=0.40]{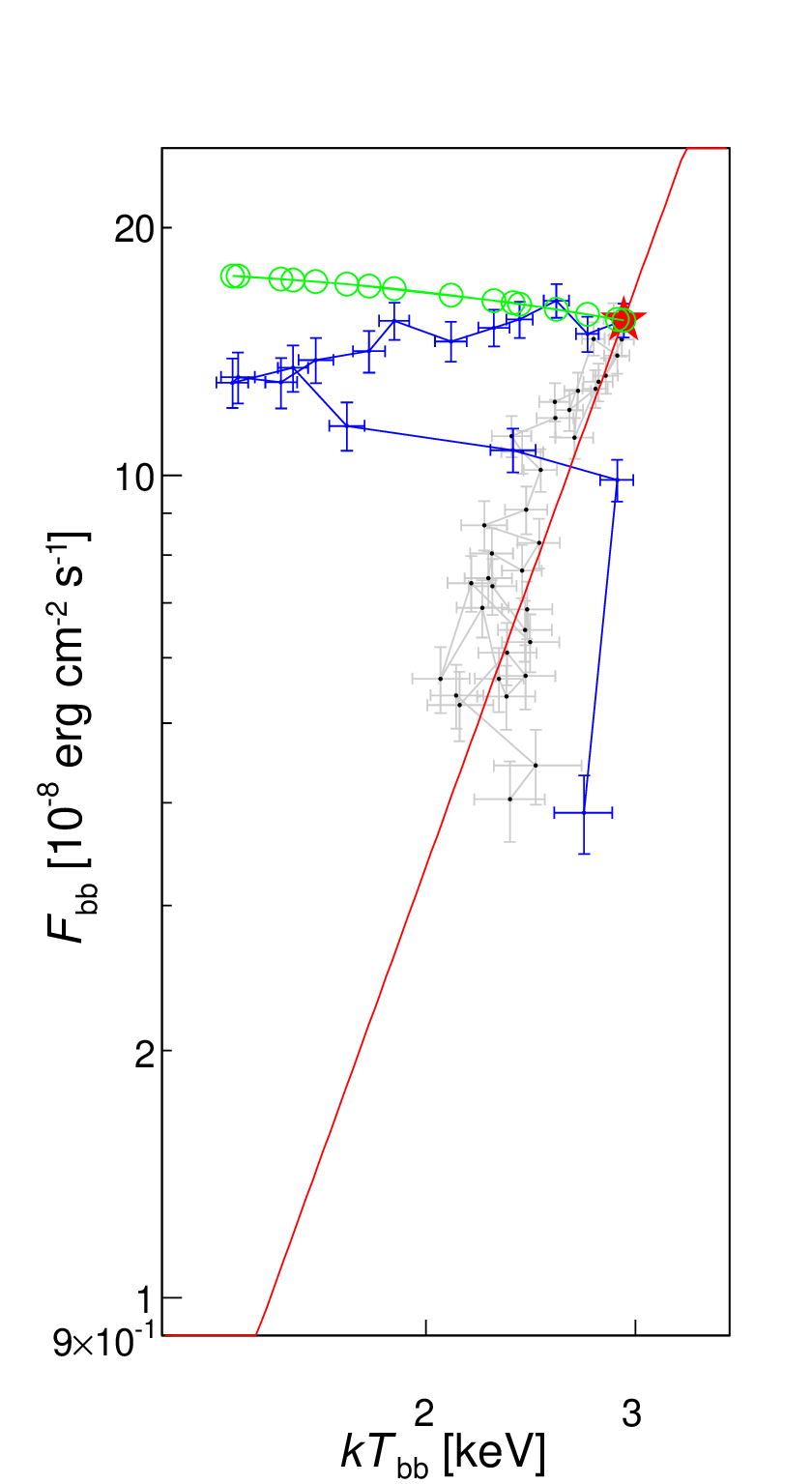}
     \caption{For burst \#2, the model predicted    the diagram  of the burst bolometric flux $F_{\rm bb}$ vs. blackbody temperature  $kT_{\rm bb}$ are marked in green; for comparison, the corresponding observed parameters,   are also given (same as Figure \ref{fig_t_f_6.918km}).
 }
\label{fig_t_f_anisotropy}
\end{figure}

The model predicted evolution of the PRE phase in the diagram of $\log F_{\rm bb}$ versus $\log kT_{\rm bb}$ is shown in Figure \ref{fig_t_f_anisotropy}; the evolution of the rise and decay PRE phases overlap.
The other high/soft state PRE bursts follow a similar evolutionary trend, but not for burst \#9, the burst detected in the low/hard state.
As shown in the right panel of Figure \ref{fig_anisotropy}, for all of them $\xi_{\rm b}<1.4$, and $\xi_{\rm b}$ is roughly equal to unity at the beginning of the rising PRE phase, which is different with other PRE bursts.
Moreover, $\xi_{\rm b}$ appears to rise as the radius expands; however, lacking the data in soft X-ray band prevents us from getting a solid conclusion on this trend.
The model above is also used in the PRE bursts of 4U~1730--22 \citep{Chen2022b}, which were detected in the high/soft state, resulting in a similar trend: $\xi_{\rm b}>1.5$ in the rising part of the PRE phase and attenuates over time in the PRE phase, which implies the universality of the evolution of $\xi_{\rm b}$. For the burst locating at the low/hard state, the absence of the flux shortage of the rising PRE could be caused by the inner accretion disk truncated far away from the NS in the low/hard state, which is consistent with the non-detection of the disk emission in the low/hard state.


In theory, the emergent radiation of a bright burst with a sub-Eddington luminosity could have significant influence on the accretion circumstance, e.g., evaporation/retreat of the inner disk appearing with a decrease of the optical depth with an order of magnitude \citep{Fragile2020}.
However, we notice that the evaporation/retreat of the inner disk is not instantaneous and the timescale is comparable with the interval of the PRE phase, e.g., it takes roughly several seconds for the inner disk to evaporate/retreat,  as shown in Figure \ref{anisotropy}.
This is at least longer by an order of magnitude than the timescale ($<$ 0.25 s) \citep{Fragile2020} of the evaporation/retreat of the inner disk due to Poynting-Robertson drag, which suggests that the evaporation/retreat of the inner disk could be governed by some other mechanisms.

Moreover, another scenario, the concave disk geometry caused by the burst emission could also block the NS surface \citep{He2006}. However, the byproduct of the concave disk--a bright iron emission line is only marginally detected around 6.4 keV even for the bright PRE burst in this work (Figure \ref{fig_spec_residual}).

{\bf Furthermore, there are another  possible explanation for the anisotropically detected during PRE bursts is that the anisotropically of the burst emission from the NS surface is intrinsic, e.g., only parts of the photosphere is lifted by the radiation force  since the rising timescale of the burst too short ($\sim$ 0.5 s,  as shown in Figure \ref{fig_burst_lc}) to spread over the whole NS surface. In addition, the variation of the hydrogen mass fraction during the PRE phase could also
results in different Eddington luminosities and leads to different observed fluxes.} 



\section{Summary and Conclusion}

In this work, we have presented a spectral analysis of nine bursts and persistent emission from 4U~1608--52 in its 2022 outburst observed Insight-HXMT.
The outburst behaves in a typical form of the atoll sources: preceding low/hard state--high/soft state--lagging low/hard state in a time sequence.
For the persistent emission in the high/soft state,
the joint spectra are well fitted by an absorbed convolution thermal-Comptonization model;
for the persistent emission in the low/hard state, the thermal component is replaced with the NS surface's emission and the Comptonization component has a higher temperature of electrons and a smaller optical depth, which indicates a truncated disk in a large distance and an optically thin spherical corona.
The variable persistent emission or the up-scattering of the burst emission by the surrounding hot electrons could reduce the residuals from the blackbody model and explain the constant Comptonization factor (percentage increase) for fluxes of bursts with different accretion rates.
Since the two groups' bursts have different accretion circumstances, e.g., a different inner disk radii, as expected, the bursts emission of the two groups should have different anisotropic degrees: the flux shortage for the bright PRE bursts detected in the high accretion rate state with a small inner disk radius is absent in the PRE burst in the low accretion rate state without the accretion disk detection. The above results imply the accretion rate dependant anisotropy of the burst emission, which was model predicted, e.g., by \citet{Galloway2021} and first observed in this work.
However, there is only one single PRE burst occurring in the low/hard state, the universality of the conclusion needs further verification.

\acknowledgements
This work made use of the data and software from the Insight-HXMT
mission, a project funded by China National Space Administration
(CNSA) and the Chinese Academy of Sciences (CAS).
This research has made use of data and software provided by of data obtained from the High Energy Astrophysics Science
Archive Research Center (HEASARC), provided by NASA’s
Goddard Space Flight Center.
This work is supported by the National Key R\&D Program of China (2021YFA0718500) and the National Natural Science Foundation of China under grants  12173103, U2038101, 12233002.
This work was partially supported by International Partnership Program of Chinese Academy of Sciences (Grant No.113111KYSB20190020).

\bibliographystyle{plainnat}

\begin{thebibliography}{99}



\bibitem[Arnaud (1996)]{Arnaud1996}Arnaud K. A., 1996, in Jacoby G. H., Barnes J., eds, Astronomical Society of the Pacific Conference Series Vol. 101, Astronomical Data Analysis Software and Systems V. p. 17
\bibitem[Armas Padilla et al. (2017)]{ArmasPadilla2017}Armas Padilla, M., Ueda, Y., Hori, T., Shidatsu, M., Munoz-Darias, T., 2017, MNRAS, 467, 290
\bibitem[Ballantyne \& Strohmayer (2004)]{Ball2004}Ballantyne, D. R., \& Strohmayer, T. E. 2004, ApJL, 602, L105

\bibitem[Blumenthal(1974)]{1974ApJ...188..121B} Blumenthal, G.~R.\ 1974, ApJ, 188, 121. doi:10.1086/152693





\bibitem[Bult et al. (2021)]{Bult2021}Bult, P., Altamirano, D., Arzoumanian, Z. et al. 2021, ApJ, 920, 59
\bibitem[Chen et al. (2012)]{Chen2012} Chen, Y. P., Zhang, S., Zhang, S. N., et al. 2012, ApJL, 752, 34
\bibitem[Chen et al. (2019)]{Chen2019} Chen, Y. P., Zhang, S., Zhang, S. N., et al. 2019, Journal of High Energy Astrophysics, 24, 23
\bibitem[Chen et al. (2022)]{Chen2022} Chen, Y. P., Zhang, S., Ji, L., Zhang, S. N., et al. 2022, ApJL, 936, L12
\bibitem[Chen et al. (2022a)]{Chen2022a} Chen, Y. P., Zhang, S., Ji, L., Zhang, S. N., et al. 2022a, ApJ, 936, 46
\bibitem[Chen et al. (2023)]{Chen2022b} Chen, Y. P., Zhang, S., Ji, L., Zhang, S. N., et al. 2023, ApJ, 942, 97
\bibitem[Cumming (2004)]{Cumming}Cumming, A. \ 2004, Nucl. Phys. B Proc. Suppl., 132, 435
\bibitem[Degenaar et al.(2015)]{Degenaar2015}Degenaar, N., Miller, J. M., Chakrabarty, D., Harrison, F. A., Kara, E., \& Fabian, A. C. 2015, MNRAS, 451, L85
\bibitem[Degenaar et al.(2018)]{Degenaar2018}Degenaar, N., Ballantyne, D. R., Belloni, T., et al. 2018, SSRv, 214, 15
\bibitem[Done et al. (2007)]{Done2007}Done, C., Gierliński, M., Kubota, A. 2007, A\&AR, 15, 1
\bibitem[Fragile et al.(2020)]{Fragile2020}Fragile, P. C., Ballantyne, D. R., \& Blankenship, A. 2020, NatAs, 4, 541

\bibitem[Galloway et al.(2008)]{Galloway}Galloway, D. K., Muno, M. P., Hartman, J. M., et al. \ 2008, ApJS, 179, 360
\bibitem[Galloway, $\ddot{o}$zel \& Psaltis (2008a)]{Galloway2008a}Galloway, D. K., $\ddot{o}$zel, F., Psaltis, D. 2008a, MNRAS, 387, 268
\bibitem[Galloway et al. (2020)]{Galloway2020}Galloway, D. K., In’t Zand, J., Chenevez, J., et al. 2020, ApJS, 249, 32
\bibitem[Galloway \& Keek (2021)]{Galloway2021}Galloway, D. K., \& Keek, L. 2021, Astrophys. Space Sci. Lib., 461, 209
\bibitem[Grindlay et al. (1980)]{Grindlay1980}Grindlay, J. E., Marshall, H. L., Hertz, P et al. 1980, ApJL, 240, L121

\bibitem[G{\"u}ver et al.(2012)]{Guver2012}G{\"u}ver, T., Psaltis, D., \& Zel, F. 2012, ApJ, 747, 76
\bibitem[G{\"u}ver et al.(2022a)]{2022MNRAS.510.1577G} G{\"u}ver, T., Boztepe, T., Ballantyne, D.~R., et al.\ 2022a, MNRAS, 510, 1577.
\bibitem[G{\"u}ver et al.(2022b)]{2022ApJ...935..154G} G{\"u}ver, T., Bostanc{\i}, Z.~F., Boztepe, T., et al.\ 2022b, ApJ, 935, 154

\bibitem[He \& Keek (2016)]{He2006}He, C.-C. \& Keek, L. 2016, ApJ, 819, 47
\bibitem[in't Zand et al. (2013)]{int2013}in't Zand, J. J. M., Galloway, D. K., Marshall, H. L., et al. 2013, A\&A,553, A83
\bibitem[Jaisawal et al. (2019)]{Jaisawal2019}Jaisawal, G. K., Chenevez, J., Bult, P., et al. 2019, ApJ, 883, 61


\bibitem[Ji et al. (2013)]{Ji2013}Ji, L., Zhang, S., Chen, Y. P., et al. 2013, MNRAS, 432, 2773
\bibitem[Ji et al. (2014)]{Ji2014}Ji, L., Zhang, S., Chen, Y. P., et al., 2014, ApJ, 791, L39


\bibitem[Kaastra \& Bleeker (2016)]{Kaastra2016}Kaastra, J. S.; Bleeker, J. A. M. 2016, A\&A, 587, A151
\bibitem[Kajava et al. (2014)]{Kajava2014}Kajava, J. J. E., Nättilä, J., Latvala, O. M., et al. 2014, MNRAS, 445, 4218
\bibitem[Kashyap et al. (2022)]{Kashyap2022}Kashyap, U., Ram, B., Guver, T., Chakraborty, M. 2022, MNRAS, 509, 3989

\bibitem[Kashyap et al.(2022)]{2022MNRAS.512.6180K} Kashyap, U., Chakraborty, M., \& Bhattacharyya, S.\ 2022, MNRAS, 512, 6180. doi:10.1093/mnras/stac908

\bibitem[Keek et al. (2014)]{Keek2014}Keek, L., Ballantyne, D. R., Kuulkers, E., \& Strohmayer, T. E. 2014, ApJL, 797, L23
\bibitem[Kuulkers et al. (2003)]{Kuulkers2003}Kuulkers, E., den Hartog, P. R., in ’t Zand, J. J. M., et al. 2003, A\&A 399, 663
\bibitem[Lewin et al.(1993)]{Lewin}Lewin, W. H. G., van Paradijs, J., \& Taam, R. E. \ 1993, Space Sci. Rev., 62, 223



\bibitem[Li et al. (2020)]{Li2020}Li, X. B., Li, X. F., Tan, Y. et al. 2020, JHEA, 27, 64

\bibitem[London, Taam, \& Howard (1986)]{London1986} London, R. A., Taam, R. E., \& Howard, W. M. 1986, ApJ, 306, 170

\bibitem[Maccarone \& Coppi (2003)]{maccarone2003}Maccarone, T. J. \& Coppi, P. S. \ 2003, A\&A, 399, 1151
\bibitem[Muno et al. (2001)]{Muno}Muno, M. P., Chakrabarty, D., Galloway, D. K., \& Savov, P. 2001, ApJ, 553, L157
\bibitem[Poutanen et al. (2014)]{Poutanen}Poutanen, J., N$\ddot{ai}$ttil$\ddot{a}$, J., Kajava, J. J. E. et al. 2014, MNRAS, 442, 3777
\bibitem[Remillard et al. (2022)]{Remillard2022}Remillard, R. A., Loewenstein, M., Steiner, J. F. et al. 2022, AJ, 163, 130
\bibitem[S{\'a}nchez-Fern{\'a}ndez et al. (2020)]{Sanchez2020}S{\'a}nchez-Fern{\'a}ndez C., Kajava J. J. E., Poutanen J., et al., 2020, A\&A, 634, A58

 \bibitem[Robertson(1937)]{1937MNRAS..97..423R} Robertson, H.~P.\ 1937, MNRAS, 97, 423. doi:10.1093/mnras/97.6.423


\bibitem[Rybicki \& Lightman (2004)]{Rybicki2004}Rybicki, G. B., \& Lightman, A. P. 2004, Radiative Processes in Astrophysics (Weinheim: Wiley-VCH)
\bibitem[Shaposhnikov \& Titarchuk (2004)]{Shaposhnikov2004}Shaposhnikov, N., \& Titarchuk, L. 2004, ApJ, 606, L57
\bibitem[Shaposhnikov et al. (2003)]{Shaposhnikov2003}Shaposhnikov, N., Titarchuk, L., Haberl, F., 2003, ApJL, 593, L35
\bibitem[Speicher et al.(2022)]{2022MNRAS.509.1736S} Speicher, J., Ballantyne, D.~R., \& Fragile, P.~C.\ 2022, MNRAS, 509, 1736. doi:10.1093/mnras/stab3087

\bibitem[Strohmayer \& Bildsten(2006)]{Strohmayer} Strohmayer, T., \& Bildsten, L.   \ 2006, New views of thermonuclear bursts (Compact stellar X-ray sources), 113, 156
\bibitem[Suleimanov et al. (2011)]{Suleimanov2011}Suleimanov, V., Poutanen, J., Werner, K. 2011, A\&A, 527, A139
\bibitem[Suleimanov et al. (2012)]{Suleimanov2012}Suleimanov, V., Poutanen, J., Werner, K. 2012, A\&A, 545, A120
\bibitem[Suleimanov et al. (2018)]{Suleimanov2018}Suleimanov, V., Poutanen, J., Werner, K. 2018, A\&A, 619, A114

 \bibitem[Thompson et al.(2005)]{2005ApJ...634.1261T} Thompson, T.~W.~J., Rothschild, R.~E., Tomsick, J.~A., et al.\ 2005, ApJ, 634, 1261. doi:10.1086/497104

 \bibitem[Titarchuk \& Shaposhnikov  (2002)]{Titarchuk2002}Titarchuk, L. \& Shaposhnikov, N.  2002, ApJL, 570, L25
 \bibitem[Titarchuk    (1994)]{Titarchuk1994}Titarchuk, L. 1994, ApJ, 429, 340
\bibitem[Walker \& Meszaros(1989)]{1989ApJ...346..844W} Walker, M.~A. \& Meszaros, P.\ 1989, ApJ, 346, 844. doi:10.1086/168065
\bibitem[Walker(1992)]{1992ApJ...385..642W} Walker, M.~A.\ 1992, ApJ, 385, 642. doi:10.1086/170969

\bibitem[Wilms et al. (2000)]{Wilms2000}Wilms, J., Allen, A., \& McCray, R. 2000, ApJ, 542, 914
\bibitem[Worpel et al. (2013)]{Worpel2013}Worpel, H., Galloway, D. K., \& Price, D. J. 2013, ApJ, 772, 94
\bibitem[Worpel et al. (2015)]{Worpel2015}Worpel, H., Galloway, D. K., \& Price, D. J. 2015, ApJ, 801, 60
\bibitem[Zdziarski et al. (2020)]{Zdziarski2020}Zdziarski, A. A., Szanecki, M., Poutanen, J., Gierlinski, M., \& Biernacki, P. 2020, MNRAS, 492, 5234
\bibitem[Zhang et al. (2000)]{Zhang2000}Zhang, S. N., et al., 2000, Science, 287, 1239
\bibitem[Zhang et al. (2020)]{Zhang2020}Zhang, S.-N., Li, T.-P., Lu, F.-J., et al. 2020, SCPMA, 63, 249502


\end{thebibliography}


\begin{table}[ptbptbptb]\tiny
\begin{center}
\caption{The bursts' parameters  of 4U~1608--52  detected by Insight/HXMT in 2022 outburst: obsid, peak time $t_{\rm p}$, fluence   $E_{\rm b}^{bb}$,  peak flux $F_{\rm p}^{bb}$  and flux $F_{\rm d}^{bb}$ at the touch-down time derived from the blackbody model; peak flux  $F_{\rm p}^{fa}$, flux $F_{\rm d}^{fa}$ at the touch-down time  and maximum $f_{\rm a}$ ($f_{\rm a}^{\rm p}$, most locate around the middle PRE phase) derived from the $f_{a}$  model, burst duration $\tau$ from fitting the decay part (after the peak time for the non-PRE bursts, and after the touch-down time for the PRE bursts) of the time evolution of the burst flux derived from  the blackbody model.
}
\label{tb_burst}
\begin{tabular}{ccccccccccc}
\\\hline
  No & obsid & $t_{\rm p}$  & $E_{\rm b}^{bb}$ & $F_{\rm p}^{bb}$ & $F_{\rm d}^{bb}$ & $F_{\rm p}^{fa}$ & $F_{\rm d}^{fa}$ &$f_{\rm a}^{\rm p}$& $\tau$  & PRE \\\hline
   &    	&  MJD   & a  &  b & b  & b   & b & & s  & \\\hline
1&P040420700407-20220718-02-01&59778.488728	   &	$92.1\pm1.0$ 	&$12.97_{-0.57}^{+0.57}$ 	&$12.16_{-0.55}^{+0.55}$ 	&$11.97_{-0.66}^{+0.67}$ 	&$10.42_{-0.70}^{+0.71}$  &$2.38_{-0.40}^{+0.40}$	&$4.1\pm0.1$  &Y \\\hline
2&P040420700504-20220720-01-01&59780.631621	   & $214.1\pm1.2$	&$17.61_{-0.65}^{+0.65}$ 	&$14.33_{-0.61}^{+0.61}$ 	&$16.08_{-0.81}^{+0.81}$ 	&$15.29_{-0.75}^{+0.75}$  &$3.66_{-0.83}^{+0.82}$	&$7.6\pm0.2$  &Y \\\hline
$^{\#}$3&P040420700601-20220721-01-01&	59781.876482	&$110.6\pm1.4$ 	&$17.23_{-0.98}^{+0.97}$        & $14.49_{-0.86}^{+0.86}$        & -       & - & -	&$4.1\pm0.1$  &Y \\\hline
4&P040420700601-20220721-01-01&59781.936919	&$28.5\pm0.7$ 	&$4.49_{-0.36}^{+0.36}$            & -        & -      & -        & -	&$6.0\pm0.5$  &N \\\hline
$^{*}$5&P040420700601-20220721-01-01&	59782.052206	&$46.3\pm1.3$ 	&$10.21_{-0.79}^{+0.80}$           & -    & -          & -        & - 	&$2.0\pm0.1$   &N \\\hline
6&P040420700701-20220724-01-01&	59784.257822	  &$128.7\pm1.1$	&$11.90_{-0.56}^{+0.56}$ 	&$10.18_{-0.52}^{+0.52}$ 	&$10.79_{-0.70}^{+0.71}$ 	&$9.56_{-0.60}^{+0.60}$ & $2.86_{-0.81}^{+0.80}$	&$8.2\pm0.3$  &Y \\\hline
7&P040420700702-20220724-01-01&	59784.384258	 &$61.1\pm0.8$	&$10.42_{-0.51}^{+0.51}$ 	&$9.27_{-0.49}^{+0.49}$ 	&$9.95_{-0.66}^{+0.66}$ 	&$8.95_{-0.59}^{+0.60}$ &$2.33_{-0.61}^{+0.59}$	&$3.5\pm0.1$  &Y \\\hline
8&P040420700901-20220727-01-01&	59787.821566	 &$133.6\pm1.0$	&$17.52_{-0.67}^{+0.67}$ 	&$16.12_{-0.65}^{+0.65}$ 	&$15.98_{-0.82}^{+0.82}$ 	&$15.18_{-0.75}^{+0.75}$ &$6.21_{-1.23}^{+1.22} $	&$4.5\pm0.1$  &Y \\\hline
$^{*}$9&P040420701503-20220809-01-01&	59800.658380	 &$198.9\pm2.4$  	&$16.87_{-5.77}^{+10.34}$       &$10.14_{-0.74}^{+0.75}$         & -      & -    & -	&$12.4\pm0.8$  &Y \\\hline
\end{tabular}
\end{center}
\begin{list}{}{}
\item[a]{In units of $10^{-8}~{\rm erg}~{\rm cm}^{2}$}
\item[b]{In units of $10^{-8}~{\rm erg}~{\rm cm}^{2}~{\rm s}^{-1}$}
\item[$^{*}$]{The bursts absented from LE data}
\item[$^{\#}$]{The bursts absented from ME data}

\end{list}
\end{table}

\begin{table}\tiny
\centering
\caption{The results of  spectral fit of the   persistent emission of burst \#1--\#8  with  cons*tbabs*thcomp*diskbb }
\label{tb_fit_thcomp_diskbb}
\begin{tabular}{cccccccccccc}
\\\hline
No &Time   & $\tau$ & $kT_{\rm e}$  & $f_{\rm sc}$ & $kT_{\rm disk}$ & $R_{\rm disk}$ & $F_{\rm disk}$ &$F_{\rm corona}$ &$F_{\rm total}$ & $\chi_{\nu}^{2}$ (d.o.f.)\\
   & MJD  &  & keV & &keV & km & $10^{-9}$  & $10^{-9}$  & $10^{-9}$ & \\
\hline
 1  & 59778.52   & $10.1_{-0.1}^{+0.2}$ & $2.92_{-0.04}^{+0.03}$ & $1.00_{-0.1}^{}$ & $0.44_{-0.04}^{+0.05}$ & $43.7_{-6.8}^{}$ & $7.84_{-0.02}^{+0.02}$ & $10.04_{-0.05}^{+0.05}$ & $17.88_{-0.04}^{+0.04}$ & $1.45(38)$ \\\hline
2  & 59780.43   & $9.0_{-0.4}^{+0.7}$ & $3.18_{-0.12}^{+0.09}$ & $0.71_{-0.1}^{+0.1}$ & $0.54_{-0.06}^{+0.06}$ & $28.5_{-4.7}^{+8.9}$ & $8.11_{-0.02}^{+0.02}$ & $6.15_{-0.04}^{+0.04}$ & $14.26_{-0.04}^{+0.04}$ & $0.81(39)$ \\\hline
3-5  & 59781.80   & $10.1_{-0.1}^{+0.1}$ & $2.99_{-0.03}^{+0.02}$ & $1.00_{-0.1}^{}$ & $0.36_{-0.07}^{+0.06}$ & $57.5_{-0.2}^{}$ & $6.30_{-0.01}^{+0.01}$ & $9.02_{-0.03}^{+0.03}$ & $15.33_{-0.03}^{+0.03}$ & $1.24(41)$ \\\hline
6-7  & 59784.62   & $10.8_{-0.8}^{+1.0}$ & $2.92_{-0.05}^{+0.05}$ & $0.48_{-0.1}^{+0.1}$ & $0.61_{-0.04}^{+0.04}$ & $19.9_{-2.0}^{+2.8}$ & $6.29_{-0.02}^{+0.02}$ & $3.91_{-0.03}^{+0.03}$ & $10.19_{-0.03}^{+0.03}$ & $0.43(38)$ \\\hline
8  & 59787.69   & $10.2_{-0.3}^{+0.6}$ & $3.06_{-0.09}^{+0.06}$ & $0.53_{-0.1}^{+0.1}$ & $0.48_{-0.02}^{+0.04}$ & $29.1_{-4.4}^{+3.6}$ & $5.15_{-0.01}^{+0.01}$ & $3.72_{-0.03}^{+0.03}$ & $8.87_{-0.02}^{+0.02}$ & $0.65(39)$ \\\hline
\end{tabular}
\begin{list}{}{}
\item[a]{: The model parameters: the optical depth $\tau$,  the  electron temperature $kT_{\rm e}$,  the cover factor $f_{\rm sc}$,  the accretion disk  temperature   $kT_{\rm disk}$ and  the inner disk radius   $R_{\rm diskb}$ at a distance of 4 kpc and  an inclination angel of $\theta$=30$^{\circ}$, the bolometric flux of the diskbb $F_{\rm diskbb}$, the bolometric flux of the corona $F_{\rm corona}$, and the total  bolometric flux  $F_{\rm total}$ are in units of $10^{-9}~{\rm erg/cm}^{2}/{\rm s}$, reduced $\chi_{\nu}^{2}$ and the degree of freedom (d.o.f.).}
 \end{list}
\end{table}

\begin{table}\tiny
\centering
\caption{The results of  spectral fit of the   persistent emission of burst \#9  with  cons*tbabs*thcomp*bb }
\label{tb_fit_thcomp_bb}
\begin{tabular}{cccccccccccc}
\\\hline
No &Time   & $\tau$ & $kT_{\rm e}$  & $f_{\rm sc}$ & $kT_{\rm bb}$ & $R_{\rm bb}$ & $F_{\rm bb}$ &$F_{\rm corona}$ &$F_{\rm total}$ & $\chi_{\nu}^{2}$ (d.o.f.)\\
   & MJD  &  & keV & &keV & km & $10^{-9}$  & $10^{-9}$  & $10^{-9}$ & \\
\hline
9  & 59800.42   & $0.7_{-0.1}^{+0.6}$ & $183.68_{-93.49}^{+57.42}$ & $1.00_{-0.3}^{}$ & $0.49_{-0.15}^{+0.11}$ & $11.3_{-2.3}^{+26.6}$ & $0.50_{-0.03}^{+0.08}$ & $0.98_{-0.05}^{+0.20}$ & $1.48_{-0.04}^{+0.18}$ & $1.55(41)$ \\\hline
\end{tabular}
\begin{list}{}{}
\item[a]{: The model parameters: the optical depth $\tau$,  the  electron temperature $kT_{\rm e}$,  the cover factor $f_{\rm sc}$,  the  blackbody temperature   $kT_{\rm bb}$ and  the blackbody radius   $R_{\rm bb}$ at a distance of 4 kpc, the bolometric flux of the blackbody $F_{\rm bb}$, the bolometric flux of the corona $F_{\rm corona}$, and the total  bolometric flux  $F_{\rm total}$ are in units of $10^{-9}~{\rm erg/cm}^{2}/{\rm s}$, reduced $\chi_{\nu}^{2}$ and the degree of freedom (d.o.f.).}
 \end{list}
\end{table}

\clearpage


\clearpage

\clearpage


\end{document}